\documentclass[12pt,a4paper]{article} 
\usepackage{amsmath}
\usepackage{amssymb}
\usepackage{empheq}
\usepackage{bm}
\usepackage{color}
\usepackage{cite} 

\usepackage{hyperref}
\usepackage{graphicx}

\catcode `@=11 \@addtoreset{equation}{section} \catcode `@=12

\def\Tr{{\rm Tr}}

\begin{document}
\setlength{\oddsidemargin}{0cm}

\begin{titlepage}

    \begin{center}
        {\LARGE
            Complex Langevin Method \\ 
            on Rotating Matrix Quantum Mechanics \\
            at Thermal Equilibrium}
    \end{center}
    \vspace{1.2cm}
    \baselineskip 18pt 
    \renewcommand{\thefootnote}{\fnsymbol{footnote}}

    \begin{center}

        Takehiro {\sc Azuma}$^{a}$\footnote{%
            E-mail address: azuma(at)mpg.setsunan.ac.jp
        }, 
        Takeshi {\sc Morita}$^{b,c}$\footnote{%
            E-mail address: morita.takeshi(at)shizuoka.ac.jp
        }, 
        Hiroki  {\sc Yoshida}$^{c}$\footnote{%
            E-mail address: yoshida.hiroki.16(at)shizuoka.ac.jp}
        
        \renewcommand{\thefootnote}{\arabic{footnote}}
        \setcounter{footnote}{0}
        
        \vspace{0.4cm}
        
        {\it
            a. Institute for Fundamental Sciences, Setsunan University, 17-8 Ikeda Nakamachi, Neyagawa, Osaka, 572-8508, Japan  
            \vspace{0.2cm}
            \\

            b. Department of Physics,
            Shizuoka University \\
            836 Ohya, Suruga-ku, Shizuoka 422-8529, Japan 
            \vspace{0.2cm}
            \\
            c. Graduate School of Science and Technology, Shizuoka University\\
            836 Ohya, Suruga-ku, Shizuoka 422-8529, Japan
        }

    \end{center}

    
    \vspace{1.5cm}
    
    \begin{abstract}
        Rotating systems in thermal equilibrium are ubiquitous in our world.
        In the context of high energy physics, rotations would affect the phase structure of QCD.
        However, the standard Monte-Carlo methods in rotating systems are problematic because the chemical potentials for the angular momenta (angular velocities) cause sign problems even for bosonic variables.
        In this article, we demonstrate that the complex Langevin method (CLM) may overcome this issue.
        We apply the CLM to the Yang-Mills (YM) type one-dimensional matrix model (matrix quantum mechanics) that is a large-$N$ reduction (or dimensional reduction) of the $(D+1)$-dimensional U$(N)$ pure YM  theory  (bosonic BFSS model).
        This model shows a large-$N$ phase transition at finite temperature, which is analogous to the confinement/deconfinement transition of the original YM theory, and our CLM predicts that the transition temperature decreases as the angular momentum chemical potential increases.
        In order to verify our results, we compute several quantities via the minimum sensitivity method and find good quantitative agreements.
        Hence, the CLM properly works in this rotating system.
        We also argue that our results are qualitatively consistent with a holography and the recent studies of the imaginary angular velocity in QCD.
        As a byproduct, we develop an analytic approximation to treat the so-called  ``small black hole" phase in the matrix model.

    \end{abstract}
    
    
\end{titlepage}

\tableofcontents

\section{Introduction and Summary}

Rotating objects are ubiquitous in our nature. If the objects are macroscopic, they may obey thermodynamics.
However, in statistical mechanics, investigating rotating interacting many body systems at thermal equilibrium are problematic.
The chemical potential for the angular momentum, which is equivalent to the angular velocity for point particles,  makes the Euclidean actions complex even for bosonic systems, and the standard Monte Carlo method (MC) does not work because of the sign problem (see Sec.~\ref{sec-chemical}).
Hence overcoming this issue is a quite important challenge in theoretical physics.

There are various approaches to the sign problem, such as the Lefschetz-thimble method \cite{1205_3996,1303_7204} and the complex Langevin method (CLM) \cite{Parisi:1983mgm,Klauder:1983sp}. The Lefschetz-thimble method handles the sign problem by deforming the integration contour to mitigate the sign problem. The CLM, which we focus on in this paper, is a stochastic process for the complexified variables. The advantage of the CLM is that it allows us to study large systems. At first, the CLM has suffered from the problem that CLM results in a wrong result without noticing it. Recent studies \cite{1101_3270,1211_3709,1508_02377,1604_07717,1606_07627} have clarified the conditions that the CLM results converge to the correct result equivalent to the path integral. The sufficient condition to obtain the correct result is that the probability distribution of the drift norm falls off exponentially or faster (see Sec.~\ref{sec-CLM}). The drift norm is a byproduct of solving the Langevin equation numerically, which can be calculated with negligible extra CPU cost.

To test whether the CLM works in rotating systems, we study the U($N$) matrix quantum mechanics whose Euclidean action is given by \cite{deWit:1988wri, Banks:1996vh},
\begin{align}
    \label{action-BFSS}
    S
    =
    \int_0^{\beta} \hspace{-2mm} dt  
    \Tr 
    \Biggl\{ 
    \sum_{I=1}^D
    \frac{1}{2}
    \left(
    D_t X^I \right)^2
    -
    \sum_{I,J=1}^D \frac{g^2}{4} [X^I,X^J]^2
    \Biggr\}. 
\end{align}
Here $X^I(t)$ ($I=1,\cdots,D$) are $N \times N$ hermitian matrices.
$D_t:= \partial_t -i [A_t,\,]$ is a covariant derivative and $A_t(t)$ is the gauge field of the U($N$). 
$g$ is a coupling constant, and we take the 't Hooft limit $N \to \infty$ and $g\to 0$ with a fixed 't Hooft coupling $\lambda := g^2N$ \cite{tHooft:1973alw}.
The model is invariant under the local U($N$) gauge transformation $X^I \to U X^I U^\dagger $ and $D_t \to U D_t U^\dagger $.
Since we will study the system at finite temperature, we have introduced the Euclidean time $t$ and the inverse temperature $\beta=\frac{1}{T}$. 
Note that $A_t$ is a vector and $X^I$ are scalars in this one dimension.

This model has an SO($D$) rotation symmetry $X^I \to {O^I}_J X^J$, and the angular momentum is conserved.
Then, we analyze this model at thermal equilibrium with a finite angular momentum chemical potential through the CLM.
Although the model is strongly coupled and the standard perturbative analysis does not work, the model without rotation has been analyzed through the MC method \cite{Narayanan:2003fc, Aharony:2004ig, Aharony:2005ew, Kawahara:2007fn, Azeyanagi:2009zf, Azuma:2012uc, Azuma:2014cfa, Filev:2015hia, Hanada:2016qbz, Bergner:2019rca, Asano:2020yry, Watanabe:2020ufk, Dhindsa:2022uqn} and other non-perturbative methods such as the minimum sensitivity \cite{Kabat:1999hp, Hashimoto:2019wmg, Morita:2020liy, Brahma:2022ifx} and the $1/D$-expansion \cite{Azuma:2012uc, Azuma:2014cfa, Hotta:1998en, Mandal:2009vz, Morita:2010vi, Mandal:2011hb}.
Particularly, the minimum sensitivity and the $1/D$-expansion would work even in the presence of the angular momentum chemical potential.
Thus, we can test the CLM by comparing these methods. 
This is one advantage of studying this model.
(Indeed, some earlier results through the $1/D$-expansion have been reported in Ref.~\cite{Morita:2010vi}.)
Actually, we will show that the CLM reproduces the results of the minimum sensitivity quantitatively, when the chemical potential is not so large.
(We investigate the model at $D=9$ and $D=16$, and we find that the minimum sensitivity is slightly better than the $1/D$-expansion there.)
Although the CLM does not provide reliable results at larger chemical potentials, it is enough to observe non-trivial phase transitions.
Hence, the CLM partially overcomes the sign problem of the angular momentum chemical potential in the model \eqref{action-BFSS}.
As far as we know, this is the first example that quantum many body systems in thermal equilibrium with a finite angular momentum chemical potential are solved through the first principle computation.

Before going to explain the details of our analysis, we present the importance of the model \eqref{action-BFSS}.
This model appears in various contexts of high energy physics.
Here we list some of them related to the current work:
\begin{itemize}
\item
This model is a large-$N$ reduction \cite{Eguchi:1982nm} (or dimensional reduction \cite{LUSCHER1983233}) of the $(D+1)$-dimensional U($N$) pure Yang-Mills (YM) theory to one dimension.
In this view, $X^I$ are the dimensional reductions of the spatial components of the original $(D+1)$-dimensional gauge fields.
It is known that this model in the large-$N$ limit is confined at low temperatures and shows a confinement-deconfinement transition at finite temperature \cite{Narayanan:2003fc, Aharony:2004ig, Mandal:2009vz, Kabat:1999hp}.
This phase structure is similar to that of the original YM theory.
Hence, this model is important as a toy model of the original YM theory.
Particularly, if we turn on the angular momentum chemical potential in our model, it may be regarded as a toy model of a rotating pure gluon system \cite{Chernodub:2020qah, Yamamoto:2013zwa, Braguta:2021jgn, Chen:2022smf, Chernodub:2022wsw, Chernodub:2022veq}, which is actively being studied to reveal natures of neutron stars and quark gluon plasma in relativistic heavy ion colliders \cite{STAR:2017ckg}.

\item 
This model at $D=9$ appears as a low energy effective theory of $N$ supersymmetric D-particles \cite{Banks:1996vh} on Euclidean $S_\beta^1 \times S^1 \times R^8$ in type IIA superstring theory \cite{Aharony:2004ig}.\footnote{	Here $S_\beta^1 $ is the temporal direction of the model \eqref{action-BFSS} and another $  S^1 $ is a Scherk-Schwarz circle whose radius is taken large.
This system is described by a supersymmetric version of the model \eqref{action-BFSS}, but the Scherk-Schwarz circle breaks the supersymmetry and makes the fermions on the D-particles massive. When the radius of the Scherk-Schwarz circle is taken large, the fermion's masses become large and they can be ignored at low energy.
Note that the diagonal components of $X^I$ represent the positions of the D-particles in the  $ S^1 \times R^8$ space, and the scalar for the Scherk-Schwarz circle, say $X^1$, is not distinguishable to other $X^I$ due to the large circle limit. 
Then, the low energy effective theory becomes the model \eqref{action-BFSS}.
}
In this picture, a gravity dual \cite{Maldacena:1997re, Itzhaki:1998dd} of the model \eqref{action-BFSS} is given by black brane geometries.
There, the aforementioned confinement/deconfinement transition corresponds to a Gregory-Laflamme (GL) transition \cite{Aharony:2004ig, Gregory:1994bj, Mandal:2011ws}.

\item This model is also a toy model of $N$ D-particles, which may be regarded as a microscopic description of a black hole \cite{Klebanov:1997kv}.\footnote{The conventional D-particles in superstring theory are described by the supersymmetric version of the model \eqref{action-BFSS} at $D=9$, and it is called the Banks-Fischler-Shenker-Susskind (BFSS) theory \cite{Banks:1996vh}.
Since the model \eqref{action-BFSS} is a bosonic version of the BFSS theory, it is called the bosonic BFSS model.}
In this picture, the diagonal components of the matrix $X_{ii}^I$ ($i=1,\cdots, N$) represent the position of the $i$-th D-particle on the $I$-th coordinate and the off-diagonal components $X_{ij}^I$ ($i \neq j$) represent the open strings connecting the $i$-th and $j$-th D-particles, which induce interactions between the D-particles.
In this way, the model \eqref{action-BFSS} describes quantum mechanics of $N$ interacting D-particles in the $D$-dimensional space.
The interactions cause attractive forces between the D-particles and they compose a bound state. This bound state is chaotic and is regarded as a toy model of a black hole.\footnote{
\label{ftnt-SYM}
Indeed, it is known that the dynamics of the model \eqref{action-BFSS} is similar to that of the ${\mathcal N} =4$ supersymmetric YM theory (SYM) on  $S_\beta^1 \times S^3 $.
For example, this SYM theory shows a large-$N$ confinement/deconfinement phase transition related to the model \eqref{action-BFSS} \cite{Sundborg:1999ue, Aharony:2003sx}.
Particularly, the SYM theory at strong coupling has a gravity dual given by AdS geometries \cite{Witten:1998zw, Aharony:1999ti}. 
There, at low temperature, a thermal AdS geometry is stable while an AdS black hole geometry is favored at high temperature, and a phase transition between these two geometries is called the Hawking-Page transition \cite{Hawking:1982dh}.
These geometries correspond to the confinement and deconfinement phases in the SYM theory, and hence they are also related to those of the model \eqref{action-BFSS}. 
Note, however, that these phase structures differ from the supersymmetric D-particles in the type-IIA superstring theory \cite{Banks:1996vh}. In particular, they do not show confinement at low temperatures \cite{Itzhaki:1998dd}.
}
Hence, when the system rotates, it may correspond to a rotating black hole.
\end{itemize}
In this article, we mainly focus on the last D-particle picture, since $X_{ii}^I$ represents the particle position and the angular momenta for these particles are easily understood intuitively.

\subsection{Summary of this work}

We summarize our findings on the model \eqref{action-BFSS} through the CLM and the minimum sensitivity analysis.
\begin{itemize}
    \item We study the model with a finite angular momentum chemical potential through the CLM, and the results agree with the minimum sensitivity analysis quantitatively, as far as the chemical potential is not so large.
          It indicates that the CLM works properly in the rotating system.
          We also observe that the transition temperature for the confinement/deconfinement transition decreases as the chemical potential increases. See Sec.~\ref{sec-result}.
          This decreasing critical temperature is consistent with the previous studies in rotating black holes in holography \cite{Gubser:1998jb, Chamblin:1999tk, Cvetic:1999ne, Hawking:1999dp, Gubser:2000mm, Basu:2005pj, Yamada:2006rx} and rotating pure gluons \cite{Chen:2022smf, Chernodub:2022veq}. 
          
    \item We develop an approximation method in the minimum sensitivity analysis, which enables us to explore the so-called ``small black hole" solution \cite{Aharony:1999ti}.\footnote{
              Analyses of the model \eqref{action-BFSS} through the minimum sensitivity were also done in Refs.~\cite{Kabat:1999hp} and \cite{Morita:2020liy}.
              Particularly, Ref.~\cite{Morita:2020liy} confirmed the existence of the small black hole solution in the model.
              However, this study merely pointed out the existence of the solution, and the solution itself was not derived.
              They also could not derive the gapped solution, which corresponds to a large black hole.
              The derivation of these solutions through the new approximation is one of our results in this article.}
          This solution has a negative specific heat similar to Schwarzschild black holes, and is important in the context of quantum gravity.
          See, for example, Fig.~\ref{fig-D=9}.
              
    \item
          By using the minimum sensitivity, we study the properties of the model with an imaginary angular momentum chemical potential, which has been employed to evade the sign problem in the MC computations \cite{Yamamoto:2013zwa, Braguta:2021jgn, Chen:2022smf, Chernodub:2022wsw, Chernodub:2022veq}.
          We find a stable confinement phase at high temperature when $D=3$.
          This is consistent with the recent study in the four-dimensional pure Yang-Mills theories \cite{Chen:2022smf}.
          We also argue a condition for the existence of the stable high temperature confinement phase in our model \eqref{action-BFSS}.
          Besides, we compute the critical temperature for the real chemical potential through the analytic continuation of the imaginary chemical potential, and find that the analytic continuation quantitatively works for finite chemical potentials.
          See Sec.~\ref{sec-Im}.

\end{itemize}

This paper is organized as follows. In Sec.~\ref{sec-review}, we review the thermodynamical properties of the model \eqref{action-BFSS} without angular momentum (readers familiar with this topic can skip this section). In Sec.~\ref{sec-chemical}, we introduce angular momentum to the model \eqref{action-BFSS}, and show that the model has a sign problem. 
In Sec.~\ref{sec-result}, we present our main results that the CLM successfully predicts non-trivial phase structure of the model \eqref{action-BFSS}, which agrees with the minimum sensitivity analysis.
In Sec.~\ref{sec-CLM}, we present the details of the application of the CLM to the model.
In Sec.~\ref{sec-minimum}, we explain the minimum sensitivity analysis.
In Sec.~\ref{sec-Im}, we study the imaginary chemical potential in our model by using the minimum sensitivity, and compare the results obtained from four-dimensional YM theories.
Sec.~\ref{Sec_discussion} is devoted to a discussion.

In this article, we take an unit $c=\hbar=k_B=1$. 
For numerical computations, we take $\lambda=1$.

\section{Review of the previous results on the non-rotating model}
\label{sec-review}

\begin{figure}
    \begin{center}
        \begin{tabular}{ccc}
            \begin{minipage}{0.33\hsize}
                \begin{center}
                    \includegraphics[scale=0.5]{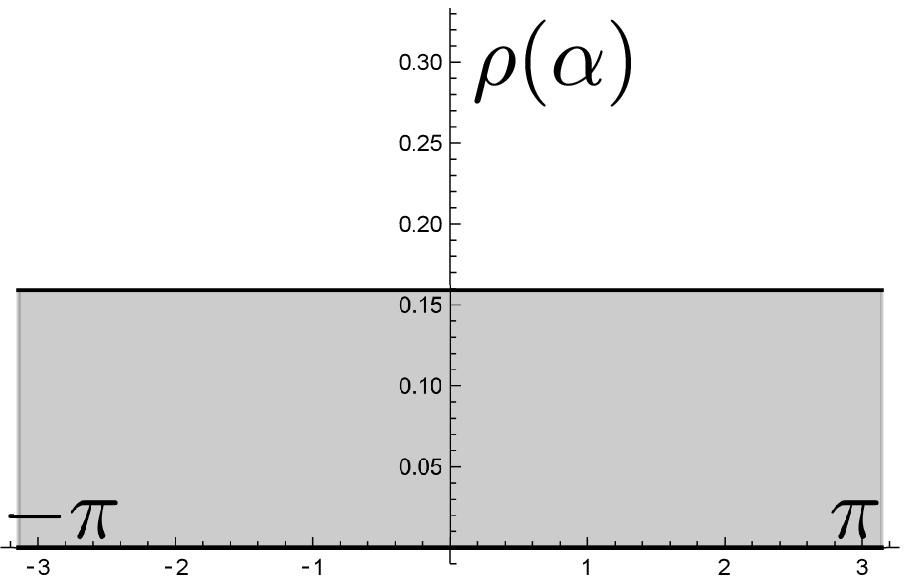}\\
                    uniform distribution
                \end{center}
            \end{minipage}
            \begin{minipage}{0.33\hsize}
                \begin{center}
                    \includegraphics[scale=0.5]{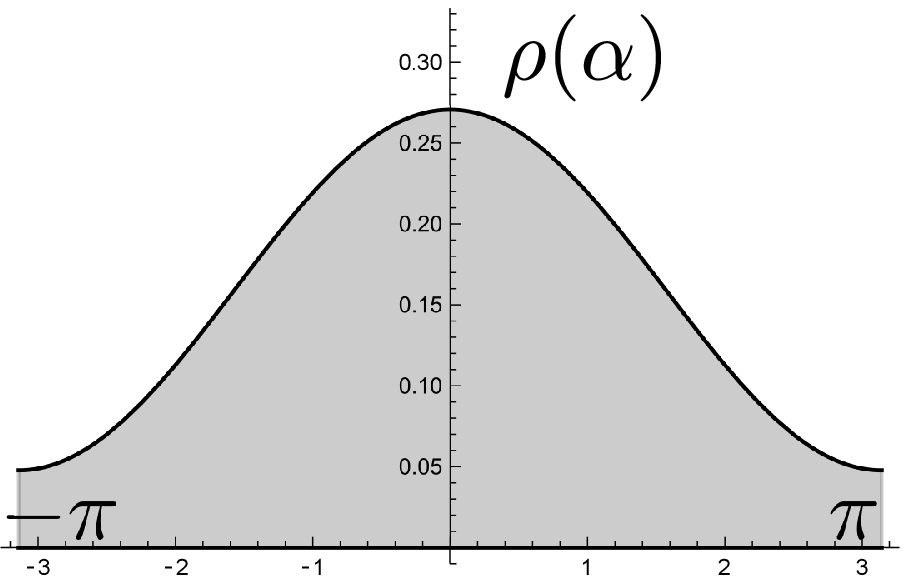}\\
                    non-uniform distribution
                \end{center}
            \end{minipage}
            \begin{minipage}{0.33\hsize}
                \begin{center}
                    \includegraphics[scale=0.5]{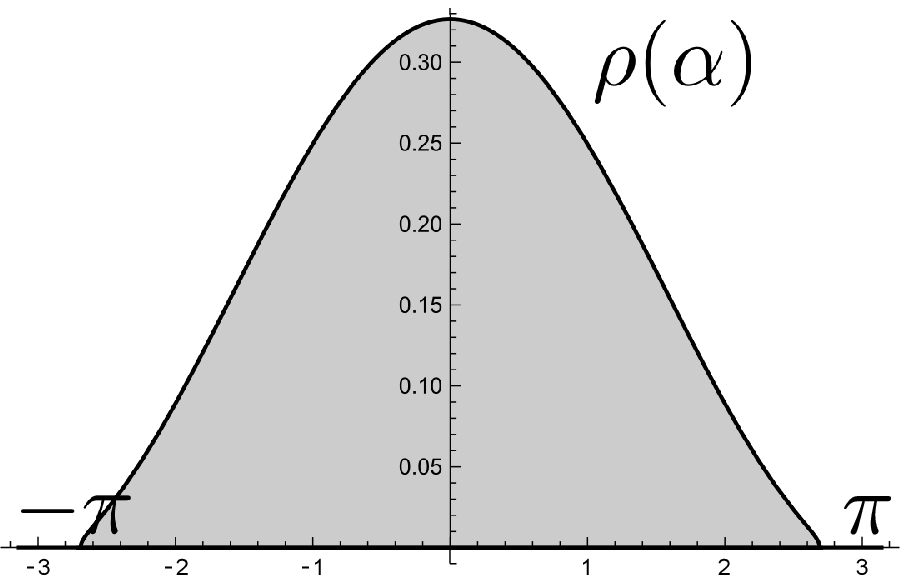}\\
                    gapped distribution
                \end{center}
            \end{minipage}
        \end{tabular}
        \caption{
            Three typical eigenvalue distributions $\rho(\alpha)$ of $A_t$ \eqref{rho}.
            The uniform distribution (left) is favored at low temperatures, while the gapped distribution (right) is favored at high temperatures.
            The non-uniform distribution might appear at middle temperatures.
        }
        \label{fig-rho}
    \end{center}
\end{figure}

In this section, we briefly review the properties of the model \eqref{action-BFSS} without rotation. 

At finite temperatures, the model has a large-$N$ confinement/deconfinement transition.
The order parameters of this transition are the Polyakov loop operators
\begin{align}
    u_n := \frac{1}{N} \Tr {\mathcal P} \exp\left( i  \int_0^{n \beta } dt A_t \right).
    \label{polyakov}
\end{align}
If $\langle u_n \rangle =0$ (${}^\forall n$), it indicates a confinement, and $\langle u_n \rangle \neq 0$ (${}^\exists n$) signals a deconfinement. 

In order to investigate the phase transition, it is useful to take the static diagonal gauge
\begin{align}
    \beta A_t = {\rm diag}(\alpha_1,\alpha_2,\cdots, \alpha_N ).
    \label{gauge-diagonal}
\end{align}
Here $\alpha_k$ ($k=1,\cdots, N$) is independent of the Euclidean time $\partial_t \alpha_k=0$.
It also satisfies $\alpha_k=\alpha_k+2\pi$.
Then, the Polyakov loops $u_n$ become
\begin{align}
    u_n= \frac{1}{N} \sum_{k=1}^N e^{in \alpha_k} .
    \label{polyakov_static_diag}
\end{align}
We also introduce the density function for $\{ \alpha_k \}$,
\begin{align}
    \rho(\alpha)=\frac{1}{N} \sum_{k=1}^N \delta(\alpha-\alpha_k)=\frac{1}{2\pi} \sum_{n \in {\mathbf Z}} u_n e^{in\alpha}, \qquad (-\pi \le \alpha \le \pi).
    \label{rho}
\end{align}
As we see soon, the profile of this function characterizes the phases of our system.

Roughly speaking, the phase structure of this system can be explained as follows.
If the scalars $X^I$ are not present, ``repulsive forces'' work between $\{ \alpha_k \}$, and they uniformly distribute on the configuration space that is a circle ($\alpha_k=\alpha_k+2\pi$).
Then, the density function $\rho(\alpha)$ becomes $\rho(\alpha)=1/2\pi$ as plotted in Fig.~\ref{fig-rho} (left).
We call this solution a uniform solution.\footnote{As we will see, in the minimum sensitivity analysis, phases of the model \eqref{action-BFSS} are obtained as saddle point solutions of an effective action. Hence, we may call phases ``solutions".
}
From Eq.~\eqref{rho}, $u_n=0$ is satisfied, and the system is in a confinement phase. 

Once we turn on the scalars $X^I$, they induce ``attractive forces" between $\{ \alpha_k \}$, which strengthen as temperature increases.
Thus, while the system remains confined at sufficiently low temperatures, the attractive forces would overcome the repulsive forces at high temperatures, and $\{ \alpha_k \}$ collapse to a cluster.
Then, the density function is depicted as shown in Fig.~\ref{fig-rho} (right).
There, a gap exists in the distribution of $\{ \alpha_k \}$ around $\alpha=\pi$, and this solution is called a gapped solution.\footnote{In this solution, we have taken a gauge that the peak of the density $\rho(\alpha)$ is at $\alpha=0$.
    Then, all $u_n$ would become real.
    We use this gauge throughout our minimum sensitivity analysis.
    However, in the MC and CLM, taking this gauge is difficult, and we evaluate $\langle |u_n| \rangle$ instead of taking this gauge.
    Note that $\langle |u_n| \rangle$ and $\langle u_n \rangle$ in the gauge are different in general, but the deviations are highly suppressed at large $N$, and we do not observe any issue when we compare these two quantities. 
    \label{ftnt-gauge}}
In this solution, $u_n \neq 0$ and it corresponds to a deconfinement phase.

There is another solution that connects the uniform solution and the gapped one as shown in Fig.~\ref{fig-rho} (center).
In this solution, any gap does not exist, and it is called a non-uniform solution.
We can easily check that $u_1 \neq 0$ in this configuration, and it is in a deconfinement phase.
(Hence, we have two deconfinement phases in large-$N$ gauge theories.)
Between these three solutions, large-$N$ phase transitions may occur.
Note that, whether the non-uniform solution arises as a stable phase depends on the dynamics of the system, and, for example, it is always unstable at $D=9$ as we will see soon.

In this way, the model \eqref{action-BFSS} has a confinement/deconfinement transition similar to higher-dimensional YM theories.
One feature of this transition is that the order of the transition would change depending on the dimension $D$ \cite{Azuma:2014cfa}. 
For small $D$, it is first order, while it would be second order for large $D$.
Indeed, MC computations have established that it is first order up to $D=25$ \cite{Azuma:2014cfa, Bergner:2019rca}.
On the other hand, the $1/D$-expansion predicts that the transition is second order at sufficiently large $D$ \cite{Mandal:2009vz}.
Also, the minimum sensitivity analysis at three-loop order indicates that it is first order up to $D=35$ and it becomes second order from $D=36$ \cite{Morita:2020liy}.
(Note that the minimum sensitivity analysis at two-loop order predicts the phase transition is always first order independently of $D$.
Also, it is not clear whether the three-loop computation is sufficiently reliable, and it has not been established if the order of the transition changes at $D=36$.)
Such a $D$-dependence of the transitions is also consistent with a holography.\footnote{
    As we mentioned in the introduction, the confinement/deconfineement transition of the model \eqref{action-BFSS} is related to a Gregory-Laflamme (GL) transition in gravity \cite{Gregory:1994bj}.
    Here, $D=9$ is taken and the dual gravity theory is in Euclidean $S_\beta^1\times S^1 \times R^{d}$ space with $d=8$.
    We can formally extend this correspondence to a general $D$ by taking $d=D-1$ \cite{Azuma:2014cfa}.
    Then, the order of the GL transition depends on the dimension $D$.
    It is first order for $D \le 11$ and it is second order for $D \ge 12$, and they are qualitatively similar to our matrix model results \cite{Sorkin:2004qq, Kudoh:2005hf}.
    Interestingly, a similar $D$-dependence has been observed in Rayleigh-Plateau instability in a fluid model, too \cite{Cardoso:2006ks, Miyamoto:2008rd}.
}

In this article, we will study the model with a finite angular momentum chemical potential through the CLM.
There, we investigate $D=9$ and $16$, where the first order transitions occur.
Hence we will employ the minimum sensitivity analysis at two-loop order rather than the $1/D$-expansion to test the CLM, since it also predicts the first order transitions.
(The minimum sensitivity analysis at three-loop order with the finite chemical potential is much more complicated than the two-loop analysis and we do not do it in this article.)

\begin{figure}
    \begin{center}
        \begin{tabular}{cc}
            \begin{minipage}{0.5\hsize}
                \begin{center}
                    \includegraphics[scale=0.95]{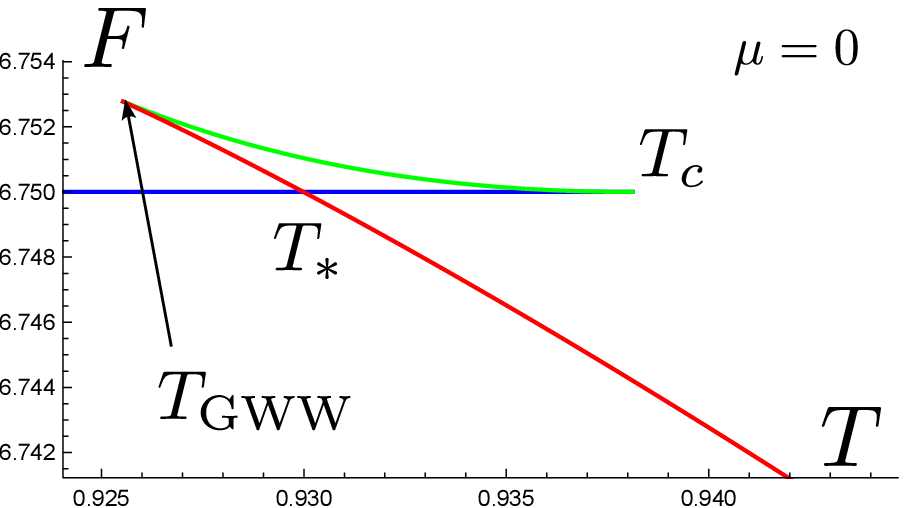}\\
                    $T$ vs. $F$
                \end{center}
            \end{minipage}
            \begin{minipage}{0.5\hsize}
                \begin{center}
                    \includegraphics[scale=0.8]{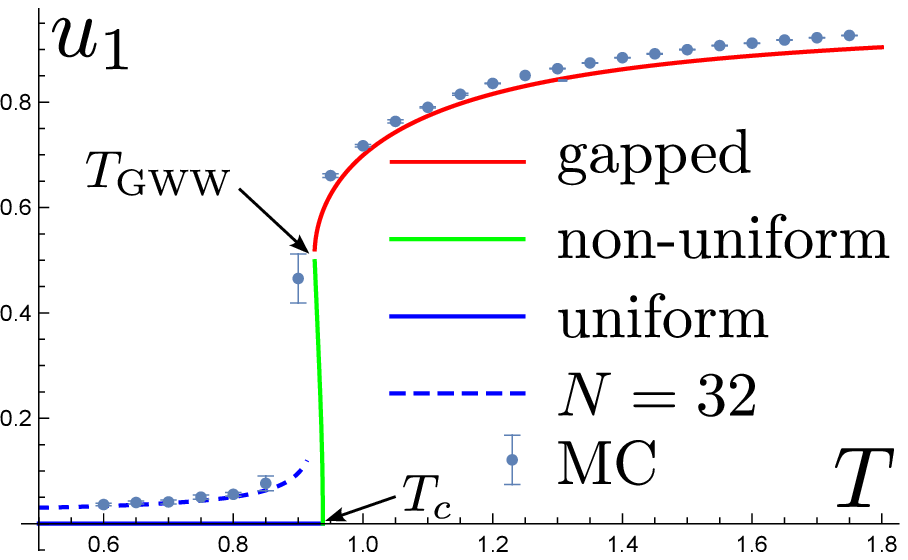}\\
                    $T$ vs. $u_1$
                \end{center}
            \end{minipage}
        \end{tabular}
        \caption{
            Temperature dependence of free energy and the Polyakov loop $u_1$ at $D=9$ at zero angular momentum chemical potentials  through the minimum sensitivity analysis at two-loop order.
            The solid curves represent the results at large $N$. 
            The blue, green and red curves are for the uniform, the non-uniform and the gapped solutions, respectively.
            (See Sec.~\ref{sec-free-energy} and Sec.~\ref{subsec-Polyakov}.)
            The dashed blue curve is the leading $1/N$ correction for $u_1$ in the confinement phase at $N=32$ \eqref{un-finite-N-main}. 
            The dots are results through MC at $N=32$, and we observe good quantitative agreements.
            The free energies of the three solutions indicate that a first-order phase transition occurs at $T=T_*$. 
        }
        \label{fig-D=9}
    \end{center}
\end{figure}

In order to test the minimum sensitivity analysis at two-loop order, we compare the results at zero chemical potential obtained through this analysis and MC (not CLM), and we find quantitative agreements.\footnote{The agreements become not so good as temperature increases. This is because we use an approximation \eqref{large-D-approximation}, and it is not reliable at higher temperatures.} (We will show the details of the minimum sensitivity analysis in Sec.~\ref{sec-minimum}.) See Fig.~\ref{fig-D=9} for $u_1$ at $D=9$.
Hence, we expect that the minimum sensitivity analysis at two-loop order may be reliable even with a finite chemical potential, and we use this method to test the CLM.

We also plot the temperature dependence of free energy through the minimum sensitivity at $D=9$ in Fig.~\ref{fig-D=9} (left).
It shows a first-order phase transition as we mentioned above.

%

\section{Introducing chemical potentials for angular momenta}
\label{sec-chemical}

We introduce angular momentum chemical potentials to the model \eqref{action-BFSS}.
For simplicity, we first consider the chemical potential in a single particle quantum mechanics in two dimensions.
We take the Hamiltonian of this system as
\begin{align} 
    H=\frac{1}{2} \left(
    p_x^2+p_y^2
    \right)+V(x,y).
\end{align}
Here, $x$ and $y$ are the two-dimensional position operators, and $p_x$ and $p_y$ are their conjugate momenta.
It is convenient to employ a complex coordinate $z:=(x+iy)/\sqrt{2}$.
Then, the angular momentum is given by
\begin{align} 
    J= xp_y-yp_x=i  \left(z p_z-\bar{z} \bar{p}_z \right).
\end{align}
Now we introduce the angular chemical potential $\mu$, and the Hamiltonian is modified as
\begin{align} 
    H-\mu J= & \frac{1}{2} \left(
    p_x^2+p_y^2
    \right) +V(x,y)- \mu J \nonumber                                                    \\
    =        & p_z \bar{p}_z -i \mu \left(z p_z-\bar{z} \bar{p}_z \right)+V(z,\bar{z}).
\end{align}
Then, the the Euclidean action for this Hamiltonian, which is used in the path-integral formalism at finite temperature, is derived as
\begin{align} 
    S=\int_0^\beta dt \left\{ \left( \dot{\bar{z}}  - \mu \bar{z}    \right)\left(  \dot{z}  + \mu z    \right) +V(z,\bar{z}) \right\} .
\end{align}
Here, the first term is complex.
Hence, a finite angular momentum causes a sign problem in MC in general.

Let us introduce the angular momentum chemical potentials to the matrix model \eqref{action-BFSS} through the same procedure.
Note that the model \eqref{action-BFSS} is a quantum mechanics in $D$ dimension, and we consider rotations on $\tilde{D}$ planes ($2\tilde{D}  \le D $).
Hence, we introduce the chemical potential $\mu_I$ ($I=1, \cdots, \tilde{D}$) for each plane.
Then, the action becomes \cite{Morita:2010vi, Yamada:2006rx}
\begin{align}
    \label{action-BFSS-J}
    S
    = & 
    \int_0^{\beta} \hspace{-2mm} dt  
    \Tr 
    \Biggl\{ 
    \sum_{I=1}^{\tilde{D}}
    \left(
    D_t -\mu_I \right) Z^{I \dagger }
    \left(
    D_t +\mu_I \right) Z^I 	+
    \sum_{I=2\tilde{D}+1}^D
    \frac{1}{2}
    \left(
    D_t X^I \right)^2
    -
    \sum_{I,J=1}^D \frac{g^2}{4} [X^I,X^J]^2
    \Biggr\} \nonumber                                                                                                                                                                                                                        \\
    = & \int^{\beta}_{0} dt \textrm{Tr } \Biggl[ \frac{1}{2} \sum_{I=1}^D (D_t X^I)^2 - \sum_{I,J=1}^D \frac{g^2}{4} [X^I,X^J]^2 - \sum_{I=1}^{{\tilde D}}  \frac{\mu_I^2}{2} \left\{ (X^I)^2+(X^{\tilde{D}+I})^2  \right\} \Biggr. \nonumber \\
      & \ \ \Biggl. + i \sum_{I=1}^{{\tilde D}} \mu_I  \left\{ (D_t X^I) X^{{\tilde D}+I} - (D_t X^{{\tilde D}+I}) X^{I} \right\} \Biggr],
\end{align}
where we have defined $Z^I:= \left( X^I+i X^{\tilde{D}+I} \right)/\sqrt{2} $, ($I=1,\cdots, \tilde{D}$).

In the following analysis, for simplicity, $\mu_I$ is taken to be a common value $\mu>0$ to all $\mu_I$.

\section{Overview of our main results}
\label{sec-result}

In this section, we show our main results for the model \eqref{action-BFSS-J} derived through the CLM and the minimum sensitivity.
The details of the CLM and the minimum sensitivity are presented in Sec.~\ref{sec-CLM} and Sec.~\ref{sec-minimum}, respectively.
We mainly investigate $D=9$ and $D=16$. 
We have chosen them because $D=9$ is the critical dimension of superstring theories and $D=16$ is suitable to be compared with the minimum sensitivity, since they agree better for larger $D$ in the $\mu=0$ case as demonstrated in Ref.~\cite{Morita:2020liy}.

\subsection{Phase diagrams}
\label{sec-result-phase}

\begin{figure}[htbp]
    \begin{center} 
        \begin{tabular}{cc}
            \begin{minipage}{0.5\hsize}
                \begin{center}
                    \includegraphics[scale=0.85]{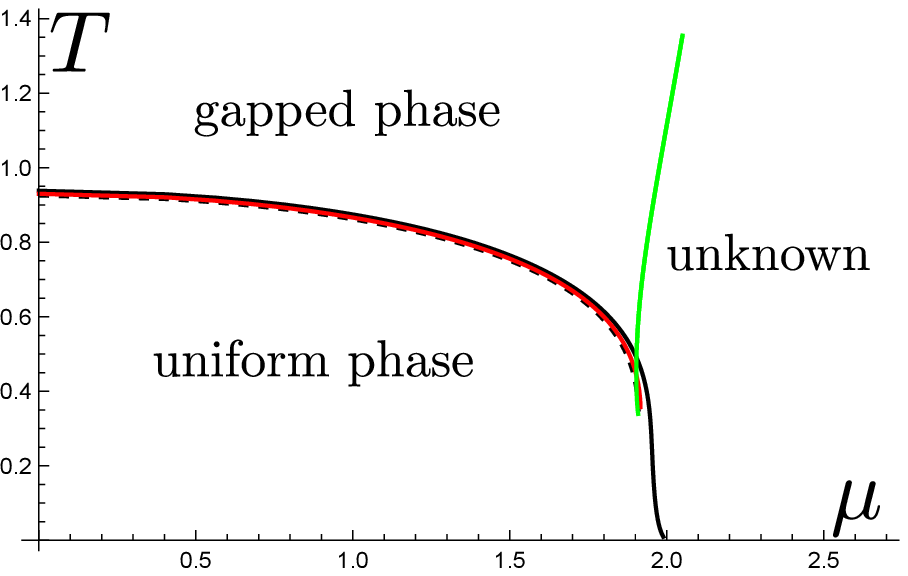}\\
                    $D=9$ with ${\tilde D}=1$
                \end{center}
            \end{minipage}
            \begin{minipage}{0.5\hsize}
                \begin{center}
                    \includegraphics[scale=0.85]{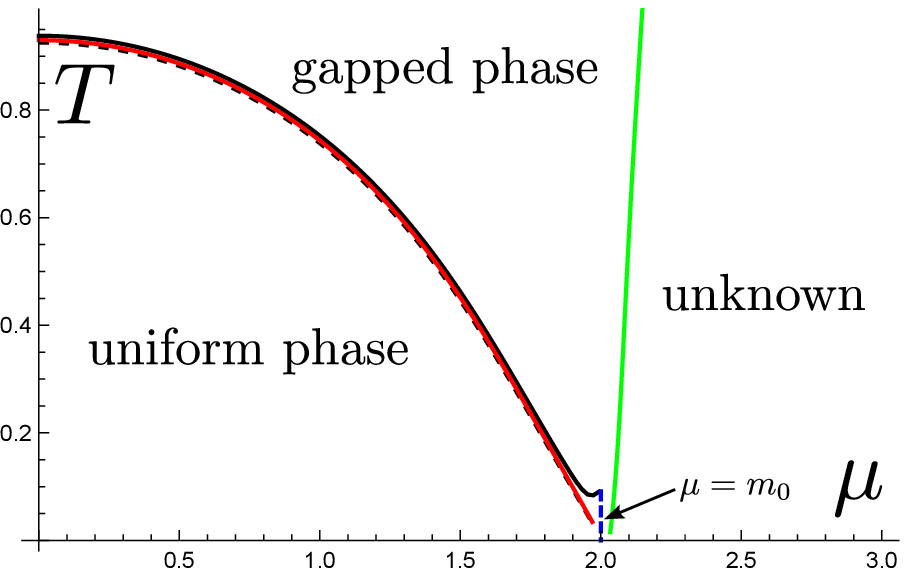}\\
                    $D=9$ with ${\tilde D}=3$
                \end{center}
            \end{minipage}
        \end{tabular}
        \caption{
            Phase diagrams for $D=9$ with $ {\tilde D}=1$ and ${\tilde D}=3$.
            The derivation is argued in Sec.~\ref{sec-free}.
            The red solid lines represent $\mu_{*}(T)$, where the first-order phase transitions between the uniform phase and the gapped phase occur.
            The green solid curves represent $\mu_{\rm unstable}(T)$, where $m_Z=\mu$ occurs and our analysis for the gapped solution is not reliable anymore beyond these curves.
            The black solid lines represent the critical points $\mu_c(T)$. 
            The black dashed lines represent the GWW point $\mu_{\textrm{GWW}} (T)$.  		
            In the ${\tilde D}=3$ case, the blue dashed line represents $\mu=m_0$ where the uniform solution becomes not reliable. 
            Although the phase structures at ${\tilde D}=1$ and ${\tilde D}=3$ are similar,
            the difference appears near $(T, \mu) \sim (0,\mu_c)$.
            While $\mu_c(T)$ reaches $T\simeq 0$ in the ${\tilde D}=1$ case, it does not in the ${\tilde D}=3$ case.
            However, our approximation near $(T, \mu) \sim (0,\mu_c)$ is not so reliable and they are not conclusive.
        }
        \label{fig-phase}
    \end{center}
\end{figure}

We draw the $\mu-T$ phase diagrams at $D=9$ with $\tilde{D}=1$ and $\tilde{D}=3$ in Fig.~\ref{fig-phase}. 
These are obtained through the minimum sensitivity analysis (see Sec.~\ref{sec-free-energy}). 
These phase diagrams show that the uniform phase is favored in a low temperature and chemical potential region, and the gapped phase is favored in a high temperature and chemical potential region. A first-order transition occurs between them.
In the ``unknown" region depicted in Fig.~\ref{fig-phase}, the minimum sensitivity analysis does not work, and we presume that this region may be thermodynamically unstable due to a large chemical potential.
(As far as we attempt, we obtain similar phase diagrams when we change $D$ and/or $\tilde{D}$.)

These phase diagrams show that a larger $\mu$ or a  larger $\tilde{D}$ makes the transition temperature lower.
Similar properties are expected in four-dimensional pure YM theories and black holes as we will argue in Sec.~\ref{sec-Im} and Sec.~\ref{sec-BHs}.

%

\subsection{Results for observables}
\label{sec-result-observables}

By using the CLM and the minimum sensitivity, we investigate $\mu$ dependence of the four observables:
the Polyakov loop $|u_1|$ and $|u_2|$, the angular momentum $J$ and the expectation values of the square of the scalars $(X^I)^2$.
The results are shown in Figs.~\ref{u1_D09}, \ref{u2_D09}, \ref{JI_D09} and \ref{r2_D09}, respectively.
There, we take $D=9$ and $D=16$, and investigate them by changing  $\tilde{D}$ and temperature $T$.
The values of  $\tilde{D}$ and $T$ that we have taken are summarized in Table \ref{critical_muc}.
In this table, the phase transition point $\mu_*$ computed by the minimum sensitivity, at which the uniform phase turns into the gapped phase, is also listed. 

In the CLM, $N$=16 and 32 are taken at $D=9$, and $N=16$ is taken at $D=16$.   
Note that we present the numerical results obtained by the CLM only in the parameter region of $\mu$ where the data are acceptable. 
(The criterion for the acceptable data is based on the drift norms argued in Sec.~\ref{sec-CLM}.)
We find that obtaining acceptable results for larger $\mu$ is harder.

In the following subsequent subsections, we argue the details of the obtained observables.

\begin{table} [htbp]
    \renewcommand{\arraystretch}{1.3}
    \begin{center} 
        \begin{tabular}{|c||c|} \hline
            parameters                                  & $\mu_*$ \\ \hline \hline
            $\displaystyle D=9, ~{\tilde D}=1,~ T=0.85$ & $1.1$   \\ \hline
            $\displaystyle D=9, ~{\tilde D}=1,~ T=0.90$ & $0.71$  \\ \hline \hline
            $\displaystyle D=9, ~{\tilde D}=3,~ T=0.85$ & $0.67$  \\ \hline
            $\displaystyle D=9, ~{\tilde D}=3,~ T=0.90$ & $0.42$  \\ \hline  
        \end{tabular} 	 \hspace*{5mm}
        \begin{tabular}{|c||c|} \hline
            parameters                                   & $\mu_*$ \\ \hline \hline
            $\displaystyle D=16, ~{\tilde D}=1,~ T=0.80$ & $1.69$  \\ \hline
            $\displaystyle D=16, ~{\tilde D}=1,~ T=0.85$ & $1.29$  \\ \hline \hline
            $\displaystyle D=16, ~{\tilde D}=5,~ T=0.80$ & $0.87$  \\ \hline
            $\displaystyle D=16, ~{\tilde D}=5,~ T=0.85$ & $0.62$  \\ \hline 
        \end{tabular}
        
        \caption{
            The list of parameters that are used in our analysis shown in Figs.~\ref{u1_D09} - \ref{r2_D09}.
            In the $D=9$ case, we take $\tilde{D}=1$ and 3 and T=0.85 and 0.90.
            In the $D=16$ case, we take $\tilde{D}=1$ and 5 and T=0.80 and 0.85.
            The transition point $\mu_*$, at which the uniform phase turns into the gapped phase for these parameters are also shown.
            ($\mu_*$ are obtained through the minimum sensitivity analysis in Sec.~\ref{sec-free-energy}.)
        }
        \label{critical_muc}
    \end{center}
\end{table}

\subsubsection{Polyakov loop $|u_1|$ and $|u_2|$}
\label{sec-result-Polyakov}

The Polyakov loop $\{ u_n \}$ \eqref{polyakov} is the order parameter of the large-$N$ confinement/deconfinment transition as mentioned in Sec.~\ref{sec-review}.  $u_1=0$ at large $N$ indicates the uniform phase, and
$u_1 \ne 0$ indicate the non-uniform phase and the gapped phase.
Besides, the minimum sensitivity predicts that the non-uniform solution satisfies $u_2 = 0$, whereas the gapped solution satisfies $u_2 \neq 0$ (see Sec.~\ref{sec-free}).
So, $u_1$ and $u_2$ may tell us which phases appear.
However, it is known that the Polyakov loops suffer relatively large O$\left( \frac{1}{N} \right)$ corrections \eqref{un-finite-N-main} at large $N$ \cite{Aharony:2003sx, Azuma:2014cfa}. (The corrections for other quantities are typically O$\left( \frac{1}{N^2} \right)$.) Hence, we take care to distinguish the signals of the gapped and non-uniform phases and the $1/N$ corrections.

\begin{figure} [htbp]
    \centering
    \includegraphics[width=0.43\textwidth]{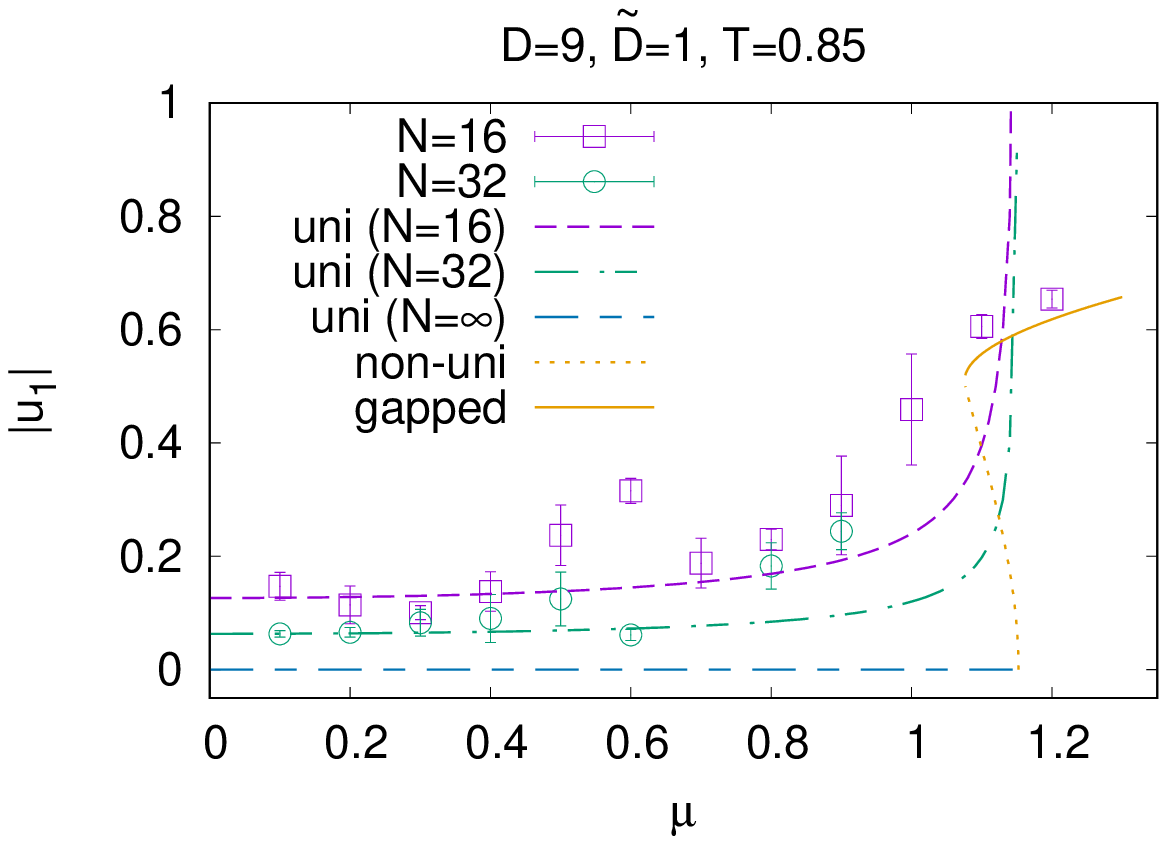}
    \includegraphics[width=0.43\textwidth]{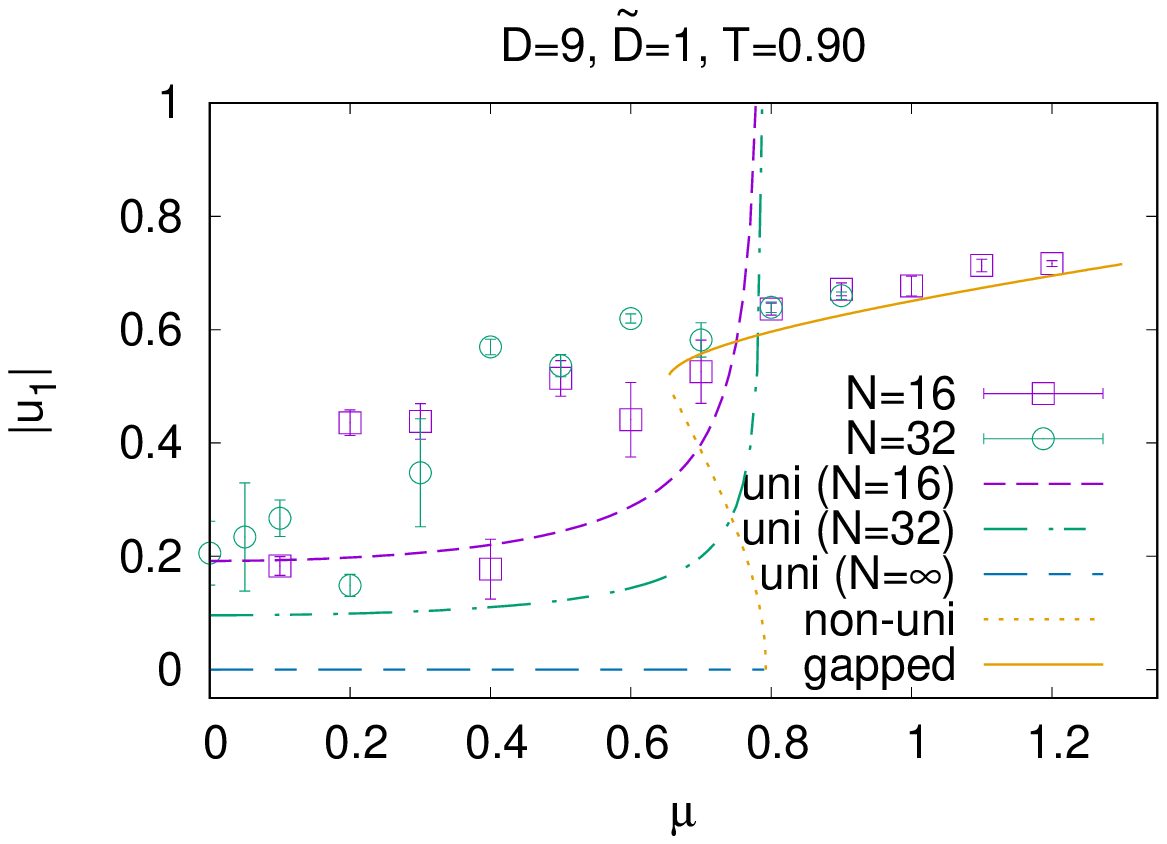}
    \includegraphics[width=0.43\textwidth]{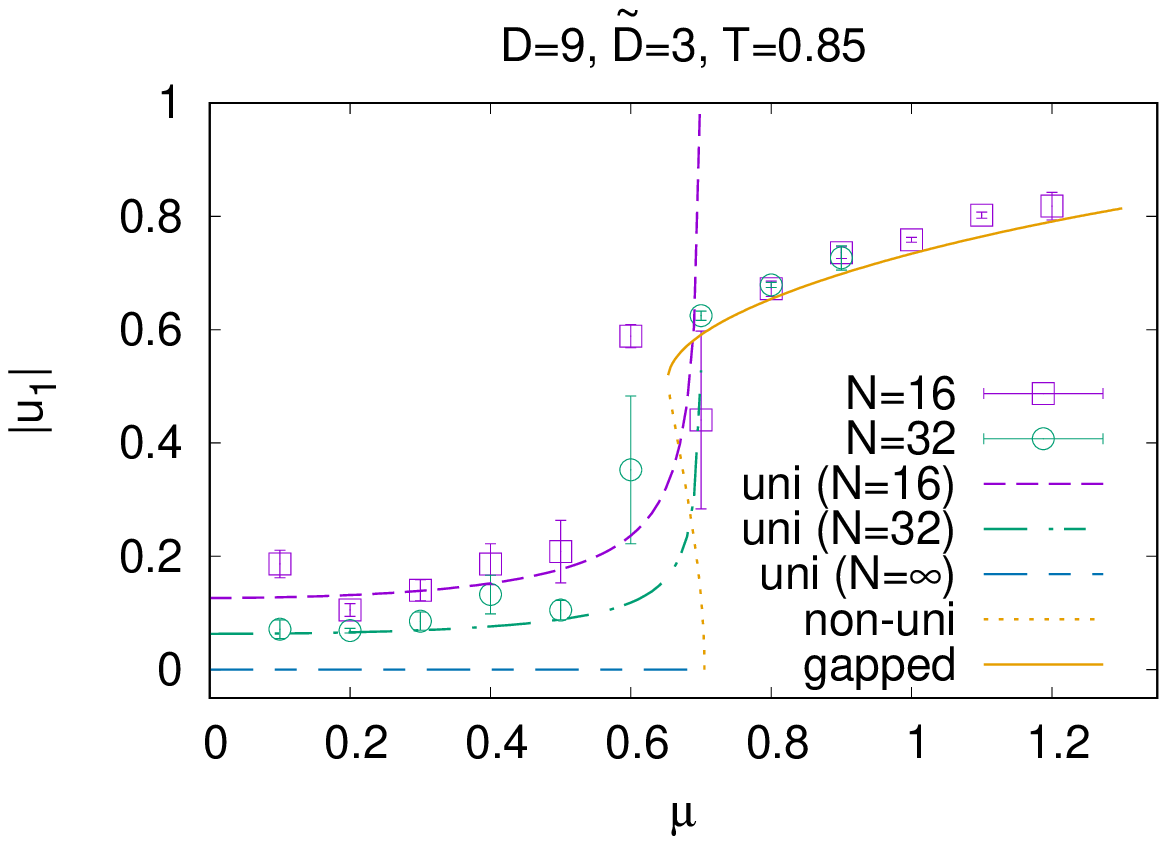}
    \includegraphics[width=0.43\textwidth]{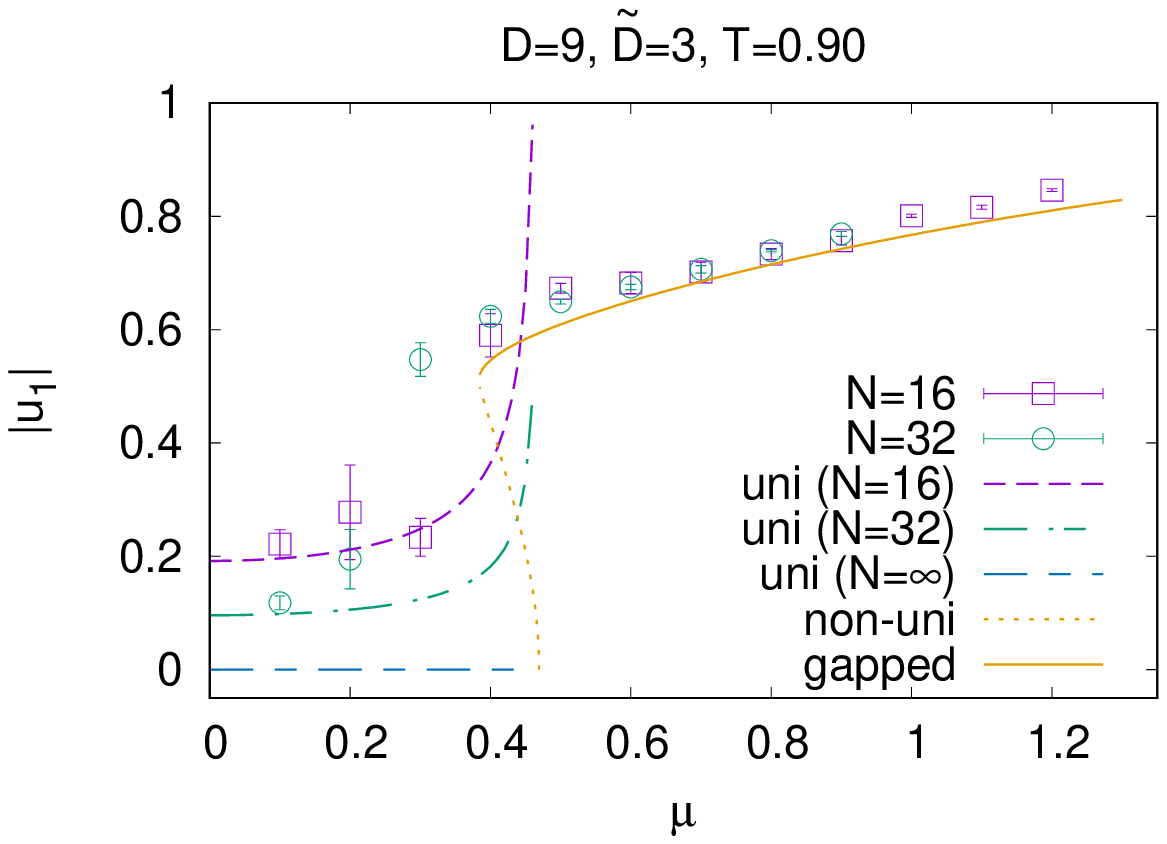}
    \includegraphics[width=0.43\textwidth]{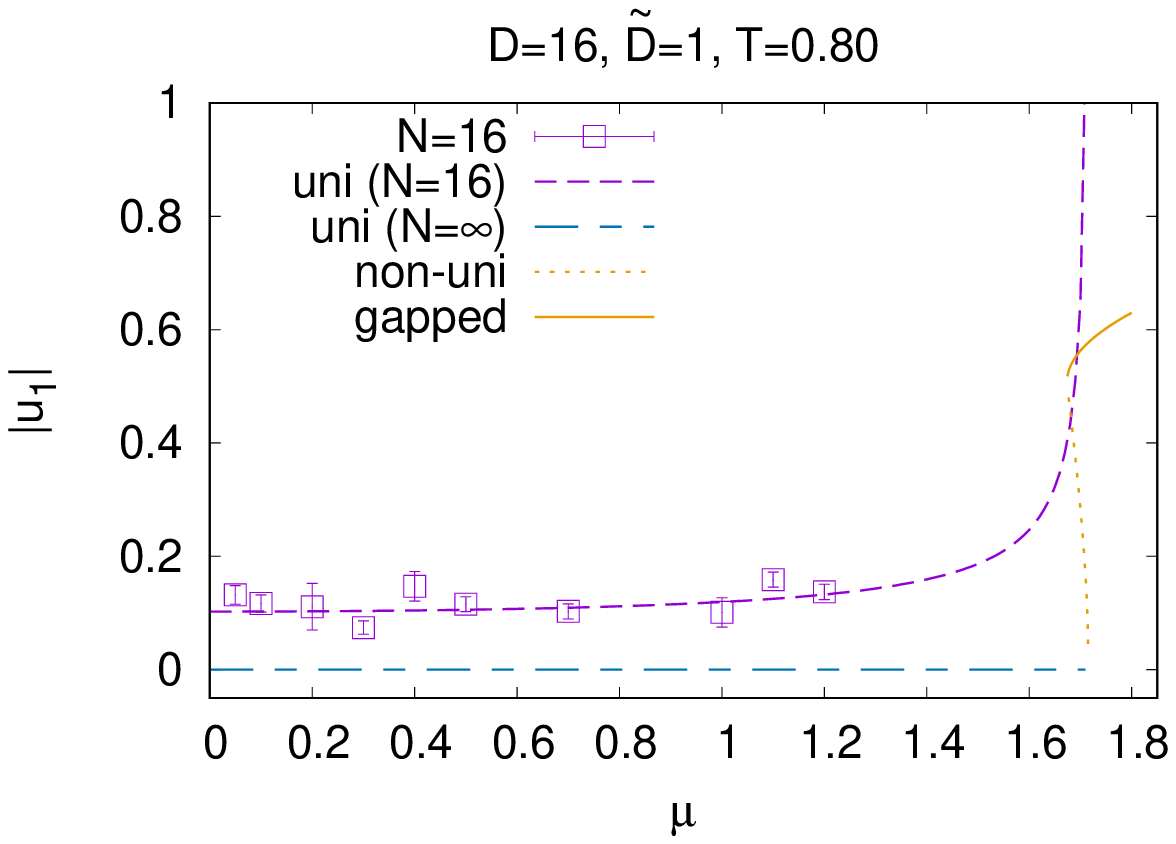}
    \includegraphics[width=0.43\textwidth]{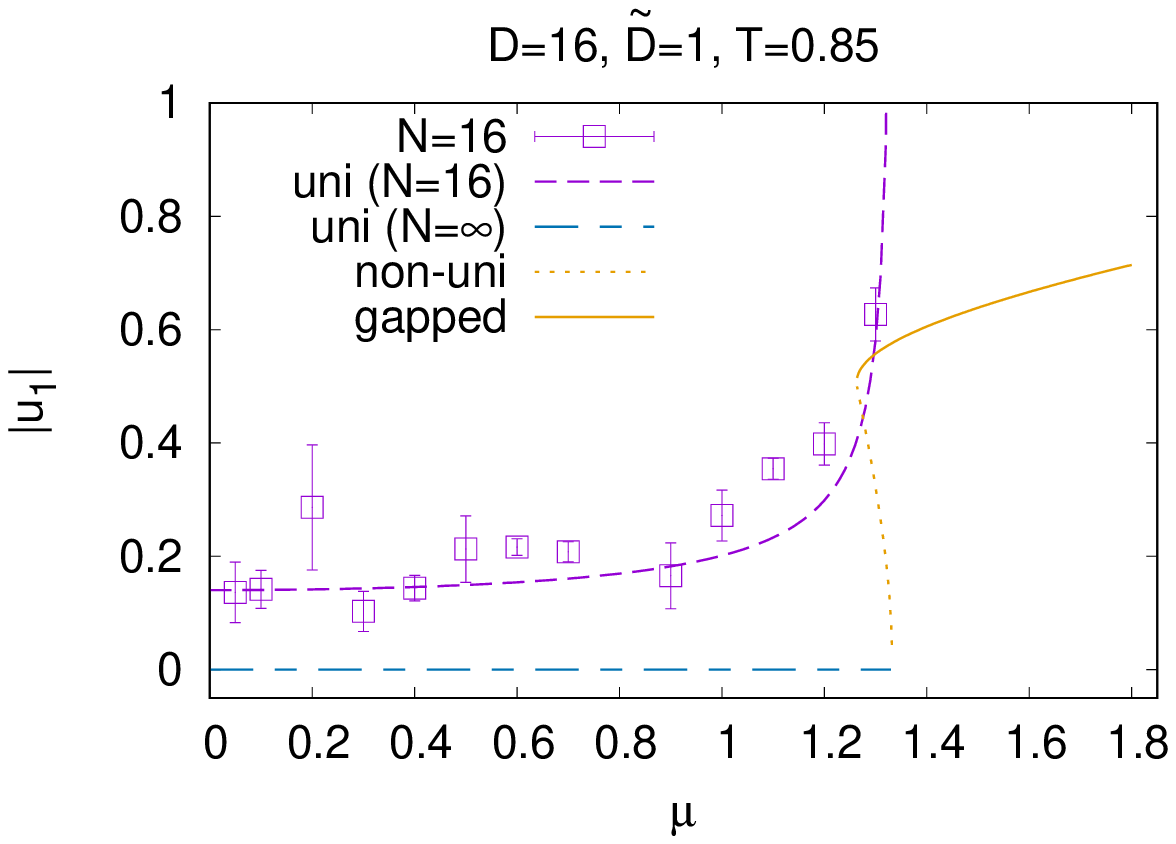}
    \includegraphics[width=0.43\textwidth]{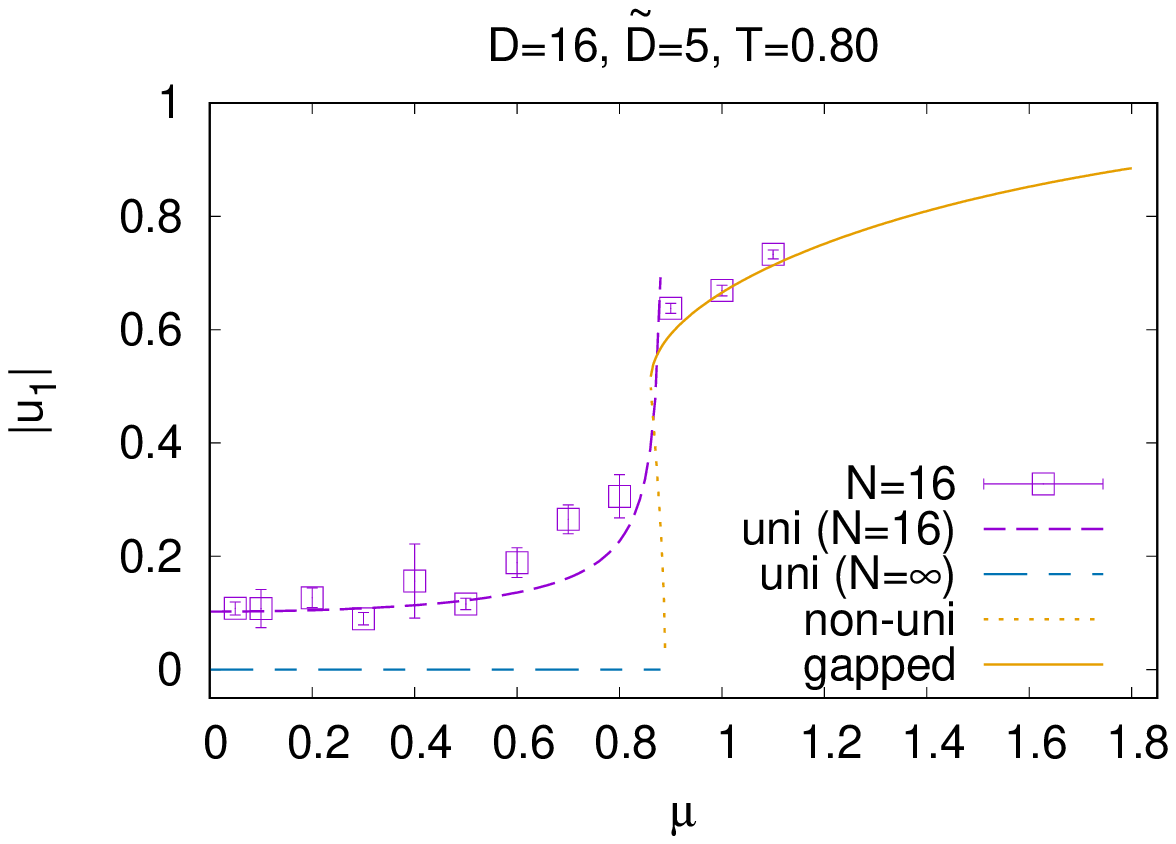}
    \includegraphics[width=0.43\textwidth]{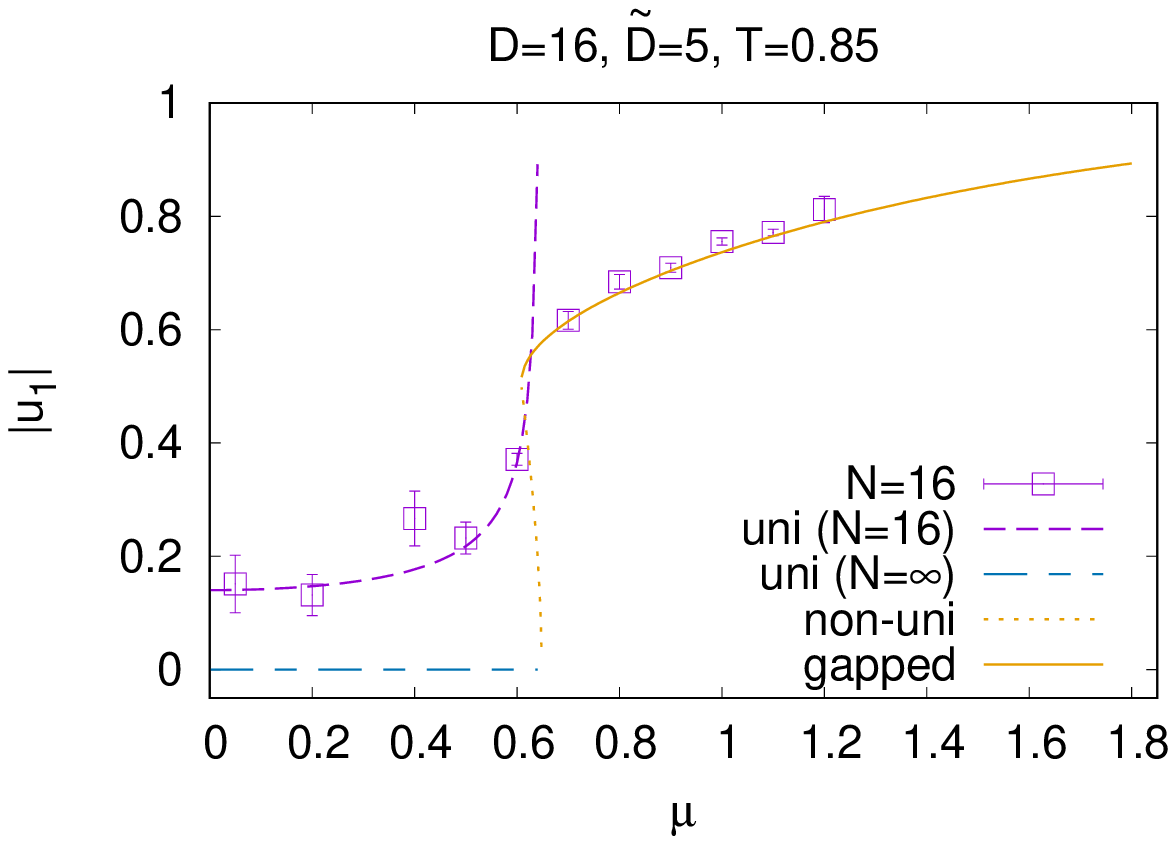}
    \caption{$|u_1|$ is plotted against $\mu$. We present $\langle |u_1| \rangle$ obtained by the CLM, at $N=16,32$ for $D=9$ and $N=16$ for $D=16$. The lines denote the results of the minimum sensitivity \eqref{un-finite-N-main}-\eqref{un-gapped-main}. The dashed lines represent those of the uniform phase at $N=16,32,\infty$, while the dotted and solid lines represent those of the non-uniform and gapped phase at $N=\infty$, respectively.}\label{u1_D09}
\end{figure}

\begin{figure} [htbp]
    \centering
    \includegraphics[width=0.43\textwidth]{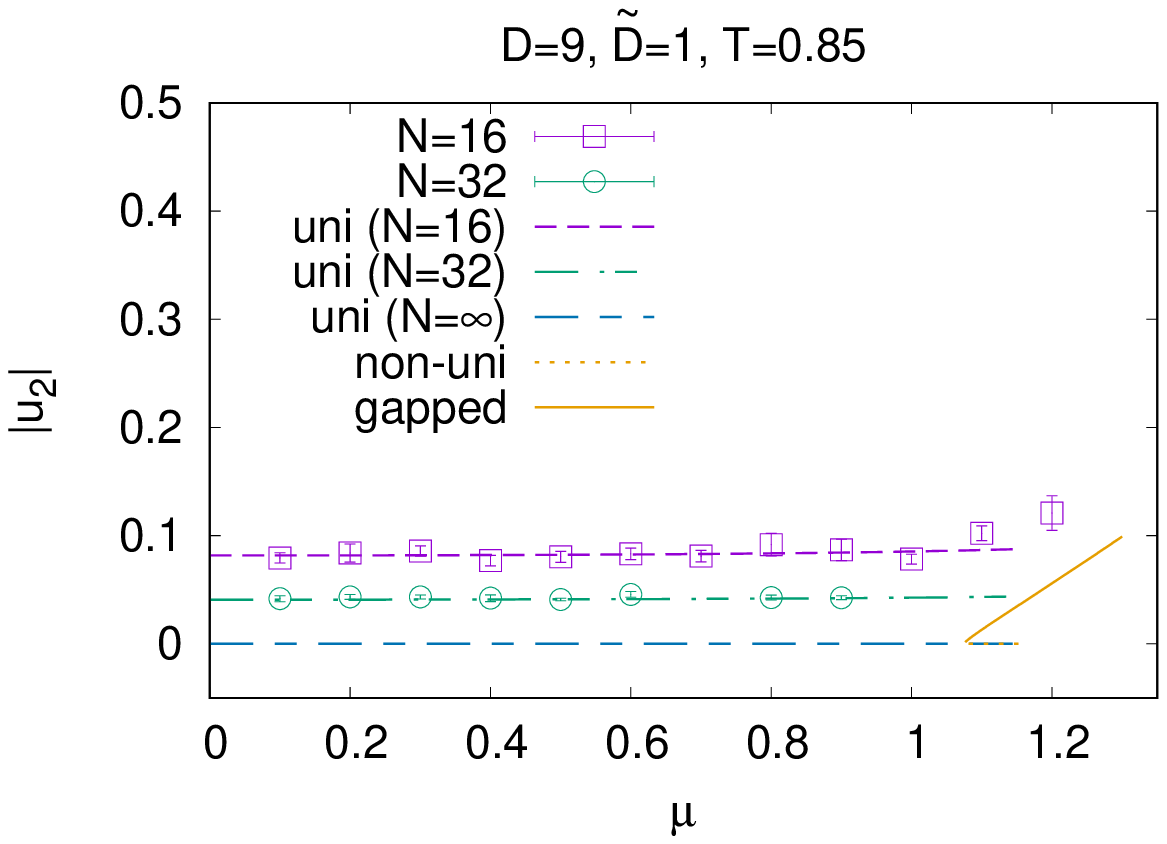}
    \includegraphics[width=0.43\textwidth]{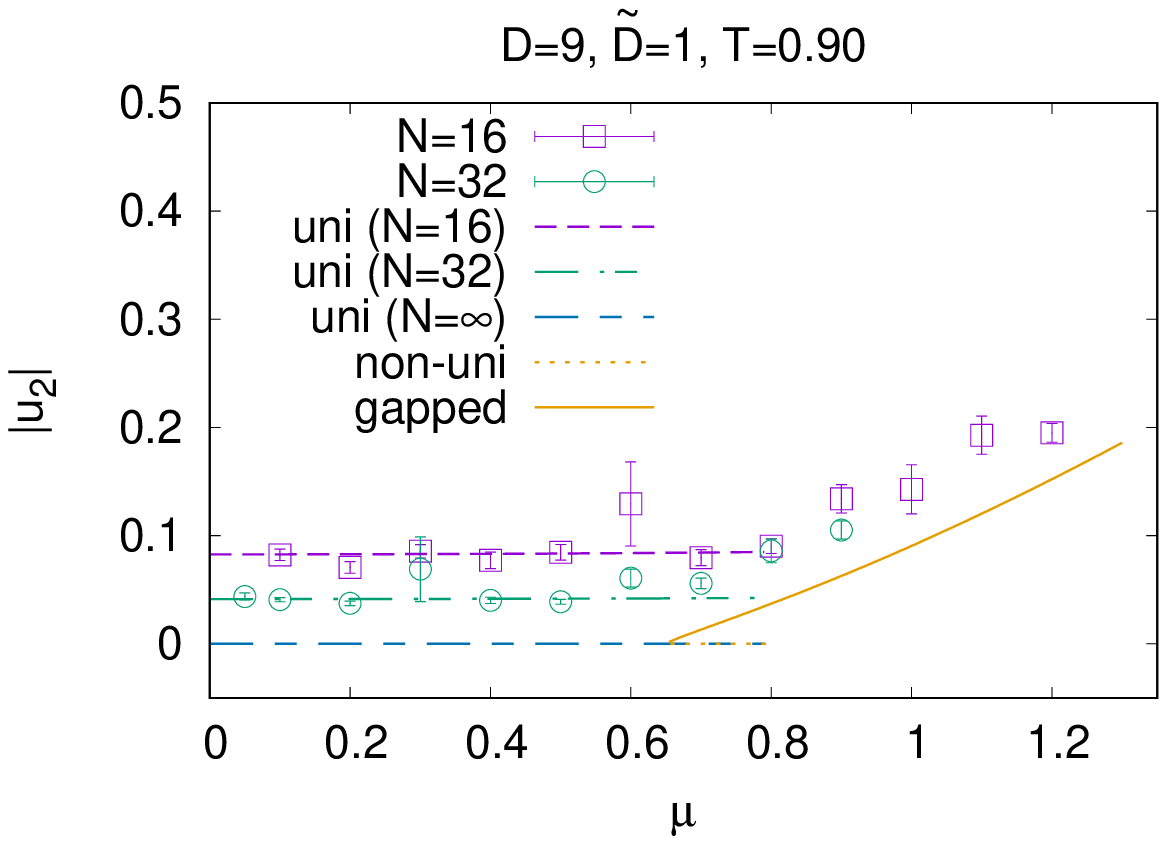}
    \includegraphics[width=0.43\textwidth]{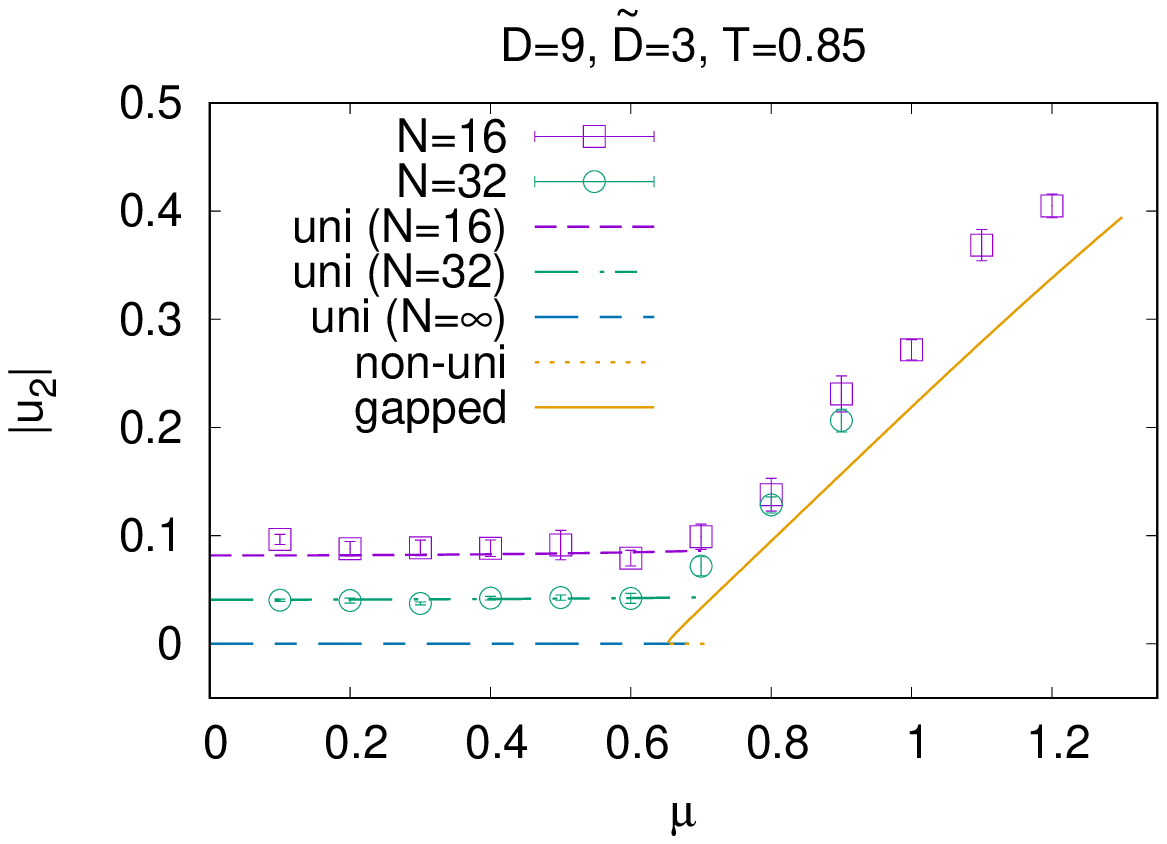}
    \includegraphics[width=0.43\textwidth]{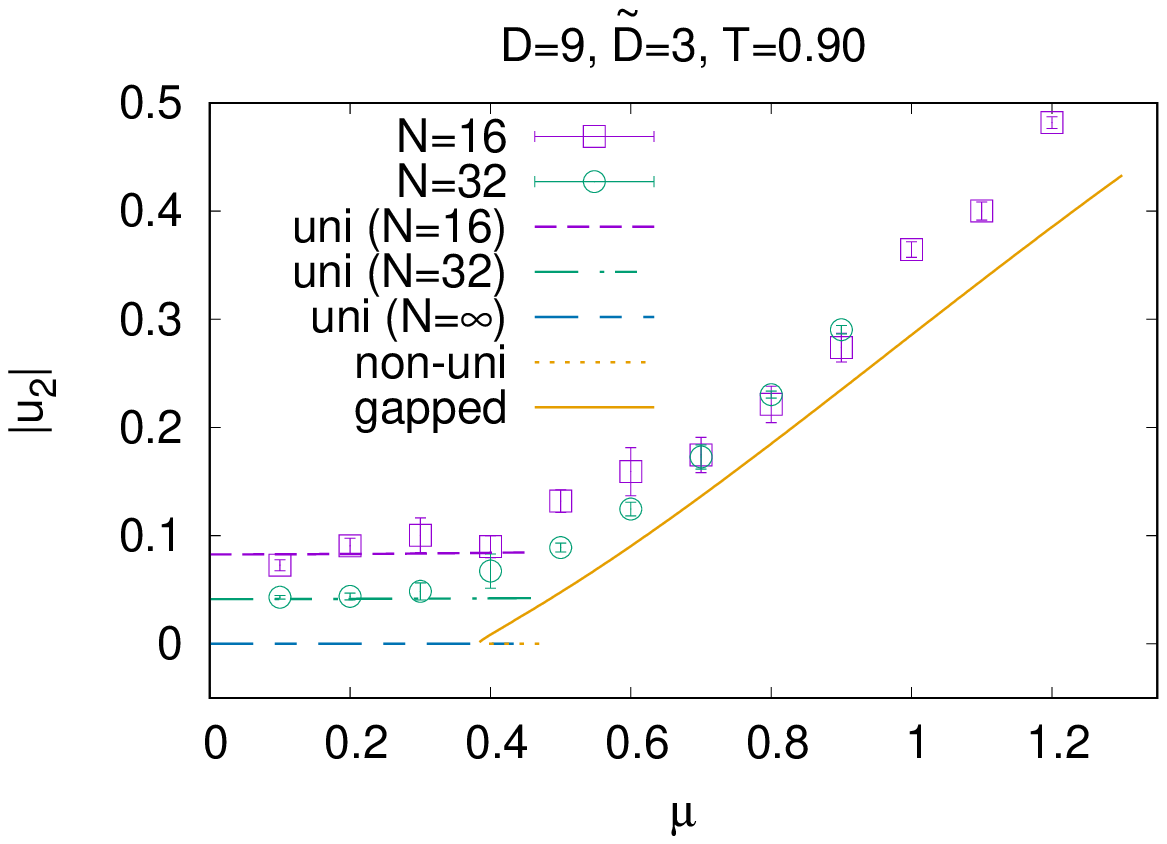}
    \includegraphics[width=0.43\textwidth]{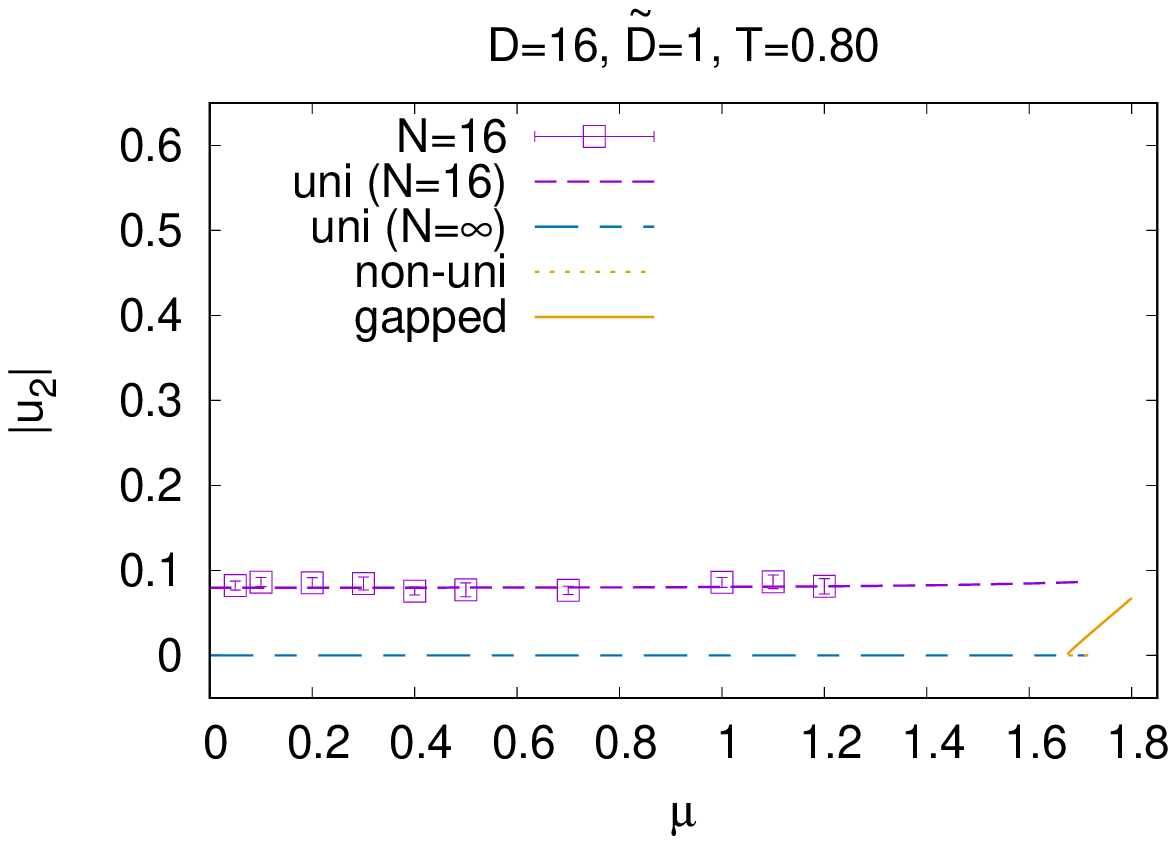}
    \includegraphics[width=0.43\textwidth]{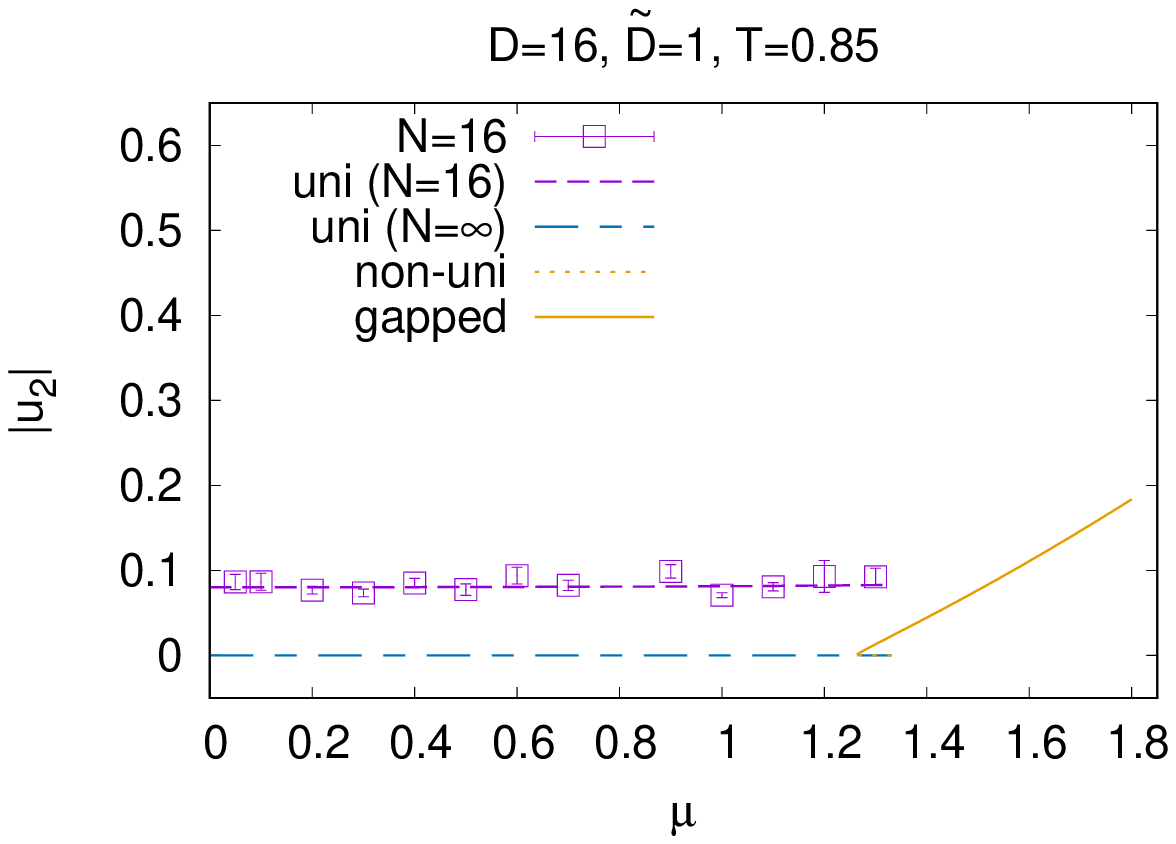}
    \includegraphics[width=0.43\textwidth]{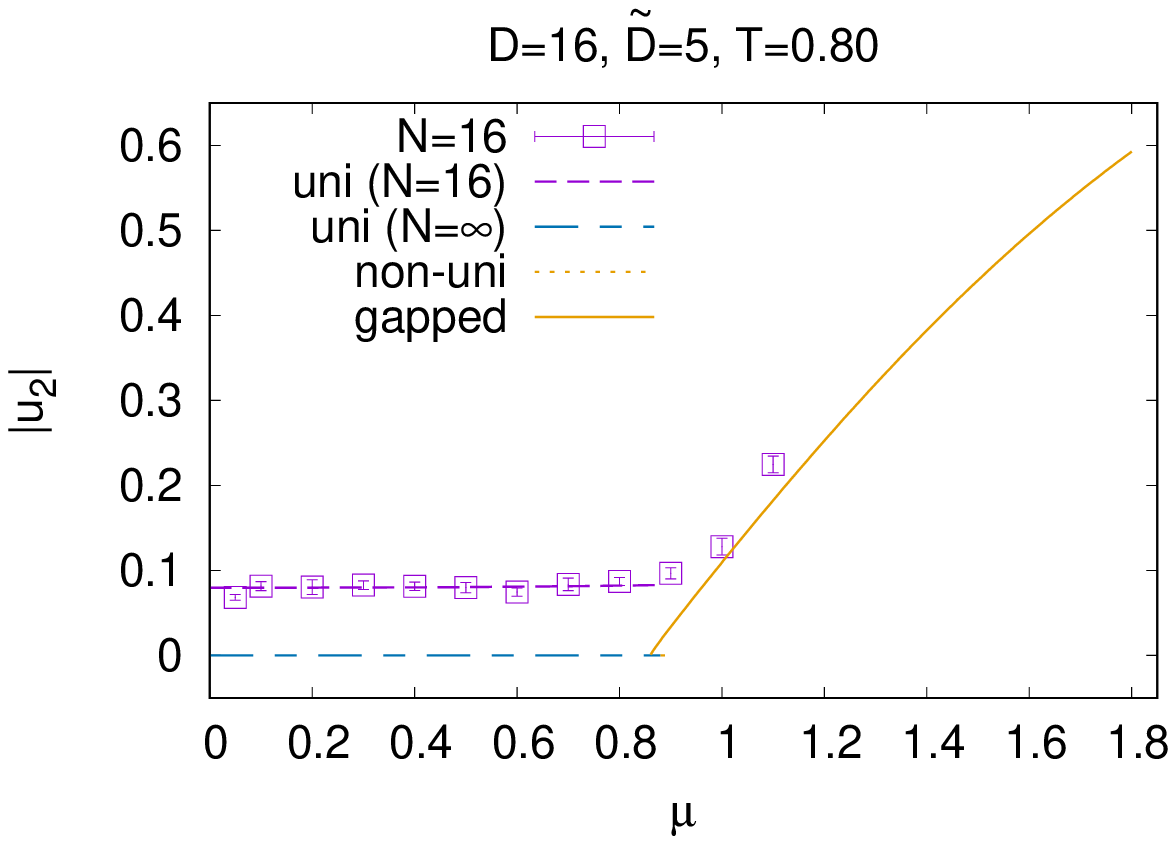}
    \includegraphics[width=0.43\textwidth]{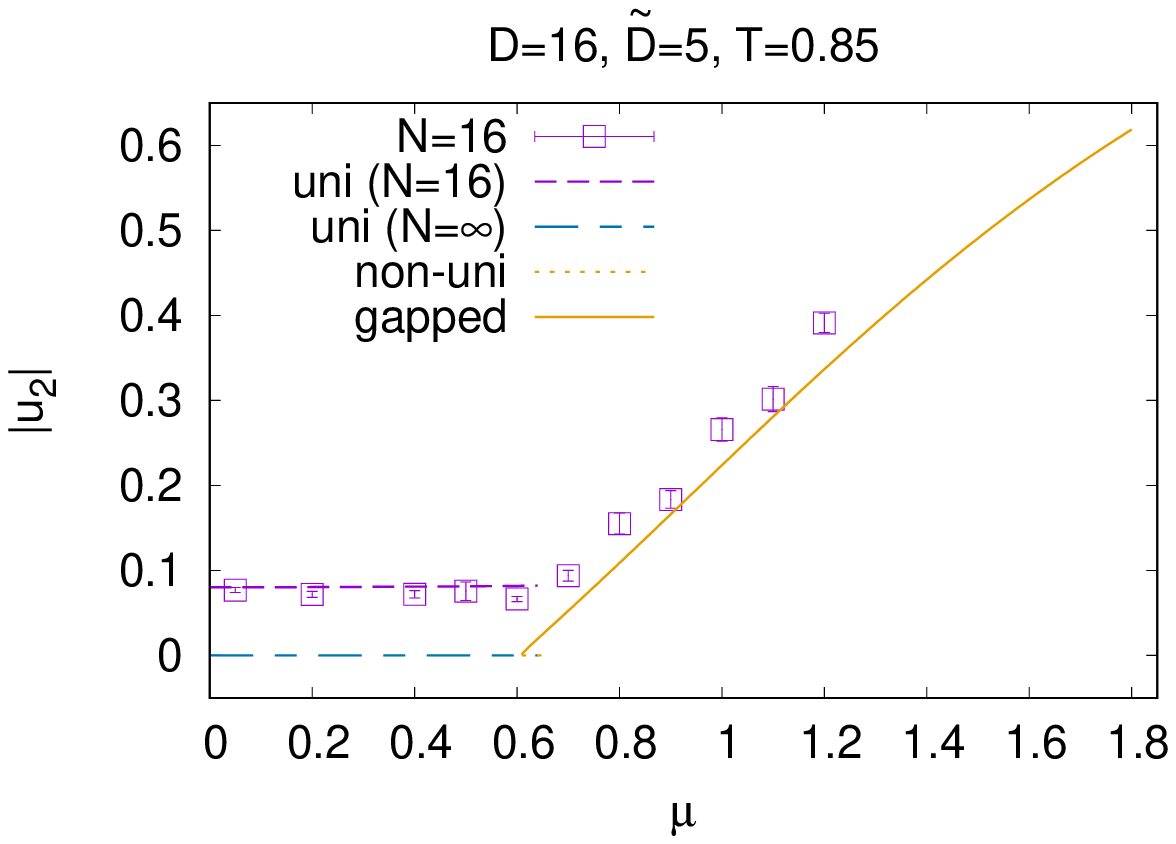}
    \caption{$|u_2|$ is plotted against $\mu$. We present $\langle |u_2| \rangle$ obtained by the CLM, at $N=16,32$ for $D=9$ and $N=16$ for $D=16$. The lines denote the results of the minimum sensitivity \eqref{un-finite-N-main}-\eqref{un-gapped-main}. The dashed lines represent those of the uniform phase at $N=16,32$ and large $N$, while the dotted and solid lines represent those of the non-uniform and gapped phase at $N=\infty$, respectively.
        Note that the non-uniform solutions overlap with the uniform solutions at $N=\infty$ because both are $u_2=0$.
    }\label{u2_D09}
\end{figure}

Our results for $|u_1|$ and $|u_2|$ against $\mu$ are plotted in Figs.~\ref{u1_D09} and \ref{u2_D09}.
The minimum sensitivity predicts the first-order transition from the uniform phase ($u_1=u_2=0$) at low $\mu$ to the gapped phase ($u_1 \neq 0$ and $u_2 \neq 0$) at high $\mu$.
The CLM results agree with the minimum sensitivity results at large $N$ in the gapped phase. In the uniform phase, $|u_n|$ has a finite-$N$ effect of order O$\left( \frac{1}{N} \right)$ \cite{Azuma:2014cfa}, and we compare the CLM results of $|u_n|$ with the minimum sensitivity result with the $1/N$ correction \eqref{un-finite-N-main}, which also agree with each other.

Among our results, the deviation between the CLM and the minimum sensitivity at $T=0.90$ in the $D=9$ with $\tilde{D}=1$ case is larger.
We presume that the fluctuation near $T=0.90$ is large in the CLM, since the phase diagram in Fig.~\ref{fig-phase} indicates that $T_c \simeq 0.9$ for $\mu \lesssim 1$.

Although we observe the phase transitions in the CLM,  
we do not attempt to determine their order.
In order to do it, we may need to evaluate,
for example, the susceptibility for $u_1$ \cite{Azuma:2014cfa}, but it requires computation at larger $N$ in the CLM, and we leave it for a future work.

\subsubsection{Angular momentum $J$}
\label{sec-result-J}

\begin{figure} 
    \centering
    \includegraphics[width=0.43\textwidth]{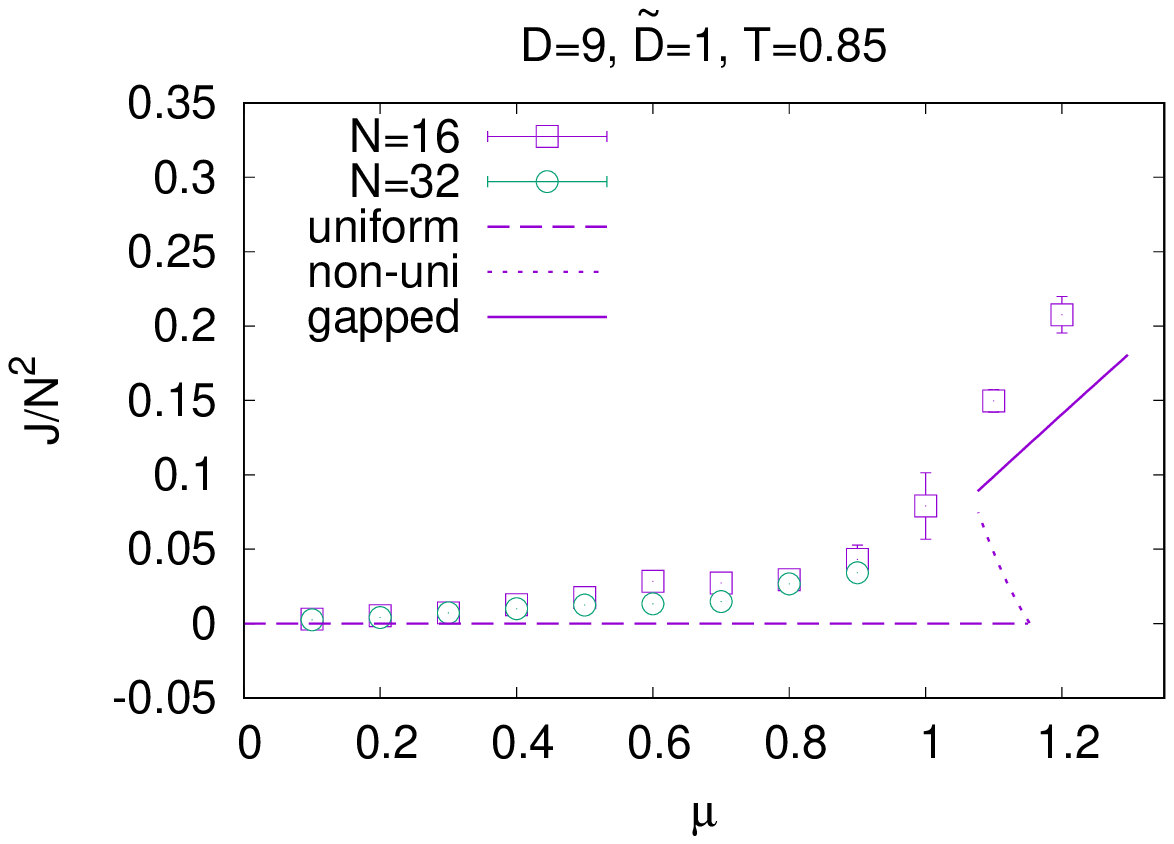}
    \includegraphics[width=0.43\textwidth]{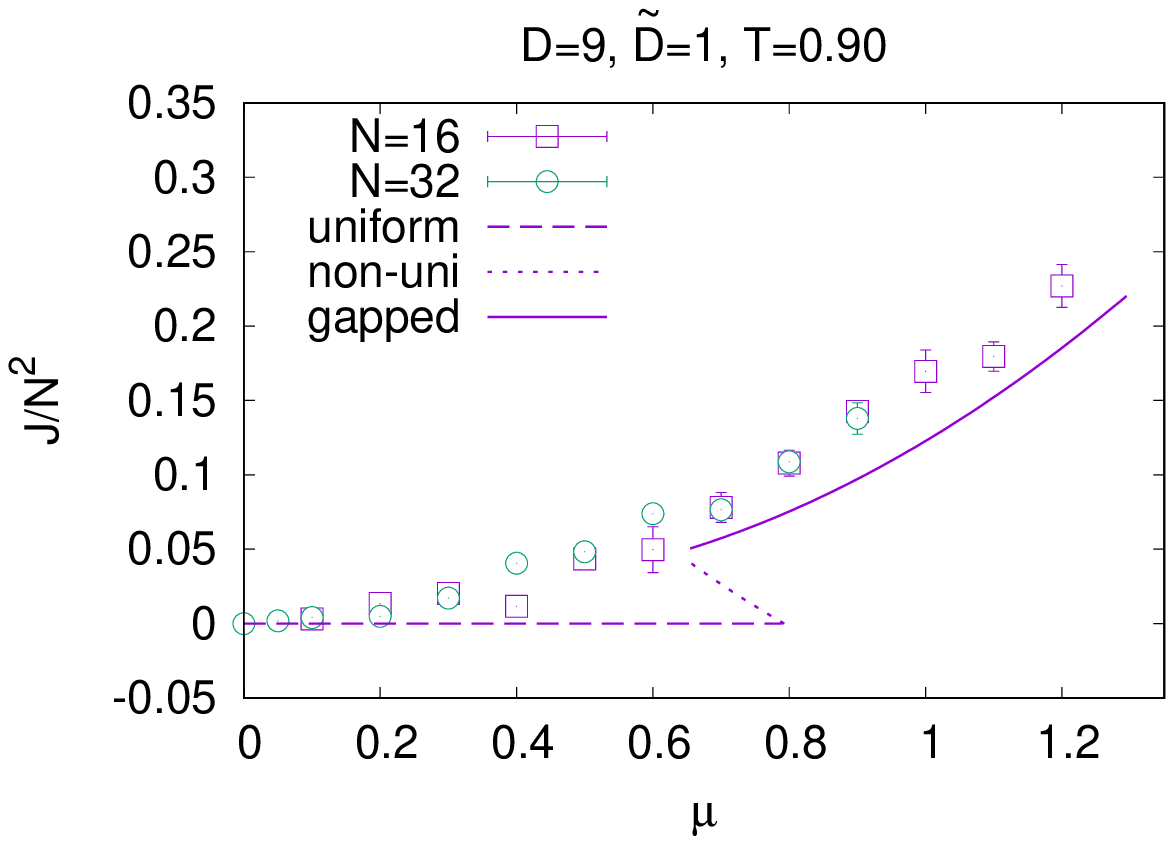}
    \includegraphics[width=0.43\textwidth]{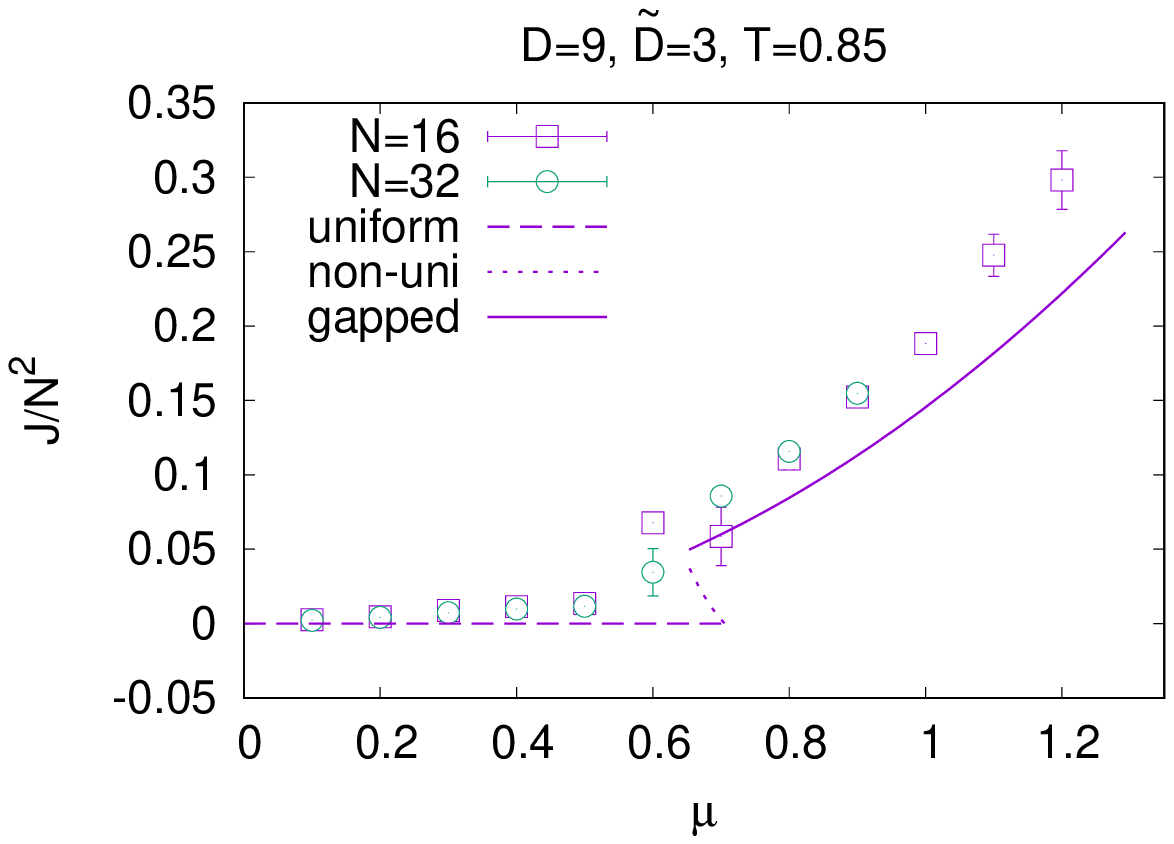}
    \includegraphics[width=0.43\textwidth]{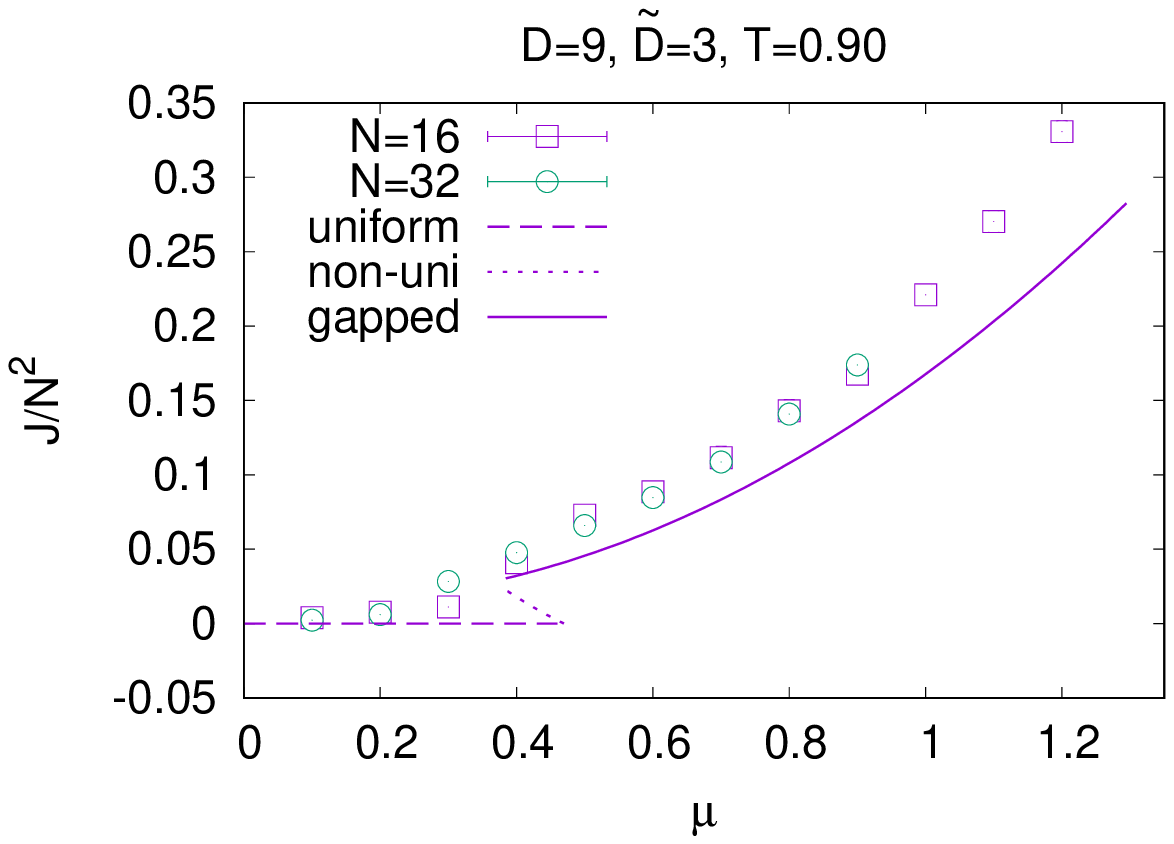}
    \includegraphics[width=0.43\textwidth]{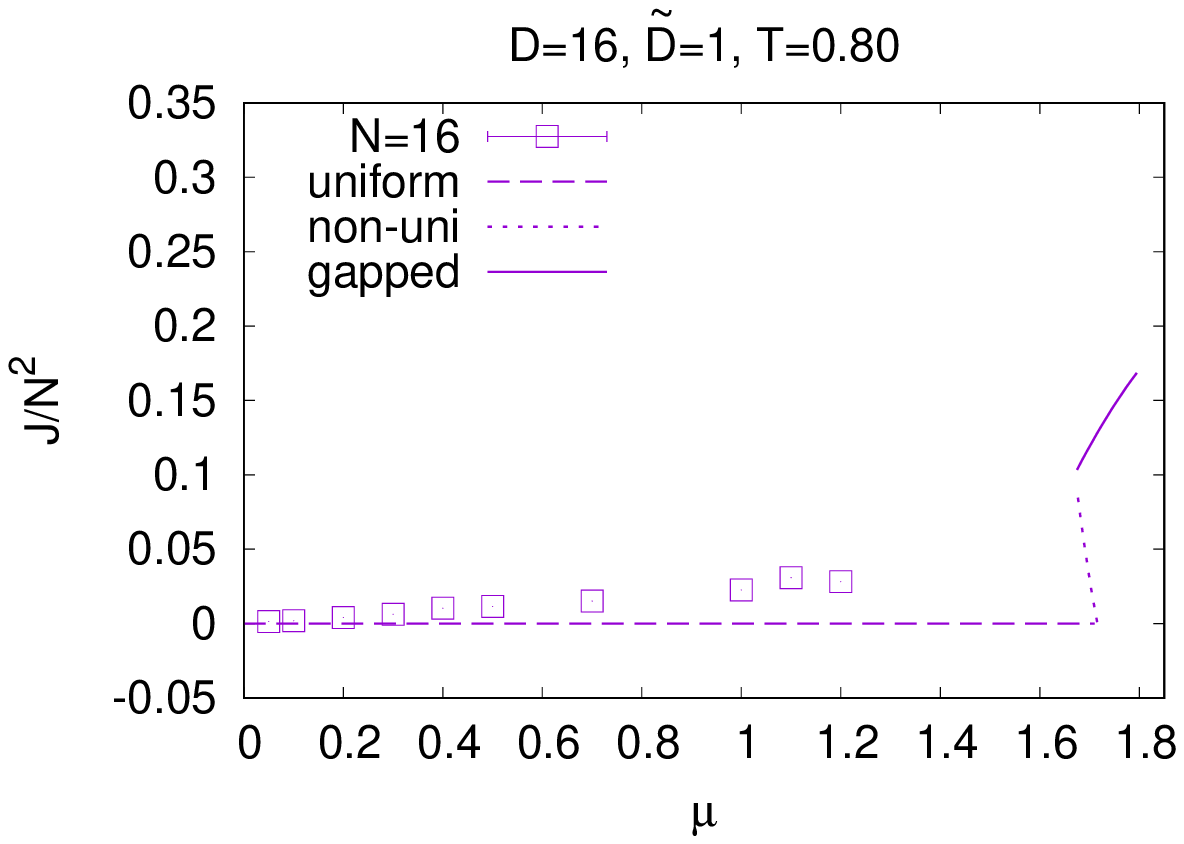}
    \includegraphics[width=0.43\textwidth]{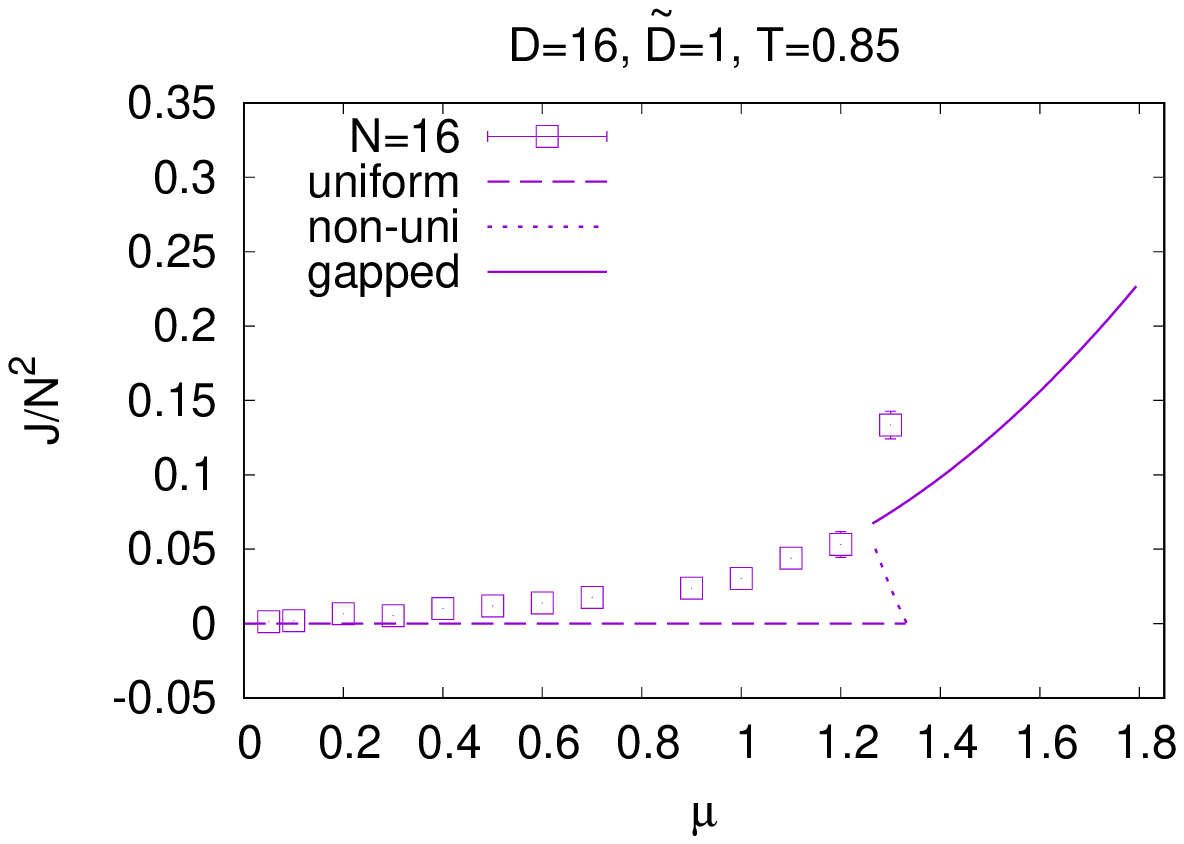}
    \includegraphics[width=0.43\textwidth]{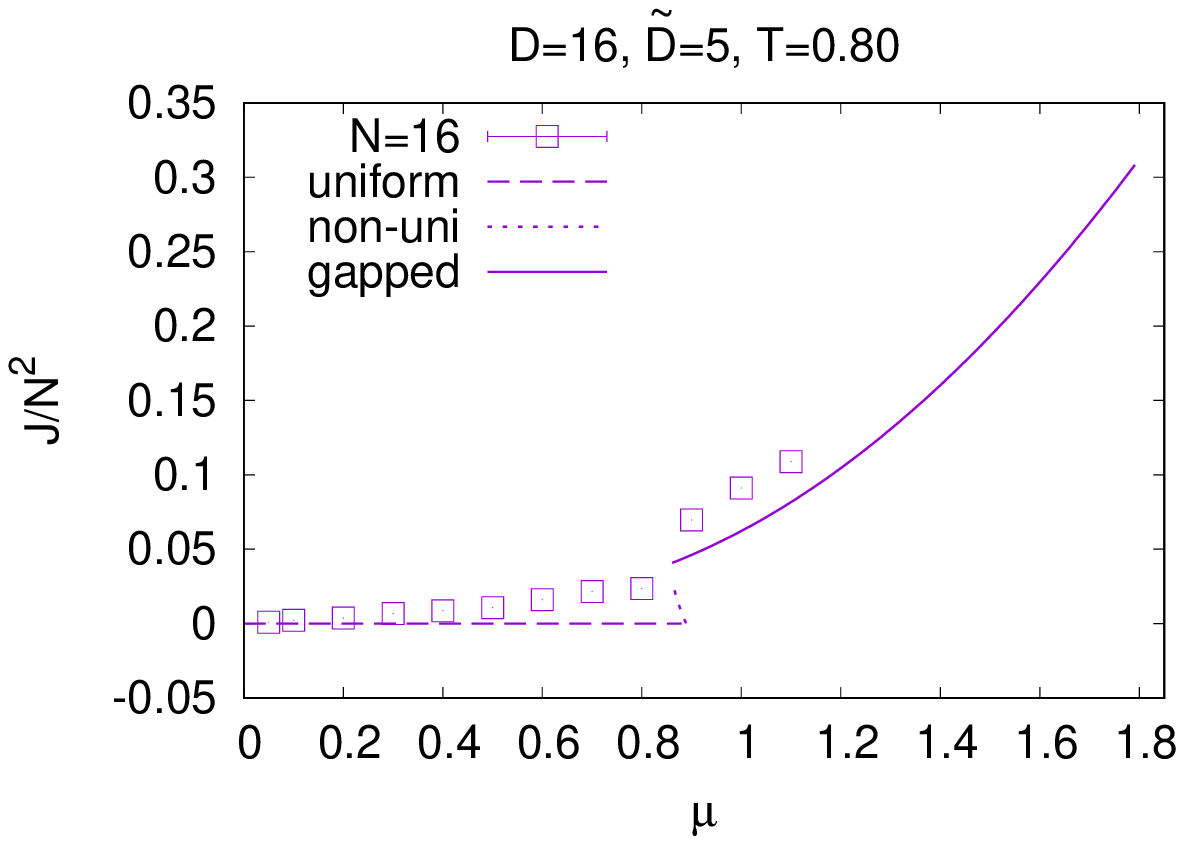}
    \includegraphics[width=0.43\textwidth]{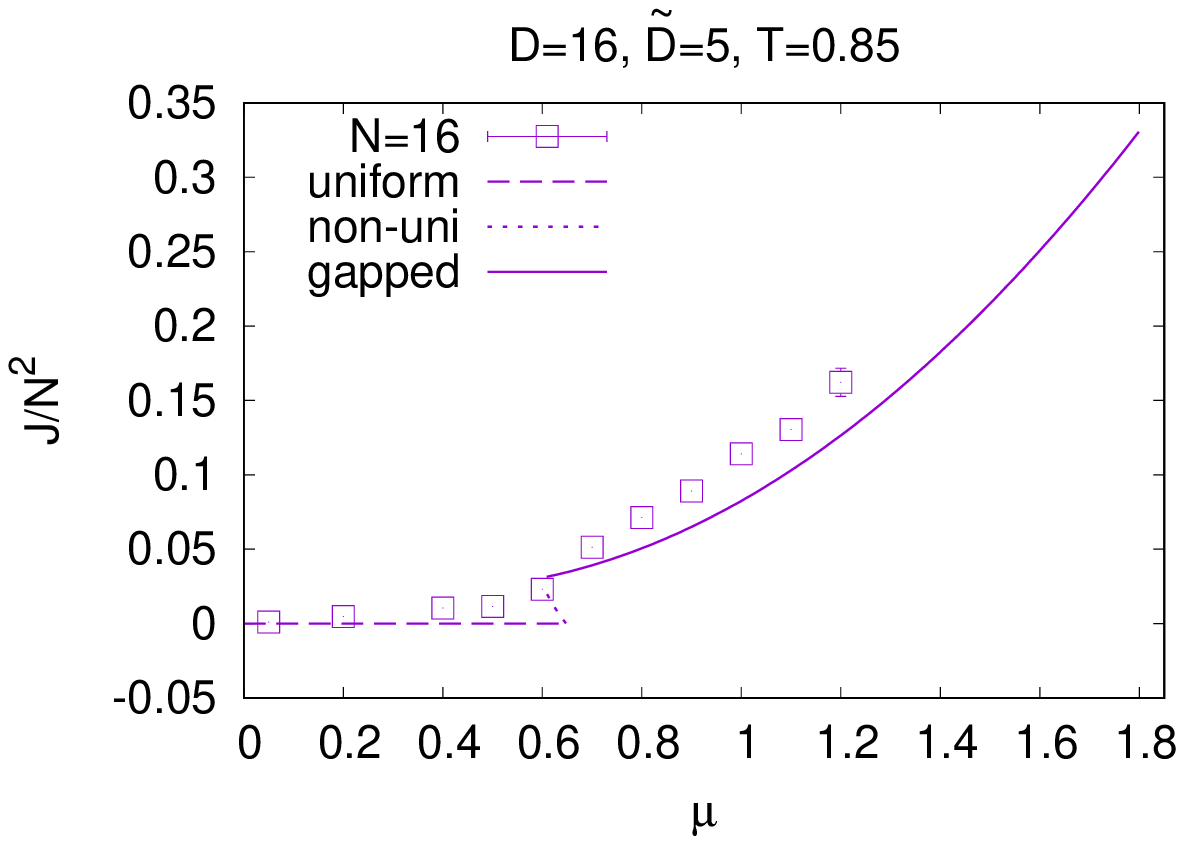}
    \caption{$J/N^2$ is plotted against $\mu$. We present $\langle J_{I=1}/N^2 \rangle$ obtained by calculating \eqref{JI_def} via the CLM, at $N=16,32$ for $D=9$ and $N=16$ for $D=16$. The lines denote the results \eqref{eq-J} of the minimum sensitivity obtained at $N=\infty$. The dashed, dotted and solid lines represent those of the uniform, non-uniform and gapped phase, respectively.}\label{JI_D09}
\end{figure}

Our results for angular momentum $J$ is shown in Fig.~\ref{JI_D09}.
Note that we have introduced the common angular momentum chemical potential $\mu$ for the $\tilde{D}$ planes in the model \eqref{action-BFSS-J}, and we evaluate the angular momentum for the single plane.
The definition of $J$ in the CLM is given in Eq.~\eqref{JI_def}.
In the minimum sensitivity, we derive $J$ through Eq.~\eqref{eq-J}.

We observe that $J$ is close to zero in the uniform phase in the CLM. This is a feature of the confinement phase (see Sec.\ref{subsec-angular}), and the CLM correctly reproduce it.
In the gapped phase, we observe slight discrepancies between the CLM results and the minimum sensitivity results.
We investigate the lattice space dependence and find that $J$ is more sensitive than other quantities.
Thus, the discrepancies may include artifacts of the CLM.

\subsubsection{Expectation values of scalars}
\label{sec-result-scalar}

The square of the scalar $(X^I)^2$ represents the distribution of the D-particles on the $I$-th direction. Since we have the rotational and the non-rotating directions, we define the averages
\begin{align}
    R_Z^2:= & \frac{1}{\tilde{D}} \frac{g^2}{N} \sum_{I=1}^{\tilde{D}} \left\langle \Tr Z^{I\dagger} Z^{I } \right\rangle
    , 
    \label{R_Z-result}
    \\
    R^2_X:= & \frac{1}{D-2\tilde{D}} \frac{g^2}{N} \sum_{I=2 \tilde{D}+1}^{D} \left\langle\Tr X^I X^{I } \right\rangle  ,
    \label{R_X-result}
\end{align}
and investigate them separately. 
(We have assumed that any spontaneous symmetry breaking on the rotation direction or the non-rotating  directions does not occur. Indeed, we do not observe any signal for it in the CLM.
We have also assumed that they are time independent.)\\

\begin{figure} [htbp]
    \centering
    \includegraphics[width=0.43\textwidth]{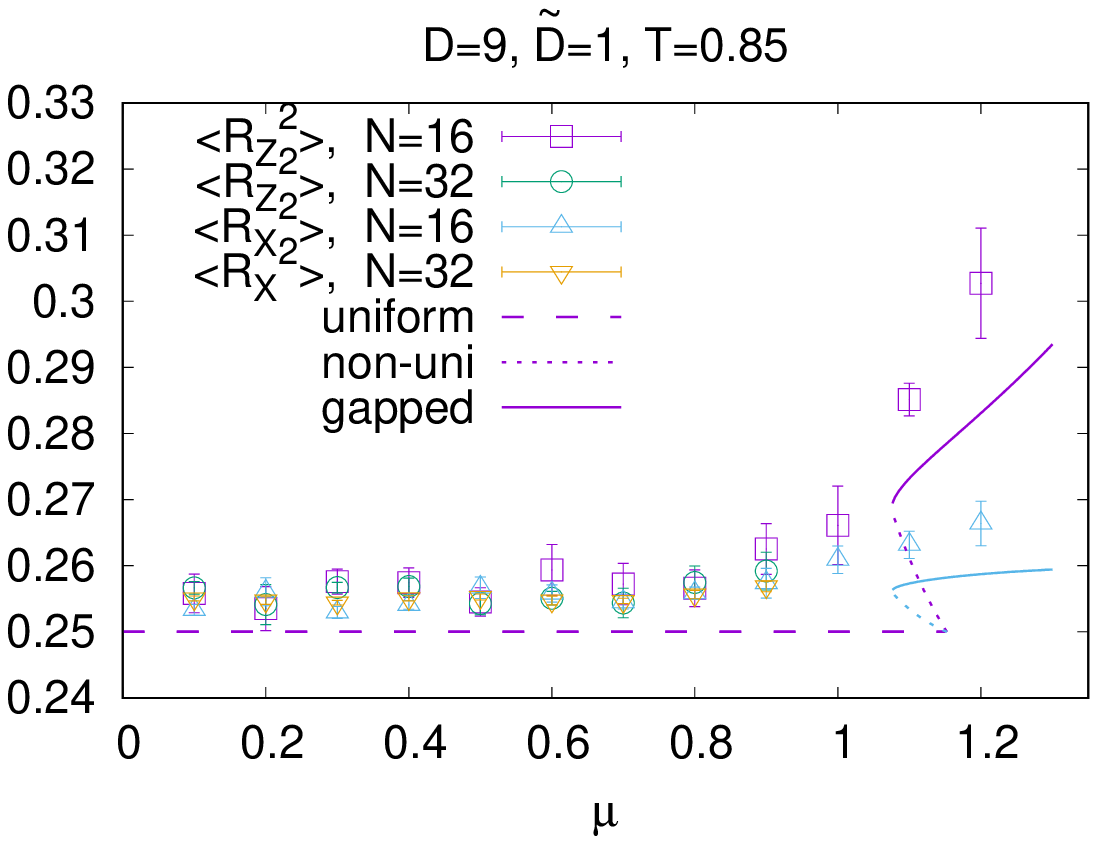}
    \includegraphics[width=0.43\textwidth]{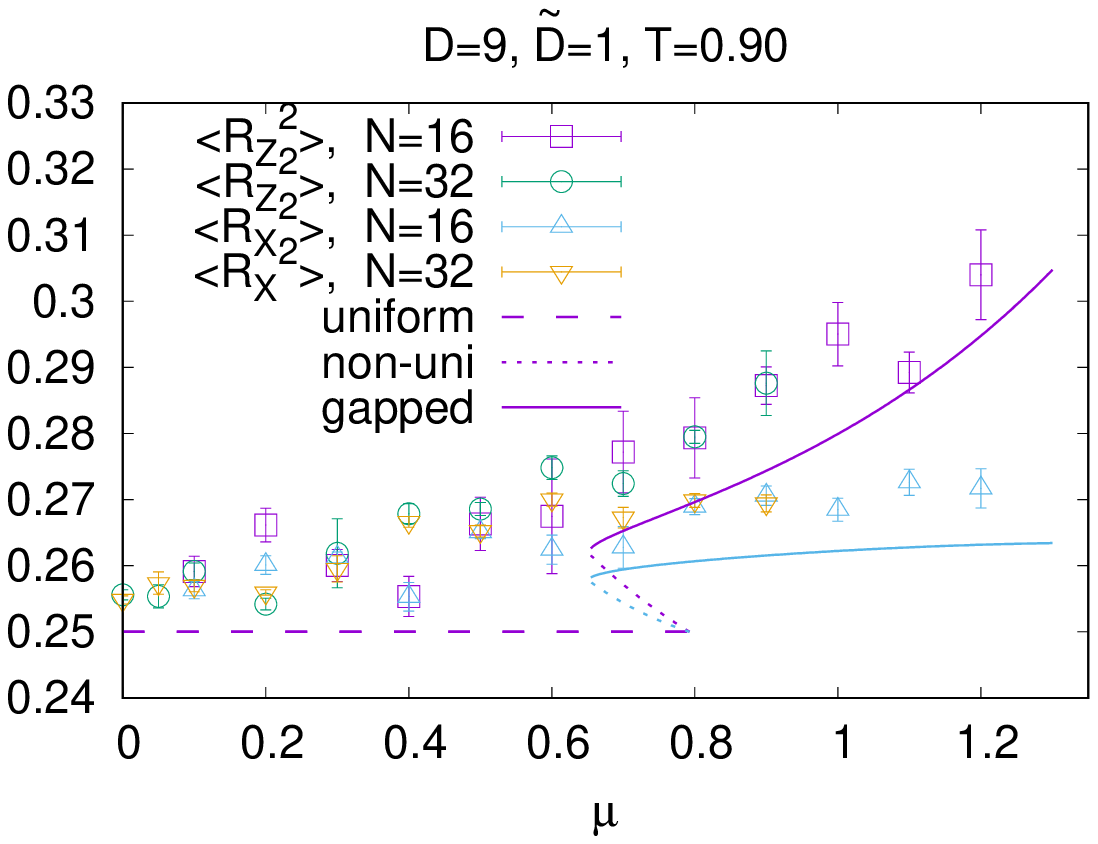}
    \includegraphics[width=0.43\textwidth]{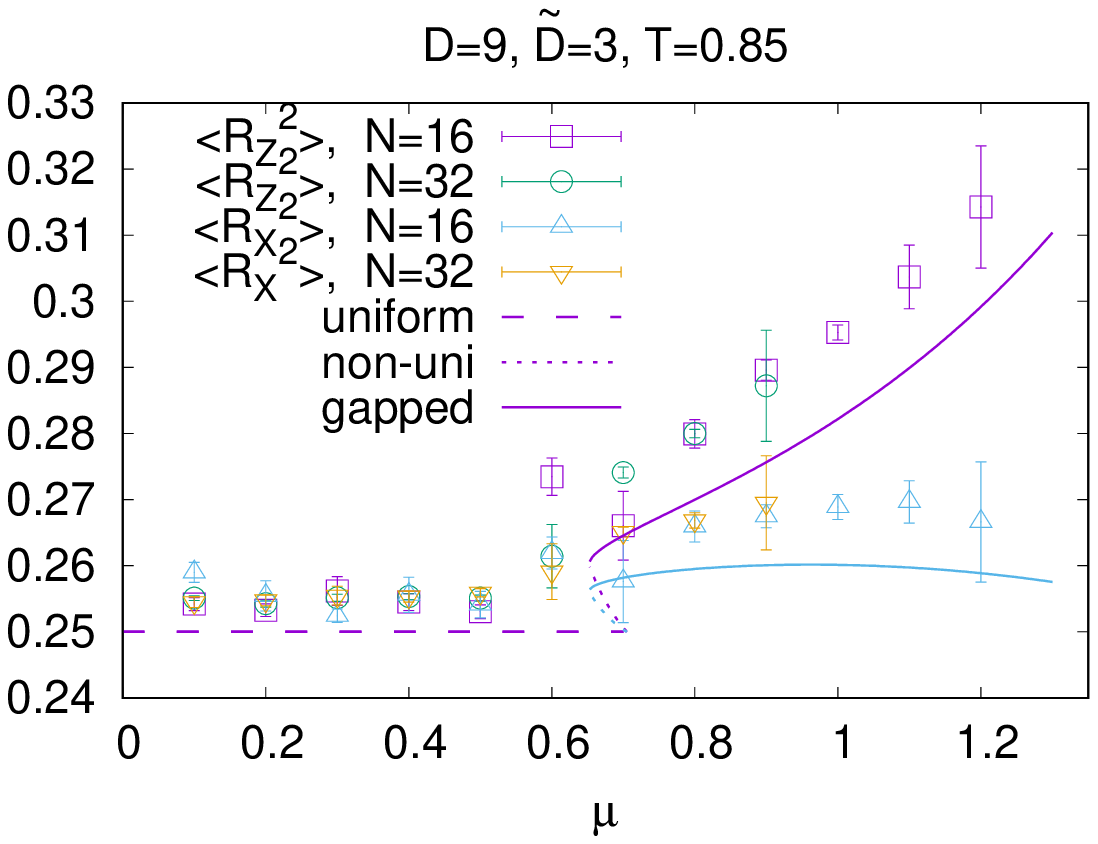}
    \includegraphics[width=0.43\textwidth]{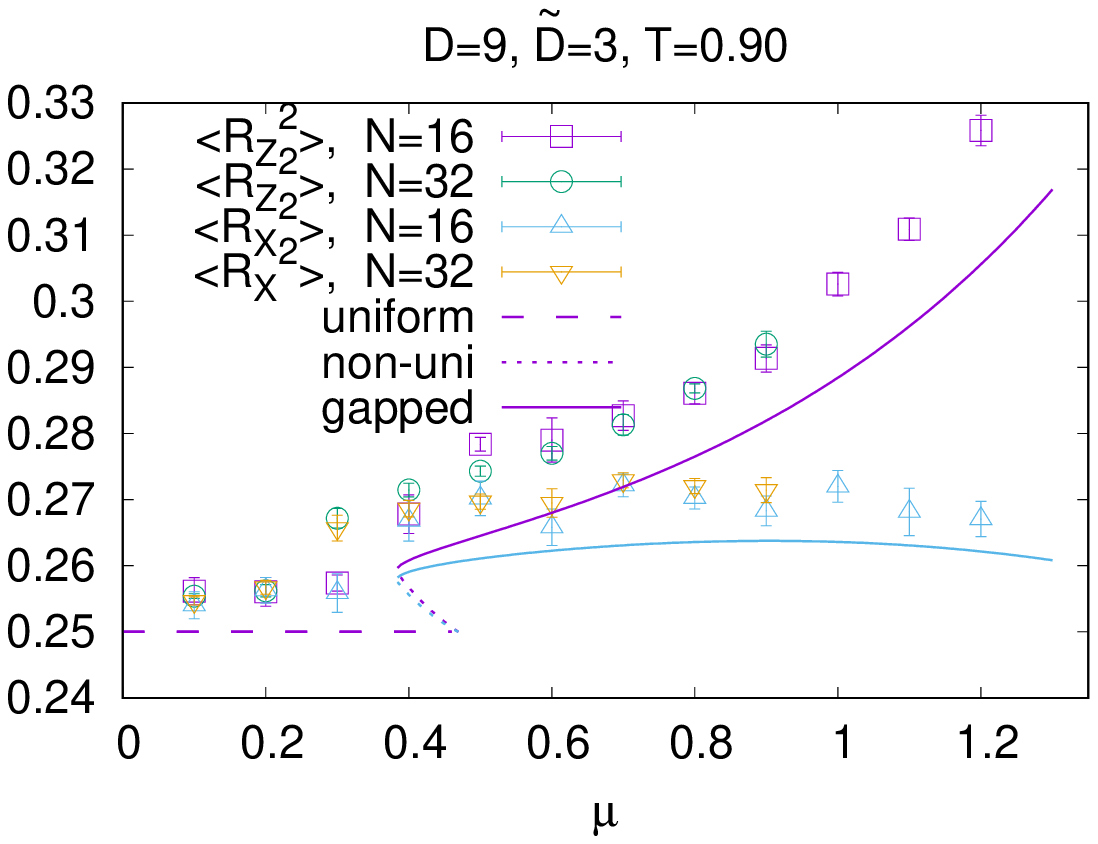}
    \includegraphics[width=0.43\textwidth]{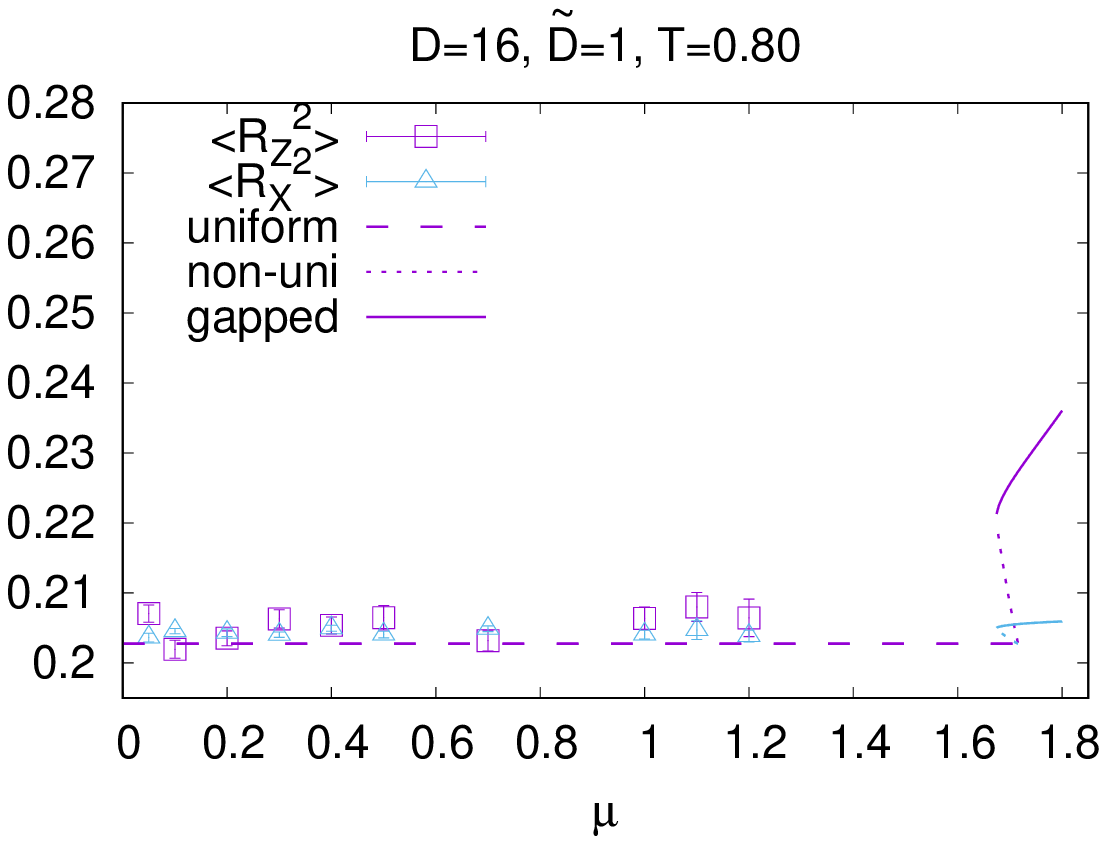}
    \includegraphics[width=0.43\textwidth]{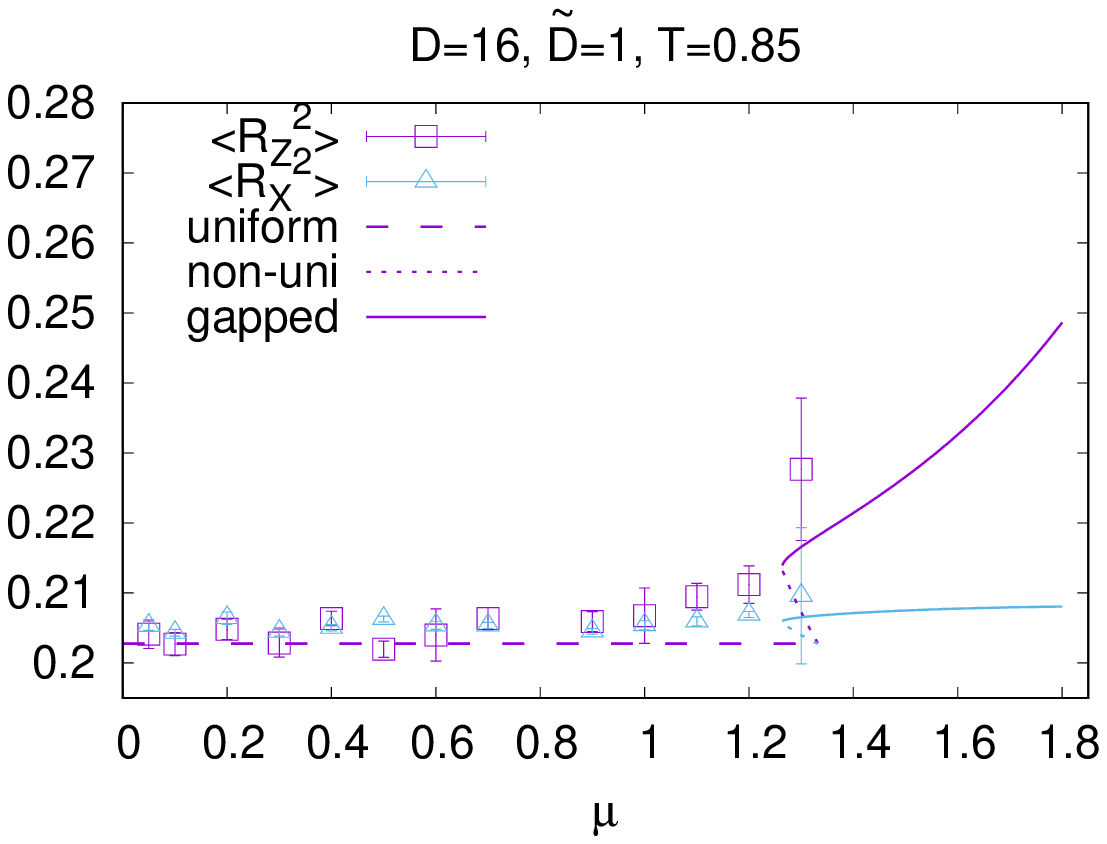}
    \includegraphics[width=0.43\textwidth]{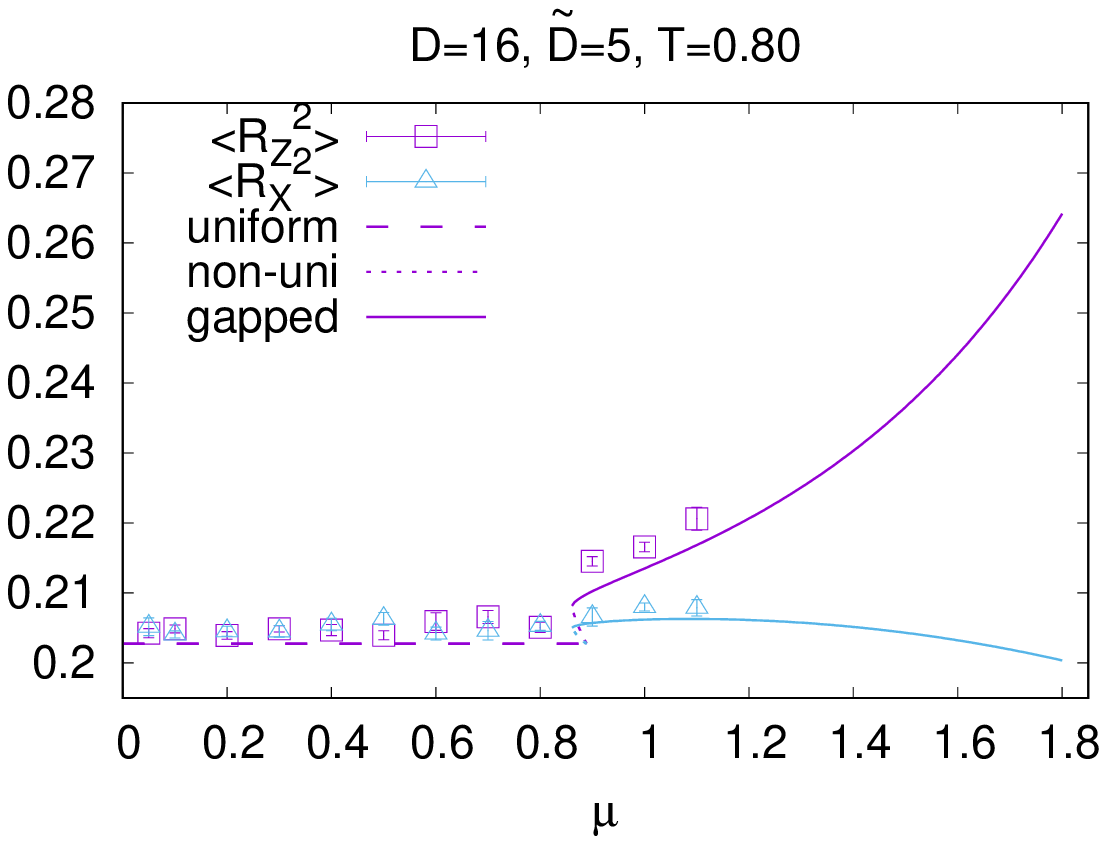}
    \includegraphics[width=0.43\textwidth]{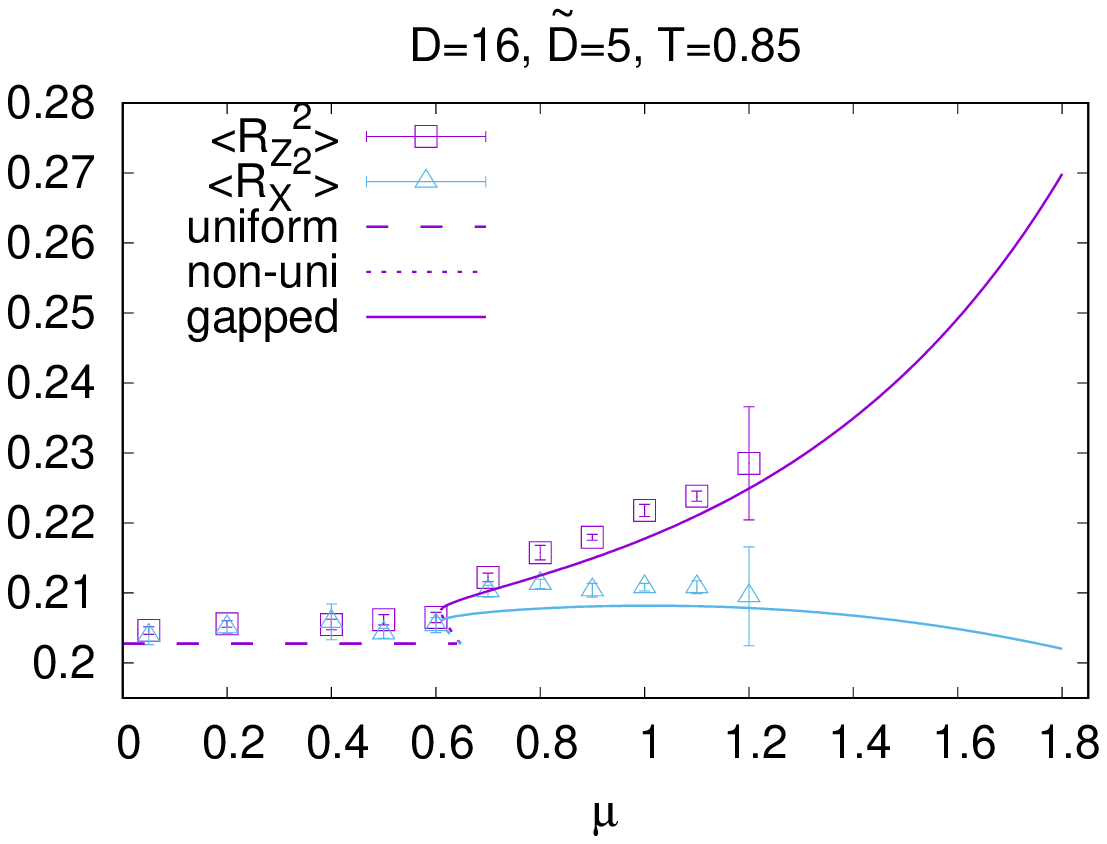}
    \caption{$R_Z^2$ and $R_X^2$ \eqref{R_X-result} are plotted against $\mu$. We present $\langle R_Z^2 \rangle$ and $\langle R_X^2 \rangle$ obtained by calculating \eqref{R_Z2} and \eqref{R_X2} via the CLM respectively, at $N=16,32$ for $D=9$ and $N=16$ for $D=16$. The lines denote the results of the minimum sensitivity (\ref{R_Z}) and (\ref{R_X}) at $N=\infty$. The dashed, dotted and solid lines represent those of the uniform, non-uniform and gapped phase, respectively.
    }\label{r2_D09}
\end{figure}

In Fig.~\ref{r2_D09}, the results of $R_Z^2$ and $R_X^2$ are plotted.
The discrepancies between the CLM results and the minimum sensitivity results are very small in the uniform phase and a few percent in the gapped phase.
They are getting worse as $\mu$ increases, where the approximation \eqref{large-D-approximation} in the minimum sensitivity is not reliable.
Again, the discrepancy is larger in the $D=9$ and $\tilde{D}=1$ at $T=0.90$ case due to the phase transition.

The both analyses capture important features of the rotating system. In the uniform phase, there is no separation between $R_Z^2$ and $R_X^2$. In the gapped phase, $R_Z^2$ gains a larger value, and $R_Z^2$ and $R_X^2$ begin to separate with each other. 
It means that the D-particles spread to the rotation directions as $J$ increases, and it is a natural consequence as rotating objects.

\section{Application of the CLM}
\label{sec-CLM}
In this section, we present the details of the application of the CLM \cite{Parisi:1983mgm,Klauder:1983sp}, which is a promising method to simulate the system with sign problem, to the action \eqref{action-BFSS-J}, and explain the derivation of the results shown in Sec.~\ref{sec-result}. 

\subsection{Complex Langevin equation}
In solving the complex Langevin equation for the action (\ref{action-BFSS-J}), we rescale it as
\begin{eqnarray}
    t = \lambda^{\frac{-1}{3}} t', \ \ A_t = \lambda^{\frac{1}{3}} {A'}_{t'}, \ \ X_{\mu} = g^{-1} \lambda^{\frac{1}{3}} {X'}_{\mu}, \ \ \mu = \lambda^{\frac{1}{3}} \mu', \label{BFSS_rescaling}
\end{eqnarray}
where $\lambda = g^2N$ is the 't Hooft coupling. This gives
\begin{eqnarray}
    S &=& N \int^{\beta'}_{0} dt' \textrm{Tr } \Biggl\{ \frac{1}{2} \sum_{I=1}^D (D_{t'} {X'}^I)^2 - \sum_{I,J=1}^D \frac{1}{4} [{X'}^I,{X'}^J]^2 - \frac{{\mu'}^2}{2} \sum_{K=1}^{2{\tilde D}} ({X'}^K)^2 \Biggr. \nonumber \\
    & & \ \ \Biggl. + \mu' i \sum_{K=1}^{{\tilde D}} \{ (D_{t'} {X'}^K) {X'}^{K+{\tilde D}} - (D_{t'} {X'}^{K+{\tilde D}}) {X'}^{K} \} \Biggr\}, \label{action-BFSS-J2}
\end{eqnarray}
where $\displaystyle D_{t'} = \partial_{t'} - i [A'_{t'}, ]$, $\beta' = \beta \lambda^{\frac{1}{3}}$
. We omit $'$ in the following. At $\mu=0$, this action is invariant under the transformations
\begin{eqnarray}
    & & X^I (t) \to X^I(t) + x^I I_N, \label{x_inv} \\
    & & A(t) \to A(t) + \alpha(t) I_N, \label{a_inv}
\end{eqnarray}
where $I_N$ is an $N \times N$ unit matrix. $x^I$ and $\alpha(t)$ are c-numbers, and $x^I$ has no dependence on $t$. The $\mu \neq 0$ case maintains the invariance under the transformation (\ref{a_inv}), but breaks the invariance under the transformation (\ref{x_inv}).

To put the action (\ref{BFSS_rescaling}) on a computer, we adopt a lattice regularization, where the number of the lattice sites is $n_t$ and the lattice space is $(\Delta t) = \frac{\beta}{n_t}$. We also adopt the periodic boundary condition $X_{\mu} (n_t+1) =X_{\mu} (1)$ and $A(n_t+1)=A(1)$. This yields the lattice-regularized action
\begin{eqnarray}
    S_{\textrm{lat}} &=& N \textrm{Tr } \sum_{n=1}^{n_t}  \Biggl\{ \frac{1}{2 (\Delta t)} \sum_{I=1}^{D} ( X^I (n+1) - V(n) X^I (n) V(n)^{-1} )^2 - \frac{(\Delta t)}{4} \sum_{I,J=1}^{D} [X^I(n), X^J (n)]^2 \Biggr. \nonumber \\
    & & \ \  - \frac{(\Delta t)}{2} \mu^2 \sum_{K=1}^{2 {\tilde D}} X^K (n)^2  + \mu i \sum_{K=1}^{{\tilde D}} (X^K (n+1) - V(n) X^K (n) V(n)^{-1} ) X^{K+{\tilde D}} (n) \nonumber \\
    & & \ \  - \Biggl. \mu i \sum_{K=1}^{{\tilde D}} (X^{K+{\tilde D}} (n+1) - V(n) X^{K+{\tilde D}} (n) V(n)^{-1} ) X^{K} (n) \Biggr\}, \label{action-BFSS-Jlat}
\end{eqnarray}
where $\displaystyle V(n) = e^{i A (n) (\Delta t)}$. We find it convenient to take a static diagonal gauge (\ref{gauge-diagonal}). 
In this gauge, we have
\begin{eqnarray}
    V(1) = V(2) = \cdots = V(n) = \textrm{diag } (e^{\frac{i \alpha_1}{n_t}}, \ e^{\frac{i \alpha_2}{n_t}}, \ \cdots, \ e^{\frac{i \alpha_N}{n_t}}).  \label{action-BFSS-Jlat-V}
\end{eqnarray}
Together with the gauge fixing term, as derived in Refs.~\cite{Aharony:2003sx, hep-th/0310286,hep-th/0601170}, we work on the action
\begin{eqnarray}
    S_{\textrm{eff}} = S_{\textrm{lat}} + S_{\textrm{g.f.}}, \textrm{ where } S_{\textrm{g.f.}} = - \sum_{k,\ell=1, \ k \neq \ell}^N \log \left| \sin \frac{\alpha_k - \alpha_{\ell} }{2} \right|, \label{S_eff}
\end{eqnarray}
The CLM consists of solving the complexified version of the Langevin equation. The complex Langevin equation is given by
\begin{eqnarray}
    \frac{d X^I_{k\ell} (n,\sigma)}{d \sigma} = - \ \frac{\partial S_{\textrm{eff}}}{\partial X^I_{\ell k} (n, \sigma)} + \eta^I_{k \ell} (n,\sigma), \ \  \frac{d \alpha_{k} (\sigma)}{d \sigma} = - \ \frac{\partial S_{\textrm{eff}}}{\partial \alpha_{k} (\sigma)} + \eta^{(\alpha)}_{k} (\sigma). \label{langevin_eq}
\end{eqnarray}
Here, $\sigma$ is the fictitious Langevin time, and the white noises $\eta^I_{k \ell} (n,\sigma)$ and $\eta^{(\alpha)}_{k} (\sigma)$ are Hermitian matrices and real numbers obeying the probability distribution proportional to $\exp \left( - \ \frac{1}{4} \int d \sigma  \sum_{n=1}^{n_t} \sum_{I=1}^{D} \textrm{Tr } \eta^I (n,\sigma)^2 \right)$ and $\exp \left( - \ \frac{1}{4} \int d \sigma \eta^{(\alpha)} (\sigma)^2 \right)$, respectively. The terms $\frac{\partial S_{\textrm{eff}}}{\partial X^I_{\ell k} (n, \sigma)}$ and $\frac{\partial S_{\textrm{eff}}}{\partial \alpha_{k} (\sigma)}$ are called ``drift terms".
The hermiticity of $X^I (n)$ and the reality of $\alpha_k$ are not maintained as the Langevin time $\sigma$ progresses, since the action $S_{\textrm{eff}}$ is complex. The expectation value of an observable ${\cal O}$ is evaluated as
\begin{eqnarray}
    \langle {\cal O}[X^I(n), \alpha_k ] \rangle =\frac{1}{\sigma_{T}} \int^{\sigma_0+\sigma_T}_{\sigma_0} d \sigma {\cal O}[X^I (n,\sigma), \alpha_k (\sigma)], \label{VEV_CLM}
\end{eqnarray}
where $\sigma_0$ is the thermalization time, and $\sigma_T$ is the time required for statistics, both of which should be taken to be sufficiently large. In Refs.~\cite{1101_3270,1606_07627,0912_3360}, it was found that the holomorphy of the observable ${\cal O}$ plays an essential role in the validity of Eq.~(\ref{VEV_CLM}). 

When we solve the Langevin equation (\ref{langevin_eq}) with a computer, we need to discretize it as
\begin{eqnarray}
    X^I_{k\ell} (n,\sigma + \Delta \sigma) &=& X^I_{k\ell} (n,\sigma) - \ (\Delta \sigma) \frac{\partial S_{\textrm{eff}}}{\partial X^I_{\ell k} (n, \sigma)} + \sqrt{\Delta \sigma} {\tilde \eta}^I_{k \ell} (n,\sigma), \label{langevin_eq2X} \\
    \alpha_{k} (\sigma +\Delta \sigma) &=& \alpha_k (\sigma) - \ (\Delta \sigma) \ \frac{\partial S_{\textrm{eff}}}{\partial \alpha_{k} (\sigma)} + {\tilde \eta}^{(\alpha)}_{k} (\sigma). \label{langevin_eq2a}
\end{eqnarray}
Here, $\Delta \sigma$ is the step size, which we take to be $\Delta \sigma = 10^{-5}$. The factor $\sqrt{\Delta \sigma}$ stems from the normalization of the discretized version of the white noises ${\tilde \eta}^I_{k \ell} (n,\sigma)$ and ${\tilde \eta}^{(\alpha)}_{k} (\sigma)$, which obey the probability distribution proportional to $\exp \left( - \ \frac{1}{4} \sum_{\sigma}  \sum_{n=1}^{n_t} \sum_{I=1}^{D} \textrm{Tr } {\tilde \eta}^I (n,\sigma)^2 \right)$ and $\exp \left( - \ \frac{1}{4} \sum_{\sigma} {\tilde \eta}^{(\alpha)} (\sigma)^2 \right)$, respectively. 

\subsection{Observables}
In the CLM, we calculate the Polyakov loop $u_n$, which is defined as Eq.~(\ref{polyakov_static_diag}), and the following observables.\footnote{In the CLM, if observables are not holomorphic, we may not obtain correct answers.
A subtle observable in our analysis is $|u_n|$ which is not holomorphic with respect to $\{ \alpha_k \}$.
However, at large $N$, $\langle |u_n| \rangle^2=\langle |u_n|^2 \rangle=\langle u_n  u_{-n} \rangle$ is held for real $\{ \alpha_k \}$, and, particularly, $\langle u_n  u_{-n} \rangle$ is holomorphic.
It turns out that this relation is approximately satisfied in our CLM in the relevant parameter region where we accept the CLM result based on the criterion described in Sec.~\ref{Subsec_CLM_test}.
Hence, evaluating $\langle |u_n| \rangle$ would not be problematic.}
\begin{eqnarray}
    J_{I} &=& J^{(1)}_{I} + J^{(2)}_{I} \ \ (I=1,2,\cdots, {\tilde D}), \ \ \textrm{ where } \label{JI_def} \\
    J^{(1)}_I &=& \frac{iN}{ \beta} \int^{\beta}_{0} \textrm{Tr} \{ X_I (t) (D_t X_{{\tilde D} +I}(t)) - X_{{\tilde D}+I} (D_t X_{I}(t)) \}, \nonumber \\
    J_I^{(2)} &=& \frac{\mu N}{ \beta}  \int^{\beta}_{0} \textrm{Tr} \{ X_I (t)^2 + X_{{\tilde D}+I} (t)^2 \}, \nonumber \\
    R_Z^2 &=& \frac{1}{2{\tilde D} N \beta} \int^{\beta}_{0}  \sum_{I=1}^{2 {\tilde D}} \textrm{Tr} X_I^2 (t) dt, \label{R_Z2} \\
    R_X^2 &=& \frac{1}{(D - 2 {\tilde D}) N \beta} \int^{\beta}_{0}  \sum^{D}_{I=2 {\tilde D}+1} \textrm{Tr} X_I^2 (t) dt. \label{R_X2}
\end{eqnarray}
In a rotating system, we cannot apply an angular momentum to the center-of-mass, which leads us to remove the trace part and implement the constraints
\begin{eqnarray}
    \frac{1}{N\beta} \int^{\beta}_0 \textrm{Tr} X^I (t) dt = 0, \ \ \textrm{Tr} A(t) = 0. \label{constraints_on_sun}
\end{eqnarray}
To this end, in the measurement routine we calculate the observables (\ref{polyakov_static_diag}), (\ref{JI_def}), (\ref{R_Z2}) and (\ref{R_X2}) in terms of the matrices
\begin{eqnarray}
    X^{I\textrm{(m)}}_{ij} (n) = X^I_{ij} (n) - \frac{1}{N n_t} \left( \sum_{n=1}^{n_t} \sum_{k=1}^N X^I_{kk} (n) \right) \delta_{ij}, \ \ \alpha^{\textrm{(m)}}_i = \alpha_i - \frac{1}{N} \sum_{k=1}^N \alpha_k. \label{constraint_on_sun_discre}
\end{eqnarray}
We solve the Langevin equation (\ref{langevin_eq}) without imposing the constraints (\ref{constraints_on_sun}), which facilitates the numerical calculation.

By using these methods, we perform the numerical computations at $n_t=60$ and obtain the results shown in Figs.~\ref{u1_D09}-
\ref{r2_D09}. As mentioned in Sec.~\ref{sec-result}, these results nicely agree with the minimum sensitivity results for finite $\mu$.
However, for larger $\mu$, we encounter several troubles in the CLM. 
We present them in the next subsection.

 \subsection{Testing the validity of our CLM} \label{Subsec_CLM_test}
The CLM faces the following two typical problems. One is the ``excursion problem", which occurs when $X^I$ and $\alpha_k$ are far from Hermitian matrix and real number, respectively.\footnote{The gauge cooling \cite{1211_3709,1508_02377,1604_07717} is a standard technique to suppress the excursion problem. However, in our complex Langevin studies we already fix the gauge as Eq.~(\ref{gauge-diagonal}), which prevents us from applying the gauge cooling. 
} The other is the ``singular drift problem", which occurs when the drift terms are too large. 
It is found in Ref.~\cite{1606_07627} that a sufficient condition to justify the CLM is that the probability distribution of the drift norms
\begin{eqnarray}
    u_X = \sqrt{\frac{1}{N^3 Dn_t} \sum_{n=1}^{n_t} \sum_{I=1}^D \sum_{k,\ell=1}^{N} \left| \frac{\partial S_{\textrm{eff}}}{\partial X^I_{\ell k} (n, \sigma)} \right|^2}, \ \ u_{\alpha} = \sqrt{\frac{1}{N} \left| \frac{\partial S_{\textrm{eff}}}{\partial \alpha_{k} (\sigma)} \right|^2} \label{drift_norms}
\end{eqnarray}
fall off exponentially or faster. If we look at the drift term, we get the drift of the CLM, and we can easily test this criterion.

\begin{figure} [h]
    \centering
    \includegraphics[width=0.43\textwidth]{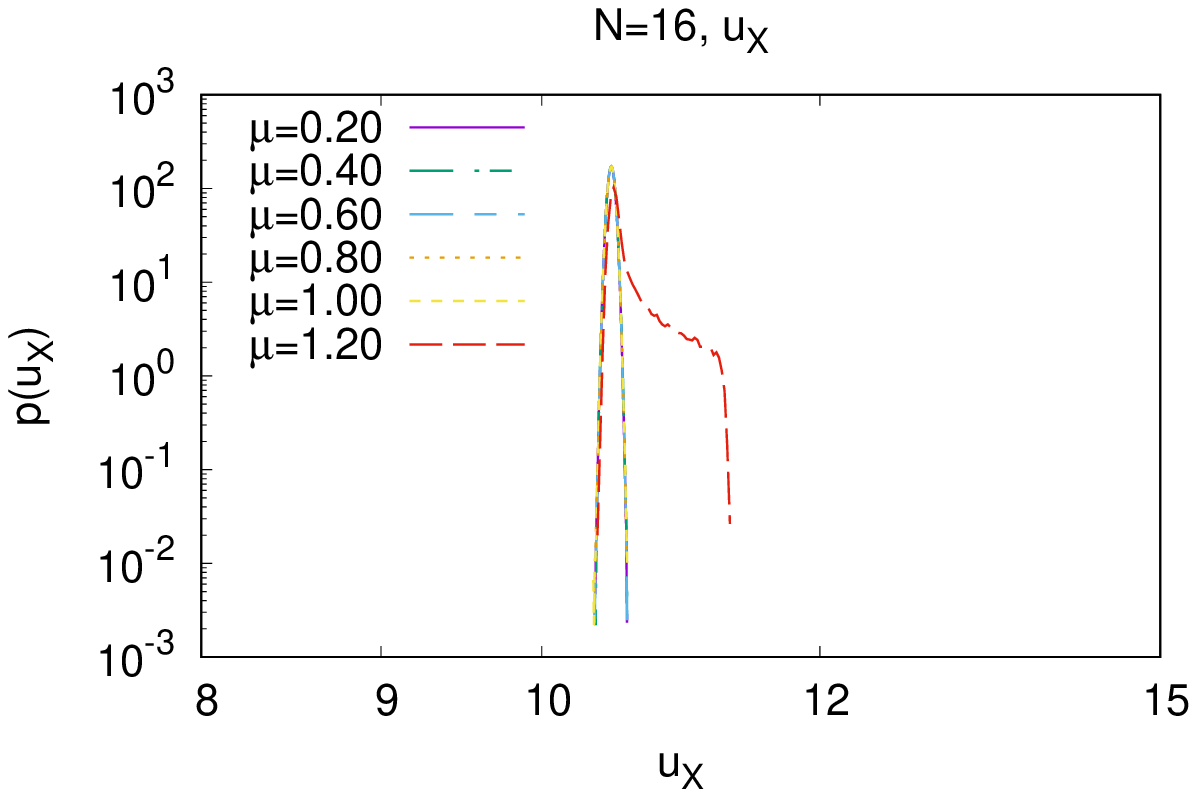}
    \includegraphics[width=0.43\textwidth]{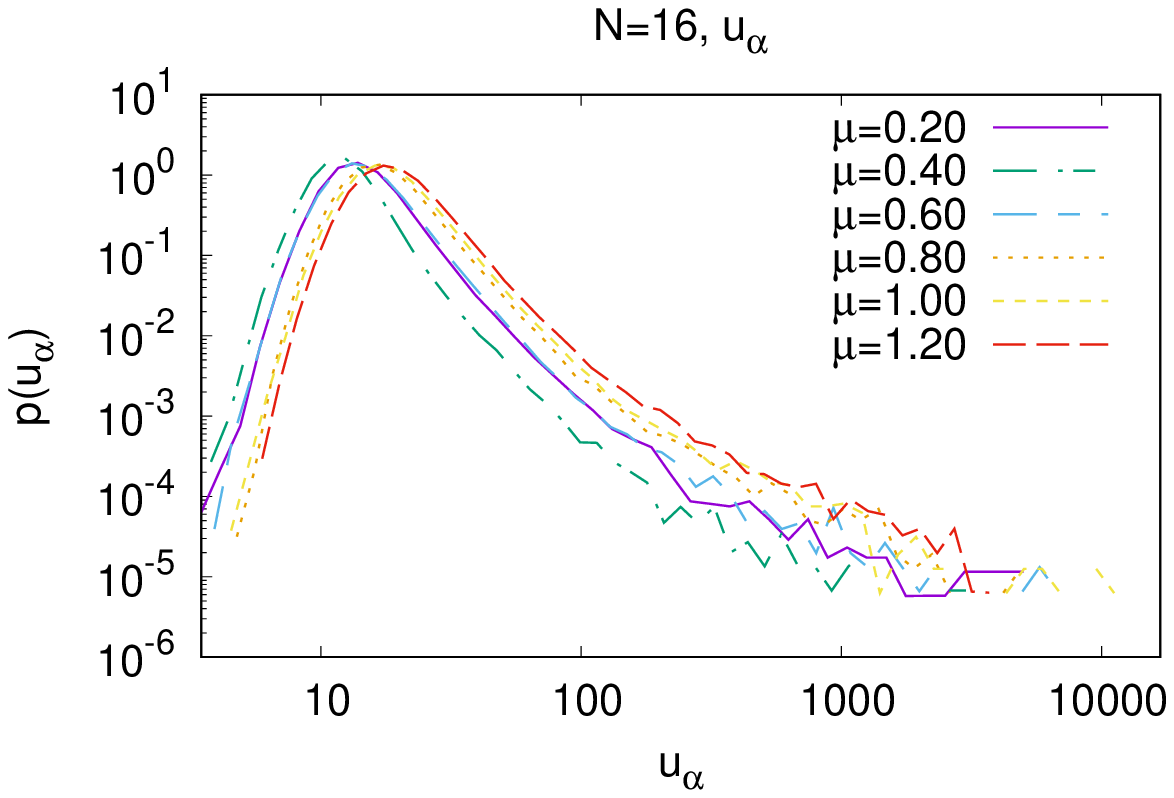}
    \includegraphics[width=0.43\textwidth]{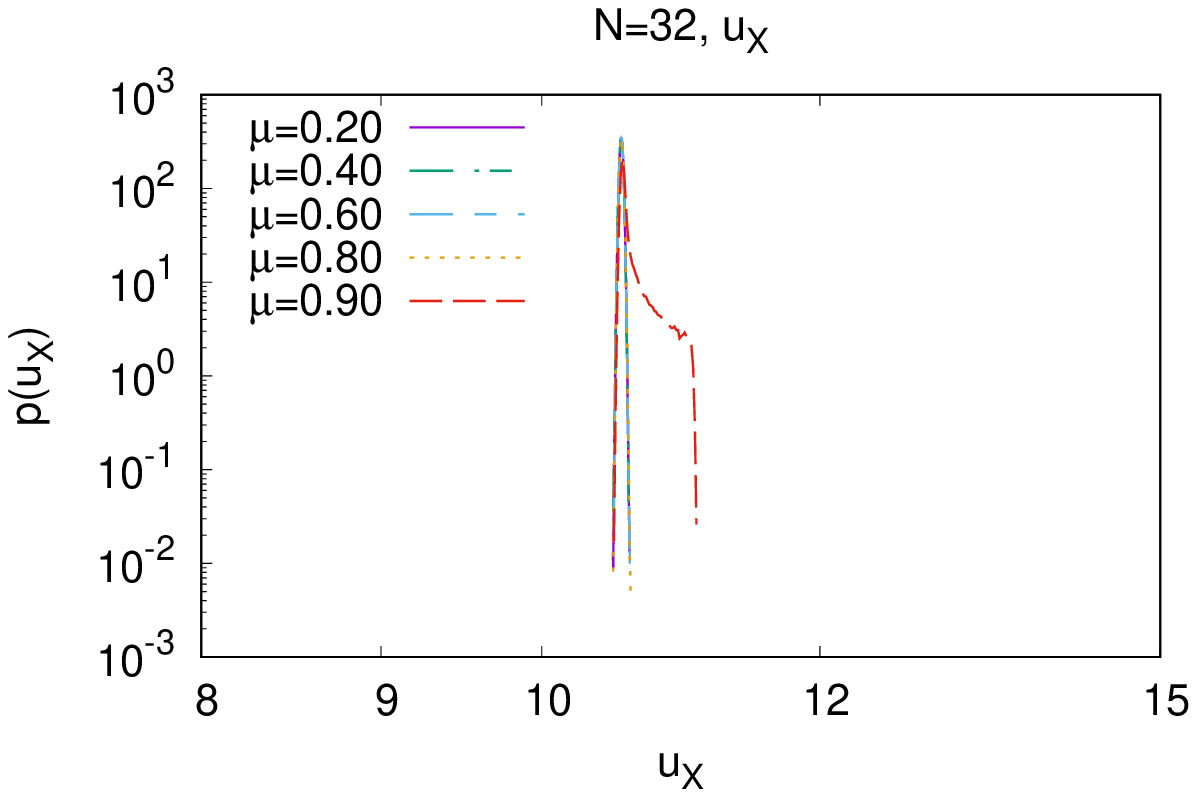}
    \includegraphics[width=0.43\textwidth]{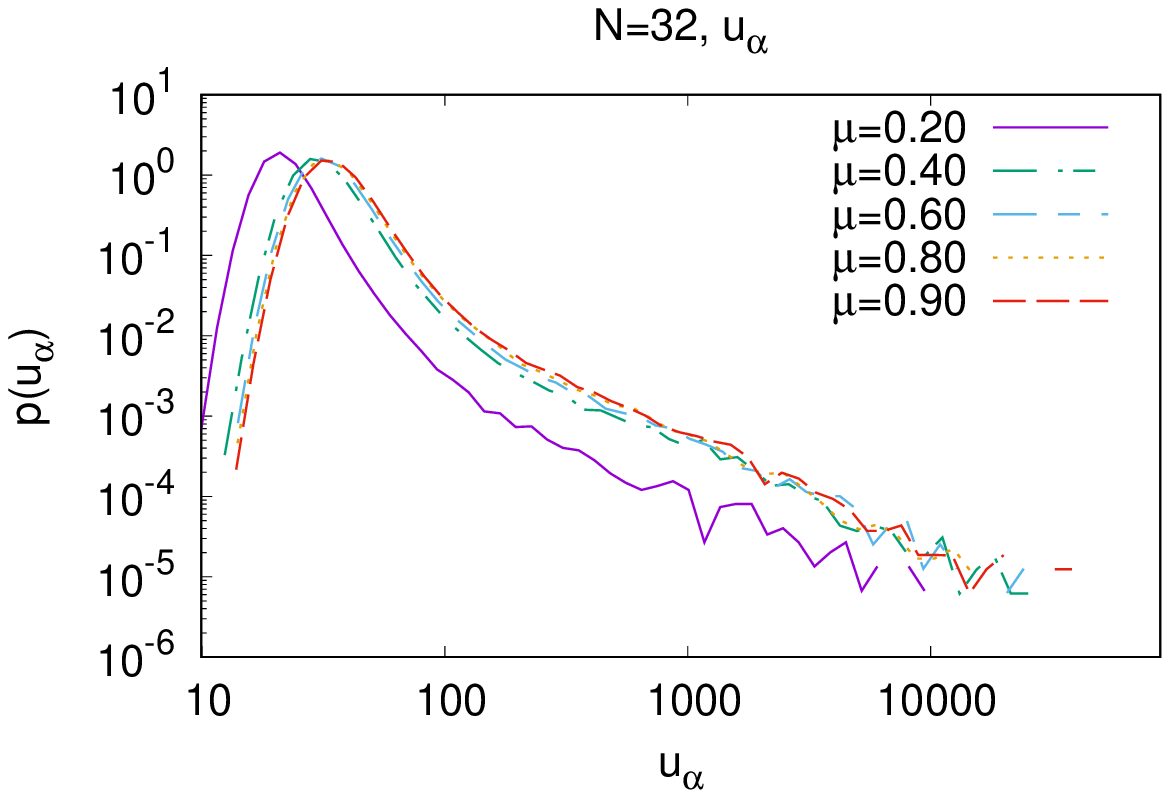}
    \caption{The log-log plot of the histogram of the drift norm $u_X$ (left) and $u_{\alpha}$ (right), for $D=9, {\tilde D}=1$, $T=0.90$ at $N=16$ (top) and $N=32$ (bottom).}\label{drift_norms_result}
\end{figure}

We work on the $D=9$, ${\tilde D}=1,3$, $T=0.85, 0.90$, $N=16,32$ and $D=16$, ${\tilde D}=1,5$, $T=0.80, 0.85$, $N=16$ cases. In these cases, we take  $n_t=60$. To probe the region of $\mu$ we can study by the CLM, we present the log-log plots of the probability distribution of the drift norms $u_X$ and $u_{\alpha}$, which we denote as $p(u_X)$ and $p(u_{\alpha})$, respectively. As a typical example, we show the $D=9, {\tilde D}=1, T=0.90$ case in Fig.~\ref{drift_norms_result}. $p(u_X)$ falls exponentially or faster for $\mu \leq 1.2$ at $N=16$ and $\mu \leq 0.9$ at $N=32$, respectively. On the other hand, $p(u_{\alpha})$ falls in a power law even for small $\mu$. As we present in Appendix \ref{appendix_drift_BFSS}, the power-law decay of $p(u_{\alpha})$ is observed even at $\mu=0$, which has no sign problem. At $\mu=0$, without the static diagonal gauge (\ref{gauge-diagonal}), the probability distribution $p(u_A)$, where $u_A$ is the drift norm without the static diagonal gauge as defined by Eq.~(\ref{drift_normV}), falls exponentially or faster. At $\mu=0$, we confirm the agreement of the observables between the cases with and without the static diagonal gauge (\ref{gauge-diagonal}). 
Also, in Ref.~\cite{1802_10381}, the static diagonal gauge (\ref{gauge-diagonal}) is taken to study the unitary matrix model, and the consistency between the CLM and analytic results is reported despite the power-law behavior of the drift norms as presented in Appendix B.
Hence, we presume that the power-law decay of $p(u_{\alpha})$ in the static diagonal gauge is harmless (Recall that the criterion in Ref.~\cite{1606_07627} is a sufficient condition). 

To save the CPU time, we take the static diagonal gauge (\ref{gauge-diagonal}) and accept the results for $\mu \leq 1.2$ at $N=16$ and $\mu \leq 0.9$ at $N=32$,
where $p(u_X)$ falls exponentially or faster while $p(u_{\alpha})$ does not.
When $\mu$ is so large as to be outside the parameter region where we accept the CLM result, the simulation gets unstable and crashes.
Similar trends are observed for other $D$, ${\tilde D}$ and $T$. In Figs.~\ref{u1_D09} - \ref{r2_D09}, we present the numerical results in the parameter region of $\mu$, where we accept the CLM result with this criterion. 

In the CLM, the hermiticity of $X_I(n)$ and the reality of $\alpha_i$ are lost, and the observables (\ref{JI_def}), (\ref{R_Z2}) and (\ref{R_X2}) are not real in general. In the parameter region where we accept the CLM result, we present their real part of the expectation values of the observables (\ref{JI_def}), (\ref{R_Z2}) and (\ref{R_X2}) obtained by the CLM, as the ensemble average of their imaginary part turns out to be close to 0. Also, the ensemble average of the imaginary part of $\alpha_i$ turns out to be close to 0.
These disappearances of the imaginary parts indicate the validity of our analysis.

\section{Minimum sensitivity analysis}
\label{sec-minimum}

In this section, we present the details of the derivation of the results shown in Sec.~\ref{sec-result} through the minimum sensitivity analysis.

\subsection{Derivation of Free energy}
\label{sec-free}

We study the thermodynamical properties of the model \eqref{action-BFSS-J} at large $N$ via a minimum sensitivity analysis \cite{Stevenson:1981vj}.
For this purpose, 
we introduce trial masses $m_Z$ and $m_X$ for $Z^I$ and $X^I$, respectively, 
and deform the action \eqref{action-BFSS-J} as \cite{Morita:2020liy}
\begin{align}
    \label{action-BFSS-MS}
    S_{\kappa}
    :=            & S_0+ \kappa S_{\rm int}, 
    \\
    S_0:=         & 
    \int_0^{\beta} \hspace{-2mm} dt  
    \Tr 
    \left\{ 
    \sum_{I=1}^{\tilde{D}}
    \left(
    D_t -\mu \right) Z^{ I \dagger}
    \left(
    D_t +\mu \right) Z^I  + m_Z^2 Z^{ I \dagger}	Z^I 
    + \sum_{I=2\tilde{D}+1}^D \frac{1}{2}
    \left(
    D_t X^I \right)^2+
    \frac{1}{2} m_X^2 X^{I2} 	\right\},      
    \label{S-free}
    \\
    S_{\rm int}:= & 
    \int_0^{\beta} \hspace{-2mm} dt  
    \Tr 
    \left\{ 
    - m_Z^2 Z^{ I \dagger}	Z^I
    -\frac{1}{2} m_X^2 X^{I2}
    -
    \sum_{I,J=1}^D \frac{g^2}{4} [X^I,X^J]^2
    \right\}. 
    \label{S-int}
\end{align}
Here $\kappa$ is a formal expansion parameter.
If we set $\kappa=1$, the mass terms are canceled and this action reproduces the original one \eqref{action-BFSS-J}.
However, if we perform a perturbative expansion with respect to $\kappa$ up to a certain order and take $\kappa=1$ after that, then the obtained quantities would depend on the trial mass parameters $m_Z$ and $m_X$.\footnote{In the action \eqref{action-BFSS-MS}, the gauge field $A_t$ interacts with $X^I$ and $Z^I$ through the covariant derivatives.
    However, thanks to the gauge fixing \eqref{gauge-diagonal}, $A_t$ does not prevent the perturbative computations with respect to $\kappa$.}
The idea of the minimum sensitivity is that we fix $m_Z$ and $m_X$ so that the dependence of a certain physical quantity on these parameters is minimized.
It has been demonstrated that this prescription works in various models\footnote{
To claim that the minimum sensitivity method works in a model, we need to compute higher loop corrections and evaluate the convergence. 
In the model \eqref{action-BFSS} at $\mu=0$, the minimum sensitivity computations in the confinement phase have been done up to three loops, and the deviations between the two-loop and three-loop calculations are a few percent \cite{Morita:2020liy}.
Thus, we expect that the minimum sensitivity at two-loop works in the model \eqref{action-BFSS} even at $\mu \neq 0$.
}.
Note that the obtained result through this method would depend on of which quantity we minimize the parameter dependence.
In our study, we investigate free energy at two-loop order, and minimize its $m_Z$ and $m_X$ dependence. 

By integrating out $X^I$ and $Z^I$ perturbatively, we obtain the effective action for $\{\alpha_k \}$ at two-loop order as shown in Appendix \ref{app-effective-action}, 
\begin{align}
    Z:= & \exp\left(-\beta F(T,\mu, m_X,m_Z) \right) \nonumber                                                                 \\
    :=  & e^{ -N^2  \beta f_0 (m_X,m_Z) } \int dU \exp\left[ -N^2 \left\{f_1(T,\mu, m_X,m_Z) -1 \right\}  |u_1|^2    \right] .
    \label{effective-action-gapped}
\end{align}
Here we have taken $\kappa=1$ and $dU$ is an integral measure  defined by \cite{Aharony:2003sx}
\begin{align}
    dU := \prod_{k} d\alpha_k e^{-S_{\textrm{g.f.}}}, \qquad
    \frac{1}{N^2}
    S_{\textrm{g.f.}}=\sum_{n=1}^{\infty}\frac{ 1}{n}|u_{n}|^{2}.
    \label{action-G.F.}
\end{align}
$f_0(m_X,m_Z) $ and $f_1(T,\mu, m_X,m_Z)$ are defined in Eqs.~\eqref{f0} and \eqref{fn}, respectively.
Whereas $f_1$ depends on $\beta$ and $\mu$, $f_0$ does not.
Note that we have used an approximation \eqref{large-D-approximation}, which is not reliable for large $T$ or $\mu$.
We have also used an assumption $\mu < m_Z$. (See Appendix \ref{app-effective-action} for the details.)
If this condition is not satisfied, the scalar $Z^I$ becomes tachyonic and the perturbative computations fail.

In order to derive the free energy $F$, we need to evaluate the $dU$ integral in Eq.~\eqref{effective-action-gapped}.
This integration at large $N$ has been studied in Ref.~\cite{Liu:2004vy}, and the result depends on the sign of $f_1$.
If $f_1>0$, the saddle point for the uniform solution (Fig.~\ref{fig-rho} [left]) dominates, while when $f_1<0$, the saddle point for a gapped solution (Fig.~\ref{fig-rho} [right]) does, and the free energy is given by \cite{Liu:2004vy}
\begin{empheq}[left={\beta F(T,\mu, m_X,m_Z)=\empheqlbrace}]{alignat=2}
    &N^2 \beta f_0 &\qquad &\text{$f_1>0$,}
    \label{F-low}
    \\
    & N^2 \left\{\beta f_0 - \frac{1}{2} \left( \frac{w}{1-w} + \log (1-w) \right)    \right\}     &       &\text{$f_1<0$}.
    \label{F-high}
\end{empheq}
Here we have defined
\begin{align}
    w:= \sqrt{\frac{-f_1}{1-f_1}}.
    \label{w}
\end{align}
Note that we treat $f_1=0$ case separately as we will explain in Sec.~\ref{sec-non-uniform}.

Correspondingly, the Polyakov loop \eqref{polyakov} at large $N$ is
computed as \cite{AlvarezGaume:2005fv}
\begin{empheq}[left={u_1=\empheqlbrace}]{alignat=2}
    & 0 &\qquad &\text{$f_1>0$,}
    \label{u1-low}
    \\
    & \frac{1+w}{2}      &       &\text{$f_1<0$} .
    \label{u1-high}
\end{empheq}
Here, we have taken the gauge mentioned in footnote \ref{ftnt-gauge}.
Similarly, for $u_n$ ($n \ge 2$), we obtain \cite{Rossi:1996hs, Okuyama:2017pil}
\begin{empheq}[left={u_n=\empheqlbrace}]{alignat=2}
    & 0 &\qquad &\text{$f_1>0$,}
    \label{un-low}
    \\
    & \frac{w^2}{n-1} P^{(1,2)}_{n-2}(2w-1)      &       &\text{$f_1<0$},
    \label{un-high}
\end{empheq}
where $P^{(1,2)}_{n-2}(z)$ denotes the Jacobi polynomial.
These results show that, when $f_1>0$, $u_n = 0$ for all $n$ and the density function $\rho(\alpha)$ becomes uniform through Eq.~\eqref{rho}.
Hence the system is confined.
On the other hand, when $f_1<0$, the gapped solution satisfies $u_n \neq 0$ and the system is deconfined.

Note that, when $f_1<0$, Eq.~\eqref{w} indicates $w>0$, and thus $u_1>1/2$.
Therefore, $u_1$ is discontinuous at $f_1=0$.
In Sec.~\ref{sec-non-uniform}, we will see that non-uniform solutions appear at $f_1=0$ and they fill this discontinuity.

We have so far evaluated the $dU$ integral in the partition function \eqref{effective-action-gapped}.
Now we fix the trial masses $m_X$ and $m_Z$ so that the dependence of the free energy $F(T,\mu, m_X,m_Z)$ on these masses is minimized.
Hence, we solve the equations
\begin{align}
    \partial_{m_X}F(T,\mu, m_X,m_Z)=0, \quad \partial_{m_Z}F(T,\mu, m_X,m_Z)=0.
    \label{minimum-F}
\end{align}
In the subsequent subsections, we evaluate these equations in each $f_1>0$, $f_1<0$ and $f_1=0$ cases.
It will determine the free energies of each solution, and they tell us their stabilities and phase structures as drawn in Fig.~\ref{fig-phase}.

\subsubsection{Uniform solution}
\label{sec-confinement}

In this subsection, we evaluate Eq.~\eqref{minimum-F} when $f_1>0$, which is in the confinement phase.
In this case, the free energy becomes $F=N^2f_0$ through Eq.~\eqref{F-low}, and we solve $\partial_{m_X}f_0=\partial_{m_Z}f_0=0 $, where $f_0$ is defined in Eq.~\eqref{f0}.
Then, we obtain
\begin{align}
    m_X=m_Z=m_0:=(D-1)^{1/3}\lambda^{1/3}.
    \label{m-2-loop}
\end{align}
By using this result, the free energy is given by
\begin{align} 
    F=N^2 f_0 (m_0,m_0)=\frac{3D}{8} (D-1)^{1/3} \lambda^{1/3}.
    \label{F-uni}
\end{align}
Hence, the free energy in this solution depends on neither temperature nor the chemical potential.
See Figs.~\ref{fig-D=9} and \ref{fig-F} for the $D=9$ case.

Note that, when we derived the effective action \eqref{effective-action-gapped}, we had assumed $m_Z < \mu $.
Hence, the uniform solution is not reliable in the region $ \mu \ge m_0 $.
For example, in the $D=9$ with $\tilde{D}=3$ case shown in Fig.~\ref{fig-phase}, such a region appears in the uniform phase, and the fate of the system there is not obvious.
It is likely that the system is unstable in this region and any stable phase does not exist.

\subsubsection{Gapped solution}
\label{sec-gapped}
\begin{figure}
    \begin{center}
        \begin{tabular}{cc}
            \begin{minipage}{0.5\hsize}
                \begin{center}
                    \includegraphics[scale=0.8]{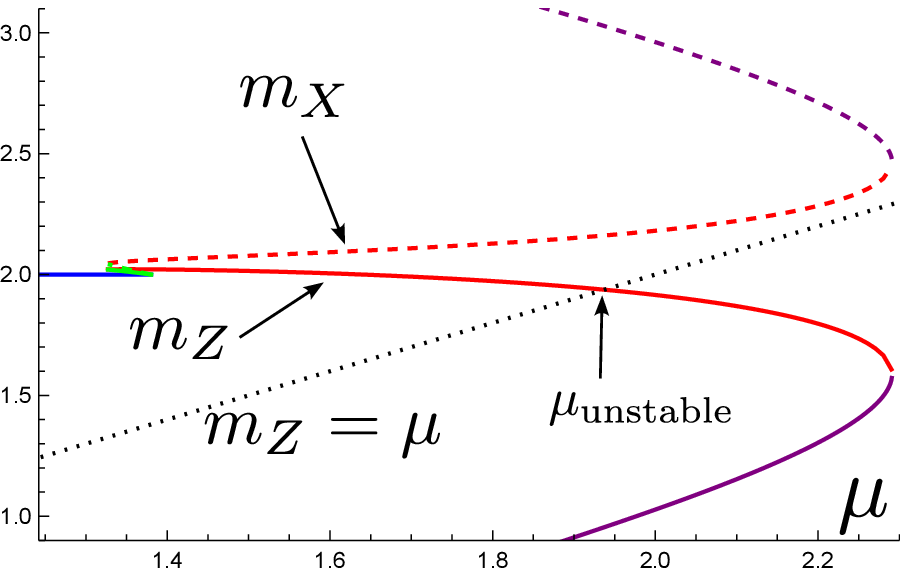}\\
                    $D=9$ with $\tilde{D}=1$ at $T=0.80$
                \end{center}
            \end{minipage}
            \begin{minipage}{0.5\hsize}
                \begin{center}
                    \includegraphics[scale=0.8]{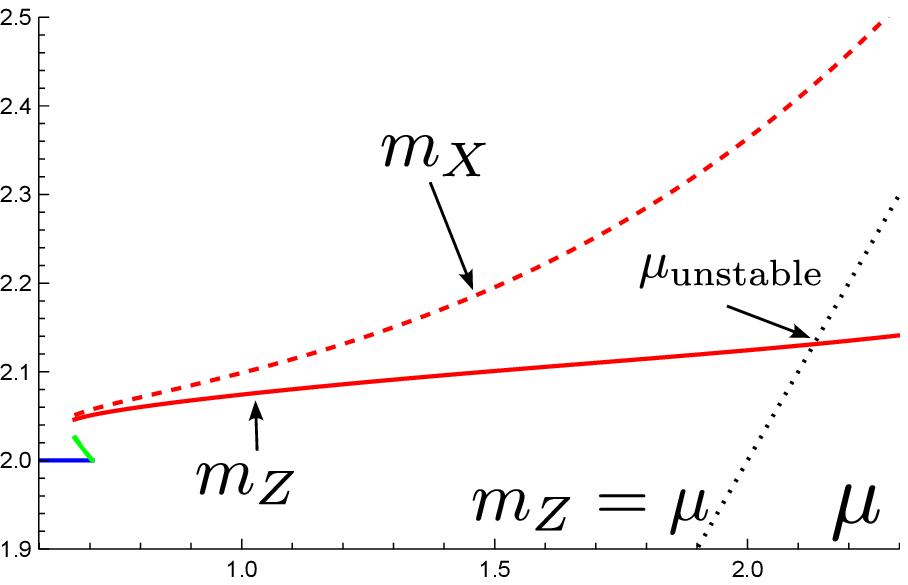}\\
                    $D=9$ with $\tilde{D}=3$ at $T=0.85$
                \end{center}
            \end{minipage}
        \end{tabular}
        \caption{
            Trial masses $m_X$ and $m_Z$ in the minimum sensitivity analysis.	
            These are obtained by solving Eq.~\eqref{minimum-F}.
            The solid lines describe $m_Z$ and the dashed lines describe $m_X$.
            The blue, green and red colors represent the uniform, non-uniform and gapped solutions, respectively.
            In the $D=9$ with $\tilde{D}=1$ case, the second gapped solutions represented by the purple lines exist for each $\mu$, while only the single gapped solution exists in the $\tilde{D}=3$ case.
            Our analysis is valid until $m_Z > \mu$, and the solutions of $m_Z$ below the black dotted line denoting $ m_Z = \mu$ are not reliable.
            The borders of the valid solutions are marked as $\mu_\text{unstable}$.
            Hence, the second gapped solution at $\tilde{D}=1$ (the purple solution) is always not reliable.
        }
        \label{fig-m}
    \end{center}
\end{figure}

\begin{figure}
    \begin{center}
        \begin{tabular}{cc}
            \begin{minipage}{0.5\hsize}
                \begin{center}
                    \includegraphics[scale=0.8]{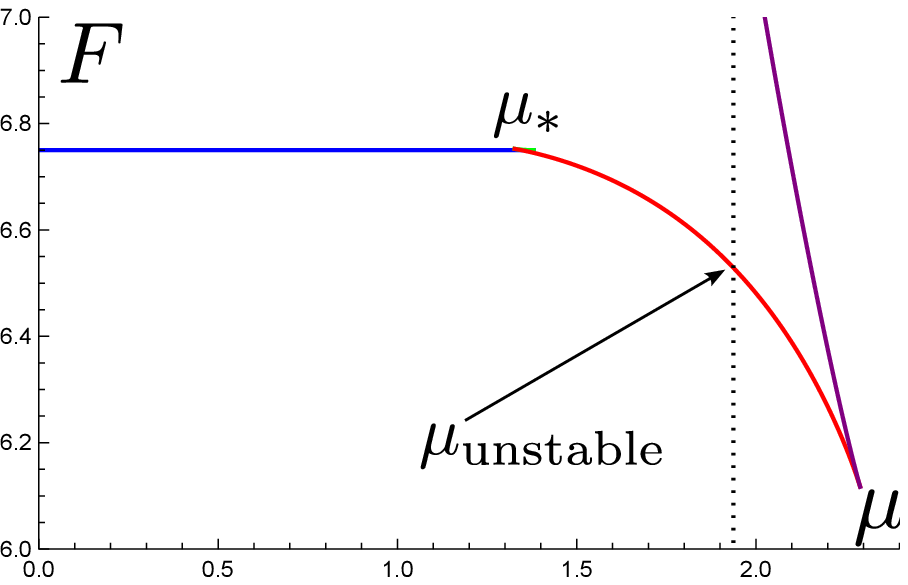}\\
                \end{center}
            \end{minipage}
            \begin{minipage}{0.5\hsize}
                \begin{center}
                    \includegraphics[scale=0.8]{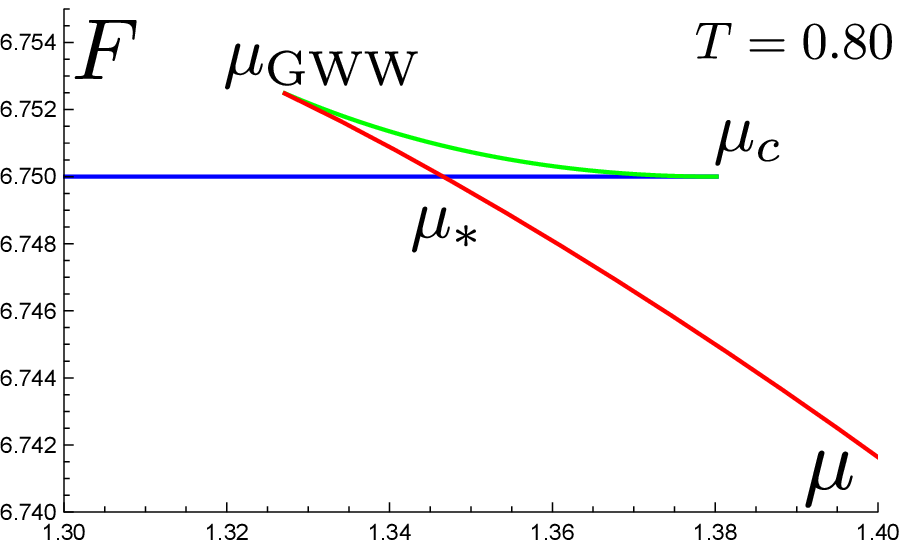}\\
                \end{center}
            \end{minipage}
        \end{tabular}
        \caption{Free energy in the $D=9$ with $\tilde{D}=1$ case at $T=0.80$.
            The right plot is an enlarged view of the left plot near the transition point $\mu=\mu_*$.
            The blue, green and red curves represent the uniform, non-uniform and gapped solutions, respectively.
            The system shows the first-order phase transition between the uniform and gapped phase at $\mu_*$.
            Our analysis for the gapped solution is reliable in the region $\mu < \mu_\text{unstable}$, where the condition $\mu < m_Z$ is satisfied.
            The purple curve is for the gapped solution shown in Fig.~\ref{fig-m}, although it appears in the unreliable region.
        }
        \label{fig-F}
    \end{center}
\end{figure}

We discuss the $f_1<0$ case.
Here, we cannot solve Eq.~\eqref{minimum-F} analytically, and we evaluate it numerically.
For a fixed $T$ and $\mu$, we may find several solutions of $m_X$ and $m_Z$.
See Fig.~\ref{fig-m} for $D=9$ with $\tilde{D}=1$ and $\tilde{D}=3$.
However, the condition $m_Z  > \mu $ had been assumed when we derived the effective action \eqref{effective-action-gapped}, and the solutions that do not satisfy it are not reliable.
As far as we investigated, only the solutions connected to the non-uniform solution at $\mu=\mu_{\rm GWW}(T)$, which we will argue in the next subsection, are reliable. 
As we increase $\mu$ with a fixed $T$, even these reliable solutions reach the point $m_Z  = \mu $, which we call $\mu=\mu_\text{unstable}$, and the fates of the systems beyond it are unclear.
These regions are presented as ``unknown" in the phase diagrams in Fig.~\ref{fig-phase}.
We presume that the systems are unstable in these regions.

%

%

\subsubsection{Non-uniform solution}
\label{sec-non-uniform}

We have seen that the uniform solutions and the gapped solutions appear when $f_1>0$ and $f_1<0$, respectively.
Thus, a transition would occur at $f_1=0$, and we investigate this case in details.
For this purpose, it is useful to rewrite the partition function \eqref{effective-action-gapped} as
\begin{align}
    \label{effective-action-un}
    Z= & \int \prod_{n=1} du_n ~\exp(-S_{\text{eff}}\left(\{ u_n \},m_X,m_Z\right)), \nonumber                                        \\  
       & S_{\text{eff}}\left(\{ u_n \},m_X,m_Z\right)=  N^2 \left(\beta f_0 +  f_1 |u_1|^2 +\sum_{n=2} \frac{1}{n} |u_n|^2  \right) .
\end{align}
Here, we have changed the integral variables from $\{ \alpha_k \}$ to $\{ u_n \}$.
(Such a change is possible in the large-$N$ limit \cite{Aharony:2003sx}.)
Note that $\{ u_n \}$ are not completely independent, since they have to satisfy the condition that the density function $\rho$ \eqref{rho} is non-negative.

The action \eqref{effective-action-un} is quadratic in $\{ u_n \}$, and their coefficients are all positive for $n \ge 2$.
Thus, $u_n=0$ ($n \ge 2$) is a stable solution.
Hereafter, we assume $u_n=0$ ($n \ge 2$) and focus on $u_1$.

By differentiating the action \eqref{effective-action-un} by $u_1$, we obtain an equation
\begin{align}
    f_1(T,\mu, m_X,m_Z) u_1^*=0 .
\end{align}
Obviously, one solution is given by $u_1^*=0$, which represents the uniform solution studied in Sec.~\ref{sec-confinement}.
The other possible solution is 
\begin{align}
    f_1(T,\mu, m_X,m_Z) =0 .
    \label{EOM-u1}
\end{align}
This is what we are interested in.
Besides, we have the conditions \eqref{minimum-F} that the dependence of the free energy on $m_X$ and $m_Z$ is minimized, 
\begin{align} 
     & \beta \partial_{m_X} f_0 (m_X,m_Z)+  \partial_{m_X} f_1(T,\mu, m_X,m_Z) |u_1|^2 =0, 
    \label{EOM-mx}
    \\ 
     & \beta \partial_{m_Z} f_0 (m_X,m_Z)+  \partial_{m_Z} f_1(T,\mu, m_X,m_Z) |u_1|^2 =0.
    \label{EOM-mz}
\end{align}
By combining these equations and Eq.~\eqref{EOM-u1}, we obtain three equations,
\begin{align} 
    u_1=\sqrt{-\frac{\beta \partial_{m_X} f_0}{\partial_{m_X} f_1}}, \qquad 	f_1 =0,
    \qquad \frac{ \partial_{m_X} f_0}{\partial_{m_X} f_1}=\frac{ \partial_{m_Z} f_0}{\partial_{m_Z} f_1}.
    \label{u_1-NU}
\end{align}
The last two equations determine $m_X$ and $m_Z$, and the first one fixes $u_1$.
These equations can be solved numerically, and the results for $D=9$ with $\tilde{D}=1$ and $\tilde{D}=3$
are shown in Fig.~\ref{fig-m}.

Then, the eigenvalue density \eqref{rho} becomes
\begin{align} 
    \rho(\alpha)=\frac{1}{2\pi} \left(1+ 2 u_1 \cos \alpha \right) .
    \label{rho-non-uniform}
\end{align}
This solution represents a non-uniform solution plotted in Fig.~\ref{fig-rho}, if $u_1 \neq 0$.
However, if $u_1 > 1/2$, the eigenvalue density \eqref{rho-non-uniform} becomes negative around $\alpha=\pi$.
Thus, the non-uniform solution is allowed only for $|u_1|\le 1/2$.
Recall that $u_1= 1/2$ is the lower bound of the gapped solution \eqref{u1-high}, and a transition to the gapped solution from the non-uniform one occurs at $u_1= 1/2$.
This transition is called a Gross-Witten-Wadia (GWW) type transition \cite{Gross:1980he, Wadia:2012fr}, and we define this transition point as $\mu=\mu_{\rm GWW}(T)$.
See Figs.~\ref{fig-D=9} and \ref{fig-F}.

On the other hand, the non-uniform solution also merges to the uniform one at $u_1 = 0$.
Through Eqs.~\eqref{m-2-loop} and \eqref{EOM-u1}, it occurs when $\mu$ and $T$ satisfies 
\begin{align}
    f_1(T,\mu, m_0,m_0)=0 .
    \label{eq-T_c}
\end{align}
We define this solution as $\mu_{c}(T)$, and call it a critical point.
Therefore the non-uniform solutions exist in the region $\mu_{\rm GWW}(T) \le \mu  \le  \mu_{c}(T) $.
See Fig.~\ref{fig-D=9} and \ref{fig-F}, again.
Note that $f_1$ may be negative beyond the critical point $\mu=\mu_{c}(T)$.
Since $f_1$ is the coefficient of $|u_1|^2$ in the effective action \eqref{effective-action-un}, the uniform solution becomes unstable in this case.

\subsubsection{Free energy and phase structure}
\label{sec-free-energy}
So far, we have obtained the three solutions: uniform, non-uniform and gapped solution. 
To see the phase structure, we evaluate their free energies.
For the uniform solution, we have derived the free energy in Eq.~\eqref{F-uni}.
For the non-uniform and gapped solutions, we obtain their free energies by substituting the numerical solutions $m_X$ and $m_Z$ into Eqs.~\eqref{effective-action-un} and \eqref{F-high}.
These results are summarized as
\begin{empheq}[left={ F/N^2=\empheqlbrace}]{alignat=3}
    & \frac{3D}{8} (D-1)^{1/3}  &\qquad &\text{(uniform solution)},
    \nonumber
    \\
    &  f_0 (m_X,m_Z) &\qquad &\text{(non-uniform solution)},
    \label{F-summary}
    \\
    &  f_0 (m_X,m_Z)- \frac{1}{2 \beta} \left( \frac{w}{1-w} + \log (1-w) \right)     &\qquad &\text{(gapped solution)} .
    \nonumber
\end{empheq}
Note that we have used $f_1=0$ and $u_n=0$ ($n \ge 2$) for the non-uniform solution.

The free energy for the $D=9$ with $\tilde{D}=1$ case at $T=0.80$ is plotted in Fig.~\ref{fig-F}.
This figure shows that a first-order transition occurs in this system. 
There, the transition point is given when the free energy of the uniform solution and the gapped one are coincident.
We define $\mu_{*}(T) $ for this point.\footnote{We also use $T_*(\mu)$, $T_\text{GWW}(\mu)$ and $T_c(\mu)$ instead of $\mu_*(T)$, $\mu_\text{GWW}(T)$ and $\mu_c(T)$.}
Fig.~\ref{fig-F} also shows that the free energy of the non-uniform solution is always higher than those of the uniform solution and the gapped one.
Actually, the free energy of the non-uniform solution is a concave function, and the specific heat is negative.
These results imply that the non-uniform solution is always unstable in the grand canonical ensemble.

One feature of this phase transition is that neither the non-uniform solution nor the gapped one exists in the region $\mu < \mu_{\rm GWW}(T)$.\footnote{We have seen that several gapped solutions may exist in our model, and the non-uniform solution is connected to one of the gapped solutions at the GWW point $ \mu_{\rm GWW}(T)$.
    Thus, other gapped solutions might exist even in the region $\mu < \mu_{\rm GWW}(T)$. 
}
This is due to our approximated effective action \eqref{effective-action-un}, and, if the action involves higher-order terms such as $u_n u_m u_{-n-m}$, the location of the GWW transition point $\mu_{\rm GWW}(T)$ would change \cite{AlvarezGaume:2005fv}. 

By combining all the results, the whole phase diagrams are obtained as drawn in Fig.~\ref{fig-phase}.

\subsection{Calculating observables}
\label{sec-observalbes}

We have investigated the phase structure of the model. 
Now, we explain the derivation of the observables shown in Figs.~\ref{u1_D09} - \ref{r2_D09} in Sec.\ref{sec-result-observables}.
The results in this section are for the large-$N$ limit, unless it is specified.

\subsubsection{Polyakov loops}
\label{subsec-Polyakov}

We evaluate the Polyakov loop operators \eqref{polyakov}, which are the order parameters of the confinement/deconfinement transition.
In the uniform solution, $u_n=0$ is obtained through Eqs.~\eqref{u1-low} and \eqref{un-low} at large $N$.
We can also derive the leading $1/N$ correction \eqref{un-finite-N} as discussed in Appendix \ref{app-un}.
The result is given by
\begin{align}
    \label{un-finite-N-main}
    u_n = \frac{1}{2N} \sqrt{\frac{\pi}{f_n(\beta,\mu, m_0,m_0)}}
    +\textrm{O} \left(\frac{1}{N^3} \right), \qquad (\text{uniform solution}).
\end{align}
(This result does not work near the critical point $f_1=0$.)

For the non-uniform solution, we have obtained 
\begin{align} 
    u_1= & \sqrt{-\frac{\beta \partial_{m_X} f_0}{\partial_{m_X} f_1}}, \quad u_n= 0 \quad (n \ge 2), \qquad   (\text{non-uniform solution}),
\end{align}
at large $N$ as argued in Sec.~\ref{sec-non-uniform}.
Here, $m_X$ and $m_Z$ are the solutions of Eq.~\eqref{u_1-NU}.

For the gapped solution, the following solution has been derived
\begin{align} 
    u_1= & \frac{1+w}{2}, \quad 	u_n= \frac{w^2}{n-1} P^{(1,2)}_{n-2}(2w-1) \quad (n \ge 2)	, \qquad   (\text{gapped solution}),
    \label{un-gapped-main}
\end{align}
in Eqs.~\eqref{u1-high} and \eqref{un-high}.
Here $w$ has been defined in Eq.~\eqref{w}, and $m_X$ and $m_Z$ are the solutions of Eq.~\eqref{minimum-F}.

The results for $u_1$ are plotted in Figs.~\ref{fig-D=9} and \ref{u1_D09}, and $u_2$ is shown in Fig.~\ref{u2_D09}.
As we have seen in Sec.~\ref{sec-result-Polyakov}, they are consistent with the CLM.

\subsubsection{Angular momentum}
\label{subsec-angular}

We have derived the free energy \eqref{F-summary} in Sec.~\ref{sec-free-energy}.
By using this result, we can read off the angular momentum via
\begin{align} 
    J=- \frac{1}{\tilde{D}} \frac{\partial F}{\partial \mu} .
    \label{eq-J}
\end{align}
(As we mentioned in Sec.~\ref{sec-result-J}, we calculate the angular momentum for the single plane. 
Hence, we have divided $-(\partial F/ \partial \mu)$ by $\tilde{D}$.)
The results are compared with the CLM as shown in Fig.~\ref{JI_D09}.
Interestingly, $J$ decreases as $\mu$ increases in the non-uniform solution.
This property would be related to thermodynamical instabilities of the non-uniform solution.
Besides, $J=0$ in the uniform phase, because the free energy \eqref{F-summary} does not depend on $\mu$, there\footnote{$J=0$ in the large-$N$ limit means that $J$ is not an O$(N^2)$ quantity. Thus, $J$ may be  O$(1)$.}.
Thus, the uniform phase does not rotate at large $N$, although the chemical potential is finite.
(Similarly, entropy in the uniform phase is zero even at finite temperatures. It means that thermal excitations are highly suppressed in the uniform phase indicating a confinement \cite{Sundborg:1999ue, Aharony:2003sx}.)

\subsubsection{ Expectation values of scalars }
\label{subsec-scalars}

We evaluate the expectation values of the scalars $R_Z^2$ and $R^2_X$ defined in Eqs.~\eqref{R_Z-result} and \eqref{R_X-result}.
Through the one-loop computation \eqref{2-looop-2pt}, we obtain
\begin{align}
    R_Z^2= & \frac{1}{\tilde{D}} \frac{g^2}{N} \sum_{I=1}^{\tilde{D}} \left\langle \Tr Z^{I\dagger} Z^{I } \right\rangle
    = \lambda	 \left\{ \frac{1 }{2m_Z}
    + \sum_{n=1}^{\infty}\frac{ 1  }{2m_Z} z^{n} (q^{n}+q^{-n}) |u_{n}|^{2} \right\} 
    , 
    \label{R_Z}
    \\
    R^2_X= & \frac{1}{D-2\tilde{D}} \frac{g^2}{N} \sum_{I=2 \tilde{D}+1}^{D} \left\langle\Tr X^I X^{I } \right\rangle
    = \lambda	 \left\{ \frac{1}{2 m_X}
    + \sum_{n=1}^{\infty}\frac{1 }{ m_X} x^{n} |u_{n}|^{2} \right\}  .
    \label{R_X}
\end{align}
To compute them, the suitable solutions for $m_X$, $m_Z$ and $u_n$ need to be substituted.
The results are plotted in Fig.~\ref{r2_D09}.

\section{Imaginary chemical potentials and relation to rotating YM theory in four dimensions}
\label{sec-Im}

Rotating quark gluon plasma (QGP) is actively being studied motivated by relativistic heavy ion colliders \cite{STAR:2017ckg} and neutron stars.
As a related problem, rotating pure YM theories are also being investigated.
Particularly, one important question is how the rotation affects the confinement/deconfinement transition temperatures.
(Rotating media are non-uniform in space, and the transition temperatures on the rotation axis is mainly studied.)
However, these theories are strongly coupled and standard perturbative computations do not work.
In addition, the sign problem prevents lattice MC computations.

To avoid these issues, the imaginary angular velocity  $\mu_{\rm Im} \in \mathbf{R} $ is considered \cite{Yamamoto:2013zwa, Braguta:2021jgn, Chen:2022smf, Chernodub:2022wsw, Chernodub:2022veq}.
(This imaginary angular velocity  $\mu_{\rm Im}  $ corresponds to the angular momentum chemical potential $\mu$ in our model \eqref{action-BFSS-J} as $\mu=i \mu_{\rm Im} $ through the dimensional reduction, and we call $\mu_{\rm Im}  $ a imaginary chemical potential, hereafter.)
The imaginary chemical potential does not cause the sign problem and MC works.
Once we obtain the results for the imaginary chemical potential,
through the analytic continuation, we may reach the results for the real chemical potential.
Such analytic continuation would work as far as the chemical potential is sufficiently small.

In this section, we review some results in pure YM theories with the imaginary chemical potential.
Then, we compute the corresponding quantities in the matrix model \eqref{action-BFSS-J} through the minimum sensitivity, and compare them with the YM theories.
We will see some similarity between the matrix model \eqref{action-BFSS-J} and the YM theories, and the matrix model provides some insights into the YM theories.

\subsection{Stable confinement phase at high temperatures}

Recently, one remarkable result on the SU(3) pure YM theory was reported in Ref.~\cite{Chen:2022smf}.
The authors investigated the high temperature regime ($T \to \infty$), where the perturbative computation is reliable, and found that the system is in a confinement phase for $ \pi/2 \le \beta \mu_{\rm Im} \le 3\pi /2 $ and in a deconfinement phase for $ 0 \le \beta \mu_{\rm Im} \le \pi/2$ and $  3\pi /2 \le \beta \mu_{\rm Im} \le 2\pi $.
Hence, the system is confined, although temperature is high.
See Fig.~4 in Ref.~\cite{Chen:2022smf}.
(Note that, when the imaginary chemical potential is turned on, the Boltzmann factor is multiplied by $\exp(i \beta \mu_{\rm Im} J)$, and the thermal partition function is periodic with respect to $\beta \mu_{\rm Im}$: $Z(\beta \mu_{\rm Im})=Z(\beta \mu_{\rm Im}+2\pi)$.)

Then, one important question is whether this high temperature confinement phase in $ \pi/2 \le \beta \mu_{\rm Im} \le 3\pi /2 $  continues to the low temperature confinement phase at $  \beta \mu_{\rm Im} =0 $.
To answer this question, we need to study the strong coupling regime in the YM theory, and it has not been understood.

This result motivates us to study the imaginary chemical potential in our matrix model \eqref{action-BFSS-J}.
Particularly, exploring the high temperature regime and investigating the fate of the confinement phase at $T \to \infty$ would be valuable.

The analysis through the minimum sensitivity is almost straightforward.
We simply need to repeat the same computations done in Sec.~\ref{sec-minimum} by using the effective action \eqref{effective-action-gapped} with  $\mu=i \mu_{\rm Im} $.
To investigate the phase structure at high temperatures, we evaluate the critical point \eqref{eq-T_c} in the limit  $T \to \infty$.
Then, through Eq.~\eqref{fn}, we obtain
\begin{align} 
    f_1(T,\mu= i \mu_{\rm Im}, m_0,m_0) \to 1- D +2 \tilde{D} \left\{1- \cos \left( \beta \mu_{\rm Im}\right) \right\}, \qquad (T \to \infty),
\end{align}
and the critical point is derived through the condition $f_1=0$.
Thus, if the relation
\begin{align} 
    1-D+4  \tilde{D} \ge 0
    \label{cond-pi}
\end{align}
is satisfied, the phase transition at $T \to \infty$ occurs at
\begin{align} 
    \beta \mu_{{\rm Im}0}  :=\arccos\left( \frac{1-D+2  \tilde{D}}{2  \tilde{D}} \right),
\end{align}
and the system is confined in $\beta \mu_{{\rm Im}0} \le \beta \mu_{{\rm Im}} \le 2\pi - \beta \mu_{{\rm Im}0} $.
On the other hand, if Eq.~\eqref{cond-pi} is not satisfied, the transition at $T \to \infty$ does not occur and the high temperature confinement phase does not exist.
Thus, the existence of the high temperature confinement phase depends on $D$ and $\tilde{D}$ via Eq.~\eqref{cond-pi}.\footnote{
    The condition \eqref{cond-pi} can be rewritten as $	  D-2 \tilde{D} \le  2 \tilde{D}+1  $.
    Here, $(D-2  \tilde{D})$ is the number of the scalar $X^I$ for the non-rotating directions and $(1+2 \tilde{D}) $ is the number of the scalars for the rotational directions plus 1 that is the contribution of the gauge fixing. 
    Ref.~\cite{Chen:2022smf} argued that the gauge fields $A^I$ for the rotational directions at high temperature behave as ghost modes.
    Therefore, the condition \eqref{cond-pi} states that, if the number of the ghost modes is greater than that of the ordinary scalars at high temperature, the system is confined.
}

Interestingly, in the $D=3$ and $\tilde{D}=1$ case that is the dimensional reduction of the four-dimensional YM studied in Ref.~\cite{Chen:2022smf},  the system is confined in $\pi/2 \le \beta \mu_{{\rm Im}} \le 3 \pi/2 $.
This is the same result as that of Ref.~\cite{Chen:2022smf}, although SU(3) is taken in Ref.~\cite{Chen:2022smf} and we have taken the large-$N$ limit.
Therefore, our model might capture the phase transition of the original model.

\begin{figure}
    \begin{center}
        \begin{tabular}{cc}
            \begin{minipage}{0.5\hsize}
                \begin{center}
                    \includegraphics[scale=0.8]{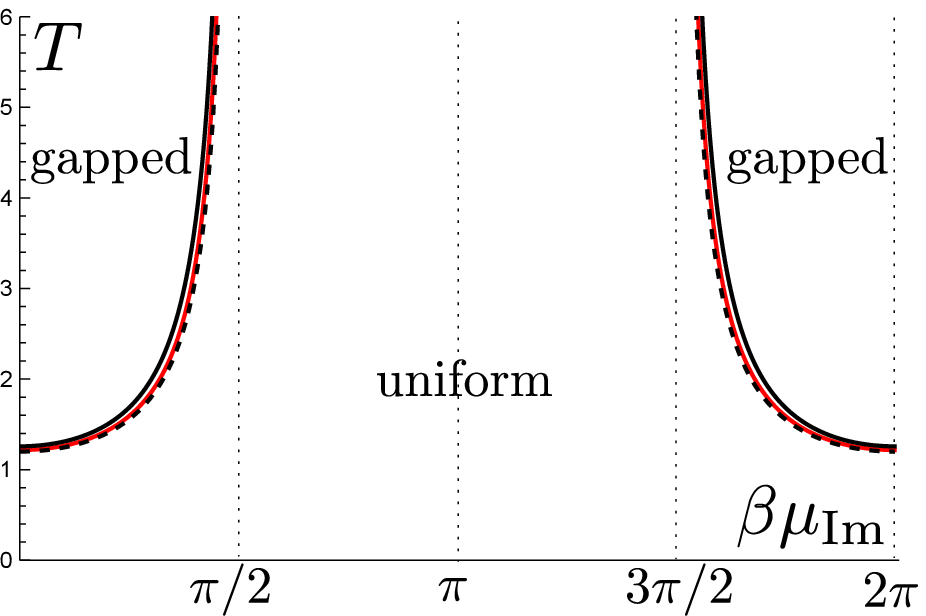}\\
                    $D=3$, $\tilde{D}=1$
                \end{center}
            \end{minipage}
            \begin{minipage}{0.5\hsize}
                \begin{center}
                    \includegraphics[scale=0.8]{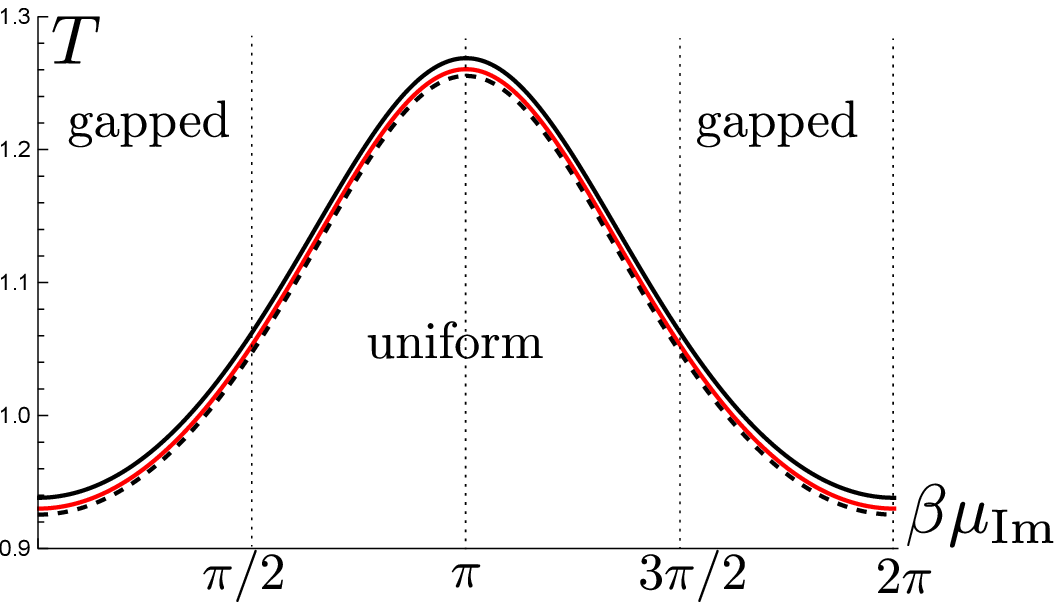}\\
                    $D=9$, $\tilde{D}=1$
                \end{center}
            \end{minipage}
        \end{tabular}
        \caption{ $(\beta \mu_{\rm Im})$-$T$ phase diagrams of the matrix model \eqref{action-BFSS-J} with the imaginary chemical potential.
            The black solid line represents the critical point $T_c(\beta  \mu_{\rm Im})$. The black dashed line represents the GWW point $T_{\textrm{GWW}} ( \beta \mu_{\rm Im})$. The red solid line represents $T_{*}(\beta \mu_{\rm Im} )$, where the first-order phase transition occurs. 
            In the $D=3$ and $\tilde{D}=1$ case, the critical point approaches to $\beta \mu_{\rm Im}=\pi/2$ and $3\pi/2$ asymptotically as $T \to \infty$.
        }
        \label{fig-Im}
    \end{center}
\end{figure}

We also derive the whole phase structures in the $D=3$ with $\tilde{D}=1$ case and the $D=9$ with $\tilde{D}=1$ case as drawn in Fig.~\ref{fig-Im}.
In the $D=9$ with $\tilde{D}=1$ case, the condition \eqref{cond-pi} is not satisfied, and the high temperature confinement phase does not appear.
In the $D=3$ with $\tilde{D}=1$ case, we observe that the high temperature confinement phase continues to the conventional confinement phase at  $ \mu_{\rm Im} = 0 $.
This suggests that the confinement phase may be continuous in the four-dimensional YM theory, too.

\subsection{Analytic continuation of the chemical potential}

Refs.~\cite{Chen:2022smf,  Chernodub:2022veq} argued that the imaginary chemical potential increases the transition temperature in the pure YM theory.\footnote{ Through a lattice computation, Ref.~\cite{Braguta:2021jgn} showed the opposite prediction that the imaginary chemical potential makes the transition temperature lower. Thus, the influence of the imaginary chemical potential is still under debate.}
Through the analytic continuation $\mu \to i \mu_{\rm Im}$, this result implies that the decreasing transition temperature under the presence of the real chemical potential at least for small $\mu$.
However, it is unclear whether such an analytic continuation provides quantitatively good results for a finite chemical potential.
Thus, it may be valuable to test the analytic continuation in our matrix model, since the transition temperatures for each the real and imaginary chemical potential can be computed.

\begin{figure}
    \begin{center}
        \begin{tabular}{cc}
            \begin{minipage}{0.5\hsize}
                \begin{center}
                    \includegraphics[scale=0.8]{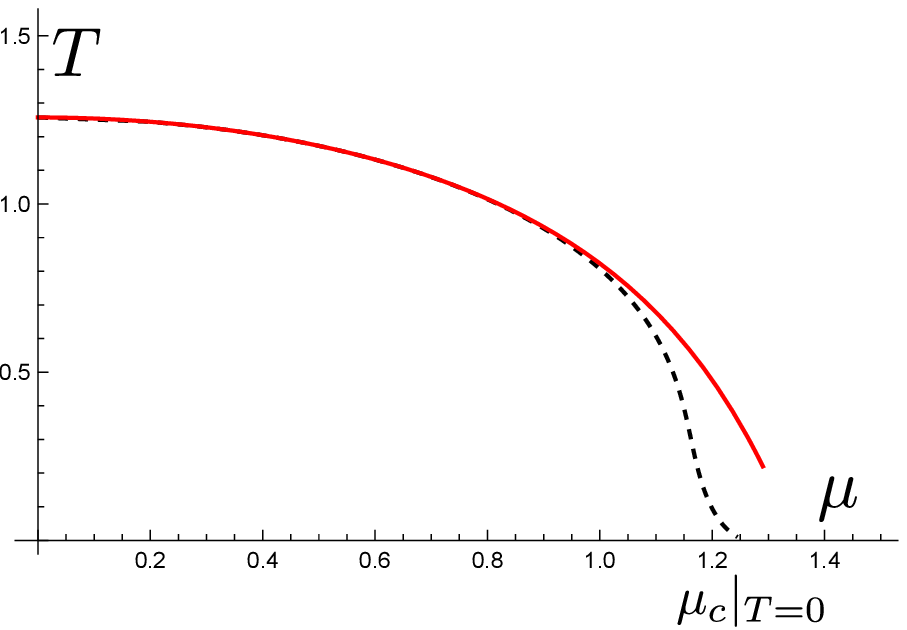}\\
                    $D=3$, $\tilde{D}=1$
                \end{center}
            \end{minipage}
            \begin{minipage}{0.5\hsize}
                \begin{center}
                    \includegraphics[scale=0.8]{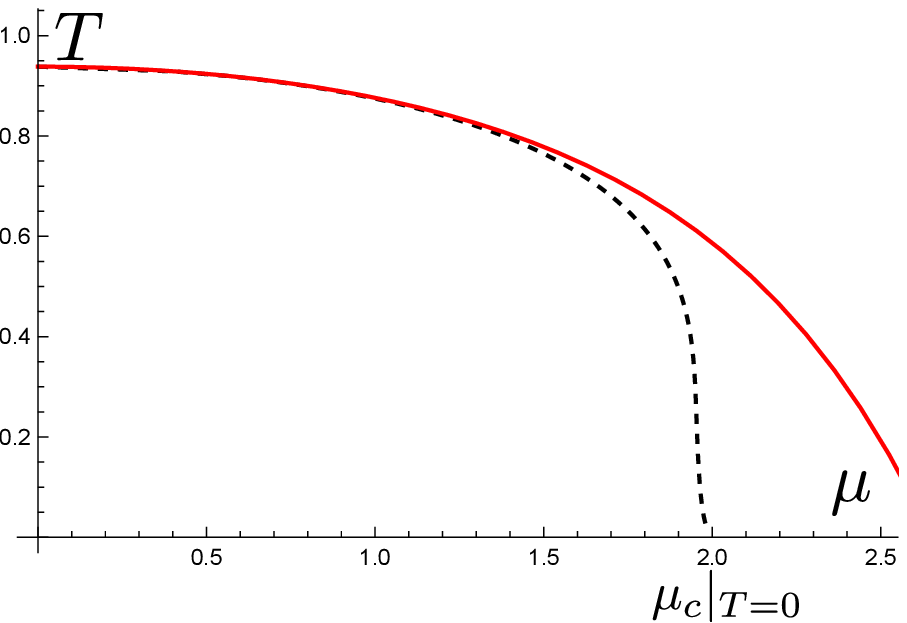}\\
                    $D=9$, $\tilde{D}=1$
                \end{center}
            \end{minipage}
        \end{tabular}
        \caption{Critical temperatures $T_c$ obtained by the analytic continuation of the imaginary chemical potential.
        $T_c$ against the {\it real} chemical potential $\mu$ are plotted.
        The black dashed curves are $T_c$ directly computed through Eq.~\eqref{eq-T_c} by using the real chemical potential.
        The red curves represent the results obtained through the analytic continuation of the imaginary chemical potential data derived in Fig.~\ref{fig-Im}. 
        In both cases, they agree for $\mu \lesssim 0.5 \times \mu_c|_{T=0} $, and  the analytic continuation works there.
        }
        \label{fig-ana-con}
    \end{center}
\end{figure}

In Fig.~\ref{fig-ana-con}, we explicitly compare the critical temperature $T_c$ against the real chemical potential $\mu$ computed through Eq.~\eqref{eq-T_c} and that of the analytic continuation of the imaginary chemical potential derived in Fig.~\ref{fig-Im}.\footnote{To obtain the analytic continuation results, we plot $T_c$ against $\mu_{\rm Im}$ by using the data employed in Fig.~\ref{fig-Im}, and fit the obtained curve by a polynomial $T_c = \sum_n c_n (\mu_{\rm Im}^2)^n $, where $c_n$ are the fitting parameters. Then, we perform the analytic continuation and obtain $T_c = \sum_n (-1)^nc_n \mu^{2n} $. This is the red curve plotted in Fig.~\ref{fig-ana-con}.}
In both $D=3$ and $D=9$ cases, we observe good agreement for $\mu \lesssim 0.5 \times \mu_c|_{T=0} $. This result suggests that the analytic continuation of the imaginary chemical potential may work in similar ranges even in the four-dimensional YM theories and QCD, too.

%

\section{Discussions}
\label{Sec_discussion}

In this article, we have studied the matrix model \eqref{action-BFSS} at finite angular momentum chemical potentials by using the CLM and the minimum sensitivity, and found the quantitative agreements. The action \eqref{action-BFSS-J}, with the chemical potential for the angular momentum added, suffers from the sign problem. This prevents us from using the conventional Monte Carlo methods, as we cannot regard $e^{-S}$ for the complex action $S$ as a probability. This leads us to study the model numerically \eqref{action-BFSS-J} using the CLM. The CLM turns out to work successfully in the parameter region of $\mu$, wide enough to elicit the behavior of the confinement and deconfinement phases. While minor discrepancies between the results of the CLM and minimum sensitivity are observed in the results presented in Figs.~\ref{u1_D09} - \ref{r2_D09}, they would be mitigated with more lattice space than $n_t=60$ and higher loop corrections \eqref{kappa-expansion} and higher order corrections \eqref{fn} in the minimum sensitivity treatment.

This is the very first result showing that a rotating quantum many body system at thermal equilibrium is analyzed through the first principle computation, as far as the authors know. 
Such rotating quantum systems are important in various topics including condensed matter and high energy physics, and our result encourages us to apply the CLM to these systems, too. 

In Sec.~\ref{sec-Im}, we have compared our matrix model and rotating pure YM theories in four dimensions under the presence of the imaginary chemical potentials.
We found the stable confinement phase at high temperature in the matrix model akin to the YM theories argued in Ref.~\cite{Chen:2022smf}.
We also found that the increasing transition temperature consistent with Refs.~\cite{Chen:2022smf, Chernodub:2022veq}.
Therefore, the natures of the matrix model \eqref{action-BFSS} is quite similar to the YM theories.
Since we can investigate the model \eqref{action-BFSS} with the real chemical potential, it may provide insights into the YM theories.
Besides, if we apply the CLM to the pure YM theories, it may shed more light on the properties of the YM theories.

\subsection{Relation to gravity}
\label{sec-BHs}

We have seen that the transition temperature decreases as the chemical potential increases in our model. 
A similar result has been obtained in the ${\mathcal N}$=4 SYM on $S_\beta^1 \times S^3 $ \cite{Basu:2005pj, Yamada:2006rx}. (See also footnote \ref{ftnt-SYM}.)
This model at strong coupling would be described by dual AdS geometries through the AdS/CFT correspondence \cite{Maldacena:1997re}, and the gravity computation \cite{Gubser:1998jb, Chamblin:1999tk, Cvetic:1999ne, Hawking:1999dp, Gubser:2000mm} also indicates a similar phase structure. (See Fig.~1 of Ref.~\cite{Yamada:2006rx}.) 
There, rotating black D3-brane solutions correspond to the rotating gapped solutions in the SYM theory.\footnote{
    The D3-branes rotate on the transverse $S^5$ in the ten dimension.
    Through the dimensional reduction of the $S^5$, the angular momenta become the Kaluza-Klein charges, and the rotating geometries reduce to the charged black branes, and Refs.~\cite{Basu:2005pj, Yamada:2006rx} studied this situation.}
Therefore, the decreasing transition temperature by a rotation is a common feature of these gauge theories and gravities.

In relation to the black holes, the non-uniform solution derived through the minimum sensitivity analysis is interesting.
As shown in Sec.~\ref{sec-free-energy}, this solution has the negative specific heat akin to black hole solutions such as Schwarzschild black holes and small black holes in AdS correspondence \cite{Aharony:1999ti}.
Hence, the non-uniform solution may explain the origin of the negative specific heat of the black holes through microscopic description.
It would be valuable to pursue this question further.

Besides, the properties of the large chemical potential regions (the ``unknown" regions in Fig.~\ref{fig-phase}) in the model \eqref{action-BFSS-J} may be understood through the gravity.
We can compute the free energy of the black branes as a function of temperature and chemical potential by using the results of Ref.~\cite{Cvetic:1999ne}. 
Then, we will see a similar result to our result shown in Fig.~\ref{fig-F}: 
Two black brane solutions appear and they merge at a large chemical potential, and, beyond this point, there is no solution. (The two black brane solutions correspond to the two gapped solutions in the matrix model.)
Although our result in Fig.~\ref{fig-F} beyond $\mu=\mu_\text{unstable}$ is not reliable, the gravity analysis indeed predicts a similar result.
It is tempting to improve our approximation in the matrix model and verify the gravity prediction.
In addition, gravity systems with angular momentum have various exotic solutions such as black rings and black Saturn, and it might be possible to find the corresponding solutions in the matrix model, too.

\paragraph*{Acknowledgment.---}
We thank P.~Basu, M.~Fukuma, Y.~Hidaka, A.~Joseph, J.~Nishimura and A.~Tsuchiya for valuable discussions and comments.
The work of T.~M. is supported in part by Grant-in-Aid for Scientific Research C (No. 20K03946) from JSPS. 
Numerical calculations were carried out using the computational resources, such as KEKCC and NTUA het clusters.

\appendix

\section{Details of the minimum sensitivity analysis}
\label{app-minimum}

\subsection{The derivation of the effective action \eqref{effective-action-gapped}}
\label{app-effective-action}

We show the derivation of the effective action \eqref{effective-action-gapped} through the minimum sensitivity analysis.
We will integrate out $X^I$ and  $Z^J$ through a perturbative calculation in the deformed  action \eqref{action-BFSS-MS} with respect to $\kappa$, and will obtain the two-loop effective action for the Polyakov loop $\{ u_n \}$,
\begin{align}
    S_{\text{eff}}(\{ u_n \},m_X,m_Z)=  \sum_{m=1}^2 \kappa^{m-1} S_{m\text{-loop}} .
    \label{kappa-expansion}
\end{align}

To perform the perturbative calculation, we take the static diagonal gauge \eqref{gauge-diagonal}.
Then the propagators of $X^I$ and $Z^I$ in the free part of the deformed action \eqref{S-free} become \cite{Mandal:2009vz, Morita:2010vi},
\begin{align} 
    \langle X_{ij}^I(t) X_{kl}^J(0) \rangle= 
    \delta_{il} \delta_{jk} \delta^{IJ} G_{Xij}(t),
    \quad
    \langle Z_{ij}^{I \dagger }(t) Z_{kl}^J(0) \rangle= & 
    \delta_{il} \delta_{jk} \delta^{IJ} G_{Zij}(t).
\end{align}
Here
\begin{align} 
    G_{Xij}(t):= & 
    \frac{1}{2m_X}	e^{ i    (\alpha_i-\alpha_j)||t||/\beta}
    \Biggl[
    e^{ -m_X ||t||	}
    \sum_{n=0}^{\infty}x^n u^i_n u_{-n}^{j}+
    e^{
            m_X ||t||
        }
    \sum_{n=1}^{\infty}x^n u^{i}_{-n} u_n^{j}
    \Biggr]
    ,
    \label{X propagator with A} \\
    G_{Zij}(t):= & 
    \frac{1}{2m_Z}	e^{ (i    (\alpha_i-\alpha_j)+\mu)||t||/\beta}
    \Biggl[
    e^{ -m_Z ||t||	}
    \sum_{n=0}^{\infty}z^n  q^{-n} u^i_n u_{-n}^{j}+
    e^{
            m_Z ||t||
        }
    \sum_{n=1}^{\infty}z^n q^n u^{i}_{-n} u_n^{j}
    \Biggr]
    ,
    \label{Z propagator with A}
\end{align}
where $x := e^{-\beta m_X}$, $z := e^{-\beta m_Z}$, $q := e^{-\beta \mu}$  and $u_n^i := e^{i  n \alpha_i}$,
which satisfies $\sum_{i=1}^{N} u_n^i =N u_n$ from Eq.~\eqref{polyakov}.
$||t||$ denotes $||t+n \beta||=t$ for $0\le t < \beta$.
In this calculation, $|\mu| < m_Z $ has been assumed, otherwise $Z^I$ becomes unstable.

By using these propagators, we obtain the one-loop term in the expansion \eqref{kappa-expansion}, \cite{Aharony:2003sx, Mandal:2009vz, Morita:2020liy}
\begin{align}
    S_{\text{1-loop}}/N^2 = & (D-2 \tilde{D}) \left\{ \frac{\beta m_X}{2}
    - \sum_{n=1}^{\infty}\frac{ x^{n}}{n}|u_{n}|^{2} \right\} \nonumber   \\
                            & + 2 \tilde{D} \left\{ \frac{\beta m_Z}{2}
    - \sum_{n=1}^{\infty}\frac{ z^{n}}{n} \frac{q^n+q^{-n}}{2} |u_{n}|^{2} \right\}
    +\frac{1}{N^2}
    S_{\text{G.F.}}.
    \label{1-loop}
\end{align}
Here $S_{\text{G.F.}}$ is the gauge fixing term \eqref{action-G.F.}, which arises when we take the constant diagonal gauge \eqref{gauge-diagonal}.

\begin{figure}
    \begin{center} 
        \includegraphics[scale=1.2]{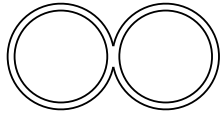}
        \caption{Planar diagrams at two loop.
        }
        \label{fig-diagram}
    \end{center}
\end{figure}

In the two-loop computation, we evaluate
\begin{align}
    S_{\text{2-loop}} & =\left\langle\int_0^{\beta} \hspace{-3mm} dt   \Tr \left( - \sum_{I,J=1}^D \frac{g^{2}}{4}\left[X^{I},X^{J}\right]^{2}- \sum_{I=1}^{\tilde{D}}m_Z^{2} Z^{I\dagger}Z^{I} - \sum_{I=2 \tilde{D}+1}^D\frac{m_X^{2}}{2}\left(X^{I}\right)^2 \right) \right\rangle.
    \label{2-loop}
\end{align}
Here, the first term can be expanded as
\begin{align}
      & \left\langle -\frac{g^{2}}{4} \sum_{I,J=1}^D  \int_0^{\beta} \hspace{-3mm} dt   \Tr \left( \left[X^{I},X^{J}\right]^{2} \right) \right\rangle \nonumber \\
    = & -\frac{g^{2}}{2}	\left\langle   \int_0^{\beta} \hspace{-3mm} dt   \Tr \left(
    \sum_{I,J=1}^{\tilde{D}} \left[Z^{I},Z^{J}\right]\left[Z^{I\dagger},Z^{J \dagger}\right]
    + \sum_{I,J=1}^{\tilde{D}} \left[Z^{I},Z^{J \dagger}\right]\left[Z^{I\dagger},Z^{J }\right]
    \right) \right\rangle
    \nonumber                                                                                                                                                   \\
      & -\frac{g^{2}}{4}	\left\langle   \int_0^{\beta} \hspace{-3mm} dt   \Tr \left(
    4 \sum_{I=1}^{\tilde{D}} \sum_{J=2 \tilde{D}+1}^D \left[Z^{I},X^{J}\right] \left[Z^{I \dagger },X^{J} \right] +	
    \sum_{I,J=2 \tilde{D}+1}^D \left[X^{I},X^{J}\right]^{2} \right) \right\rangle.
    \label{2-loop-4pt}
\end{align}
In the large-$N$ limit, the planar diagram depicted in Fig.~\ref{fig-diagram} dominates and each 
term in Eq.~\eqref{2-loop-4pt} is calculated as follows.
\begin{align}
     & -\frac{g^{2}}{2}	\left\langle   \int_0^{\beta} \hspace{-3mm} dt   \Tr \left(
    \sum_{I,J=1}^{\tilde{D}} \left[Z^{I},Z^{J}\right]\left[Z^{I\dagger},Z^{J \dagger}\right]
    + \sum_{I,J=1}^{\tilde{D}} \left[Z^{I},Z^{J \dagger}\right]\left[Z^{I\dagger},Z^{J }\right]
    \right) \right\rangle \nonumber                                                                                                                \\
     & = \frac{g^{2} \beta}{2} \sum_{i,j,k=1}^N \left\{
    (2\tilde{D}^2-4 \tilde{D} ) G_{Zji}(0)G_{Zik}(0) +(\tilde{D}^2+\tilde{D} ) \left( G_{Zij}(0)G_{Zik}(0) + G_{Zji}(0)G_{Zki}(0) \right)
    \right\} ,                                                                                                                                     \\
     & -\frac{g^{2}}{4}	\left\langle   \int_0^{\beta} \hspace{-3mm} dt   \Tr \left(
    4 \sum_{I=1}^{\tilde{D}} \sum_{J=2 \tilde{D}+1}^D \left[Z^{I},X^{J}\right] \left[Z^{I \dagger },X^{J} \right]  \right) \right\rangle \nonumber \\
     & = g^{2} \beta \tilde{D}(D-2\tilde{D})\sum_{i,j,k=1}^N \left\{G_{Zij}(0)G_{Xik}(0) + G_{Zji}(0)G_{Xik}(0)
    \right\} ,                                                                                                                                     \\
     & -\frac{g^{2}}{4}	\left\langle   \int_0^{\beta} \hspace{-3mm} dt   \Tr \left(
    \sum_{I,J=2 \tilde{D}+1}^D \left[X^{I},X^{J}\right]^{2} \right) \right\rangle \nonumber                                                        \\
     & = \frac{g^{2} \beta}{2} (D-2\tilde{D})(D-2\tilde{D}-1)  \sum_{i,j,k=1}^N  G_{Xij}(0)G_{Xik}(0).
\end{align}
The terms involving the products of the propagators can be computed by using the expressions \eqref{X propagator with A} and \eqref{Z propagator with A},
\begin{align}
      & \sum_{i,j,k=1}^N  G_{Zji}(0)G_{Zik}(0) \nonumber \\
    = & \frac{N^3}{4 m_Z^2} \Biggl(
    1+ 2\sum_{n=1}^{\infty} z^n(q^n+q^{-n}) |u_n|^2
    +\sum_{n,m=1}^{\infty} z^{n+m}(q^{n+m}+q^{-n-m}) u_{n} u_{m-n} u_{-m}
    \nonumber                                            \\
      & +\sum_{n,m=1}^{\infty}
    z^{m+n}q^{m-n} \left(
    u_{-m-n} u_{n}  u_{m}
    + u_{m+n} u_{-n}  u_{-m}
    \right)  \Biggr),
    \label{propagator-z}
\end{align}
\begin{align}
      & \sum_{i,j,k=1}^N  G_{Zij}(0)G_{Zik}(0) = \sum_{i,j,k=1}^N  \left(
    G_{Zji}(0)G_{Zki}(0)
    \right)^*   \nonumber                                                 \\
    = & \frac{N^3}{4 m_Z^2} \Biggl(
    1+ 2 \sum_{n=1}^{\infty} z^n(q^n+q^{-n}) |u_n|^2
    +2\sum_{n,m=1}^{\infty}
    z^{n+m}q^{n-m} u_{n} u_{m-n} u_{-m}
    \nonumber                                                             \\
      & 
    +\sum_{m,n=1}^{\infty} z^{m+n}q^{-m-n} u_{m+n}  u_{-m} u_{-n}
    + \sum_{m,n=1}^{\infty} z^{m+n}q^{m+n}  u_{-m-n} u_{m} u_{n}
    \Biggr),
\end{align}
\begin{align}
      & \sum_{i,j,k=1}^N G_{Zij}(0)G_{Xik}(0) =
    \sum_{i,j,k=1}^N \left( G_{Zji}(0)G_{Xik}(0)\right)^* 
    \nonumber                                   \\
    = & \frac{N^3}{4 m_Z m_X} \Biggl(
    1+ 2 \sum_{n=1}^{\infty} \left( 2x^n+ z^n(q^n+q^{-n}) 
    \right) |u_n|^2 
    +\sum_{m,n=1}^{\infty}  z^n q^{-n} x^m \left(
    u_{m+n}  u_{-m} u_{-n}+ u_{n-m}  u_{m} u_{-n}
    \right)
    \nonumber                                   \\
      & 
    +\sum_{m,n=1}^{\infty}  z^n q^{n} x^m \left(
    u_{m-n}  u_{-m} u_{n}+ u_{-n-m}  u_{m} u_{n}
    \right)
    \Biggr),
\end{align}
\begin{align}
    \sum_{i,j,k=1}^N  G_{Xij}(0)G_{Xik}(0) 
    = & \frac{N^3}{4  m_X^2} \Biggl(
    1+ 4 \sum_{n=1}^{\infty} x^n |u_n|^2 
    +2 \sum_{m,n=1}^{\infty}  x^{m+n}
    u_{m-n}  u_{-m} u_{n}
    \nonumber                        \\		&
    +\sum_{m,n=1}^{\infty}  x^{m+n} \left(
    u_{m+n}  u_{-m} u_{-n}+ u_{-n-m}  u_{m} u_{n}
    \right)
    \Biggr)	.
    \label{propagator-x}
\end{align}
By substituting these equations into Eq.~\eqref{2-loop-4pt}, we reach
\begin{align}
     & \left\langle -\frac{g^{2}}{4} \sum_{I,J=1}^D  \int_0^{\beta} \hspace{-3mm} dt   \Tr \left( \left[X^{I},X^{J}\right]^{2} \right) \right\rangle 
    = N^2 \left( A+ \sum_{n=1}^\infty B_n |u_n|^2 + \textrm{O} (u_n u_m u_{-n-m} ) \right),
    \label{2-loop-4pt-2}
\end{align}
where
\begin{align}	
    A=   & 
    \frac{\beta \lambda}{8}
    \left[
        \frac{2\tilde{D}(2 \tilde{D}-1)}{m_Z^2}
        +\frac{4 \tilde{D}(D-2 \tilde{D})}{m_X m_Z}
        +\frac{(D-2 \tilde{D})(D-2 \tilde{D}-1)}{m_X^2}
        \right] ,                                                                         \\
    B_n= & \frac{\beta\lambda}{4m_Z^2}
    \left[ 2 \tilde{D} (\tilde{D}+1) z^{2n} + \tilde{D}(\tilde{D}-2) z^{2n}(q^{-2n }+q^{2n }) 
    +	2 \tilde{D} (2 \tilde{D}-1) z^{n}(q^{-n }+q^{n }) 
    \right]
    \nonumber                                                                             \\
         & + \frac{\beta \lambda}{2m_X m_Z} \tilde{D}(D-2 \tilde{D})
    \left[ 2 x^n + (1+x^n) z^{n}(q^{-n }+q^{n }) 
    \right]
    \nonumber                                                                             \\
         & +\frac{\beta \lambda}{4m_X^2}(D-2 \tilde{D})(D-2 \tilde{D}-1)  (x^{2n}+2x^{n})
    \label{B_n}.
\end{align}
The second and third terms in Eq.~\eqref{2-loop} can be calculated from the one-loop result \eqref{1-loop},
\begin{align}
      & \left\langle\int_0^{\beta} \hspace{-3mm} dt   \Tr \left(
    - \sum_{I=1}^{\tilde{D}}m_Z^{2} Z^{I\dagger}Z^{I}
    - \sum_{I=2 \tilde{D}+1}^D\frac{m_X^{2}}{2}\left(X^{I}\right)^2 
    \right) \right\rangle \nonumber                              \\
    = & 
    - m_Z^2 \frac{\partial}{\partial (m_Z^2)} S_{\text{1-loop}}
    - m_X^2 \frac{\partial}{\partial (m_X^2)} S_{\text{1-loop}}
    \nonumber                                                    \\
    = & 
    2 N^2 \tilde{D} \left\{ -\frac{\beta m_Z}{4}
    - \sum_{n=1}^{\infty}\frac{ \beta m_Z }{4} z^{n} (q^{n}+q^{-n}) |u_{n}|^{2} \right\} 
    +N^2(D-2 \tilde{D}) \left\{ -\frac{\beta m_X}{4}
    - \sum_{n=1}^{\infty}\frac{\beta m_X }{2} x^{n} |u_{n}|^{2} \right\} 
    .
    \label{2-looop-2pt}
\end{align}

By combining these results and take $\kappa=1$, we obtain the effective action for $\{ u_n \}$,\footnote{A subtle issue is when we take $\kappa=1$.
    One option is taking $\kappa=1$ after integrating out all the dynamical variables $X^I$, $Z^I$ and $\{ u_n \}$ in the action \eqref{action-BFSS-MS}.
    However, we have taken $\kappa=1$ after integrating out $X^I$ and $Z^I$ only and keep $\{ u_n \}$ as variables in our analysis.
    This prescription makes the computations simpler.
    Since the minimum sensitivity method is not a systematic approximation and there is no guiding principle, we prioritize computational simplicity.}
\begin{align}
    S_{\text{eff}}(\{ u_n \},m_X,m_Z)= & N^2 \left(\beta f_0 (m_X,m_Z)+ \sum_{n=1} f_n(\beta,\mu, m_X,m_Z) |u_n|^2 
    + \textrm{O} (u_n u_m u_{-n-m} ) \right) .
    \label{effective-action-app}
\end{align}
Here
\begin{align}
    f_0 (m_X,m_Z)= & \frac{(D-2\tilde{D}) m_X}{4}+\frac{\tilde{D} m_Z}{2} \nonumber \\
                   & +	\frac{\lambda}{8}
    \left[
        \frac{2\tilde{D}(2 \tilde{D}-1)}{m_Z^2}
        +\frac{4 \tilde{D}(D-2 \tilde{D})}{m_X m_Z}
        +\frac{(D-2 \tilde{D})(D-2 \tilde{D}-1)}{m_X^2}
        \right],
    \label{f0}
\end{align}
\begin{align}
    f_n(\beta,\mu, m_X,m_Z)= & \frac{1}{n}
    - \tilde{D} \left(
    \frac{1}{n}+\frac{\beta m_Z}{2}
    \right) z^{n} (e^{n\beta \mu}+e^{-n\beta \mu})
    - (D-2\tilde{D}) \left(
    \frac{1}{n}+\frac{\beta m_X}{2}
    \right) x^{n} +B_n,
    \label{fn}
\end{align}
where $B_n$ has been defined in Eq.~\eqref{B_n}.

In principle, $\{ u_n \}$ can be integrated out in this effective action by using the technique proposed in Ref.~\cite{Alvarez-Gaume:2006fwd}, and we may obtain the free energy.
However, the computation would be complicated.
In order to reduce the calculations, we use the following drastic approximation,
\begin{align} 
    f_n(\beta,\mu, m_X,m_Z) \simeq & \frac{1}{n}  , \qquad (n\ge 2), 
    \label{large-D-approximation}
\end{align}
and ignore the interaction term $\textrm{O} (u_n u_m u_{-n-m} ) $.
In this approximation, the contribution of $X^I$ and $Z^I$ integrals to $f_n$ ($n \ge 2$) are neglected, and only the terms coming from the gauge fixing \eqref{action-G.F.} survive.
Then, we obtain the effective action \eqref{effective-action-gapped}, which we use in the main analysis of the phase structure of the model \eqref{action-BFSS-J}.
(The advantage of this approximation is that the analysis in Ref.~\cite{Liu:2004vy} is available.)

This approximation may be justified in the low temperature and low chemical potential regime until $T \simeq T_c(\mu)$ by assuming that $D$ is large.
This is because $m_X = m_Z \sim(\lambda D)^{1/3}$ for a large $D$ in the confinement phase \eqref{m-2-loop}, and the equation \eqref{eq-T_c}\footnote{
    We can easily show that the approximation \eqref{large-D-approximation} affects the quantities in the non-uniform phase and gapped phase, and not in the confinement phase.
    This is because $f_0$ and $f_1$ are kept in the approximation and the equations \eqref{m-2-loop} and \eqref{eq-T_c}, which determine the quantities in the confinement phase and the critical temperature $T_c(\mu)$, remain.
    However, when we evaluate the finite-$N$ corrections in the confinement phase, the approximation \eqref{large-D-approximation} will affect the results as argued in Appendix \ref{app-un}.
} for $T_c(\mu)$ leads to $x \sim 1/D$ (if $D \gg \tilde{D}$) or $zq \sim 1/\tilde{D}$ (if $D \sim 2 \tilde{D}$).
Then we obtain $f_n \sim 1/n + \textrm{O} (1/D^{n-1}) $ if $D \gg \tilde{D}$ or $f_n \sim 1/n + \textrm{O} (1/\tilde{D}^{n-1}) $ if $D \sim 2 \tilde{D}$, and the approximation \eqref{large-D-approximation} is verified in this region.
Similarly, the interaction terms $u_n u_m u_{-n-m} $ are also suppressed, since the coefficients for these terms in Eqs.~\eqref{propagator-z} - \eqref{propagator-x} involve $x^{n+m}$ or $(zq)^{n+m}$ which are small in this regime.

\subsection{$1/N$ corrections on $\langle u_n \rangle$ in the uniform solution}
\label{app-un}

We show the derivation of the leading $1/N$ corrections to $\langle  u_n \rangle$ in the uniform solution.
As we argued in Sec.~\ref{sec-non-uniform}, $u_n$ can be regarded as independent variables when $ u_n \sim 0$ for all $n$.
Since the action \eqref{effective-action-app} is approximately quadratic with respect to $\{ u_n \}$, we can easily evaluate the expectation values through the saddle point method, where the saddle is given by $u_n=0$ in the uniform solution.
Then, we obtain
\begin{align}
    \label{un-finite-N}
    \langle u_n \rangle= \frac{1}{2N}\sqrt{ \frac{\pi}{f_n(\beta,\mu, m_0,m_0)}}+\textrm{O} \left( \frac{1}{N^3} \right) .
\end{align}
Here we have taken the gauge introduced in footnote \ref{ftnt-gauge}.
Note that this result for $\langle u_1 \rangle$ is not reliable near the critical point \eqref{eq-T_c} where $f_1=0$.

We compare this result with the MC and the CLM as shown in Figs.~\ref{fig-D=9}, \ref{u1_D09} and \ref{u2_D09}.
In this computation, we have employed Eq.~\eqref{fn} for $f_n$ and have not used the approximation \eqref{large-D-approximation}.
They nicely agree.

\section{Drift norm of the bosonic BFSS model}\label{appendix_drift_BFSS}
In this section, we discuss the behavior of the drift norm in the {\it real} Langevin simulation of the bosonic BFSS model (\ref{action-BFSS}), which is free from the sign problem. In the following, we discuss the action (\ref{action-BFSS-J2}), which is lattice-regularized as (\ref{action-BFSS-Jlat}), with $\mu=0$.
We compare the behavior of the drift norms, as defined by Eqs.~(\ref{drift_norms}) and (\ref{drift_normV}), for the cases with and without the static diagonal gauge (\ref{gauge-diagonal}), respectively. 

\subsection{Langevin equation without the static diagonal gauge}
Here, we describe how to solve the real Langevin equation for $\mu=0$ without the static diagonal gauge (\ref{gauge-diagonal}). 
The scalar fields $X^I$ are updated as (\ref{langevin_eq2X}), which makes no difference from the simulation with the static diagonal gauge (\ref{gauge-diagonal}). The gauge field now depends on the temporal direction, which is updated in terms of $V(n) = e^{i A(n) (\Delta t)}$ as
\begin{eqnarray}
    V(n, \sigma+\Delta \sigma) &=& \exp \Biggl\{ i \sum_{a=1}^{{\cal G}} \lambda^a \left( - (\Delta \sigma) \nu^a (V(n,\sigma)) + \sqrt{\Delta  \sigma} \eta^a (n, \sigma) \right) \Biggr\} V(n, \sigma), \label{langevin_eq2a_nostat}
\end{eqnarray}
instead of (\ref{langevin_eq2a}).
\footnote{Here, we present the result $\mu=0$, which has no sign problem from the outset, and there is no need to implement the gauge cooling \cite{1211_3709,1508_02377,1604_07717}. When we work on the $\mu \neq 0$ case, the gauge cooling is useful to suppress the excursion problem by minimizing the hermiticity and unitary norm for $X^I$ and $V$ defined by
\begin{eqnarray}
    {\cal N}_X = - \sum_{n=1}^{n_t} \sum_{I=1}^D \textrm{Tr} [(X^I(n) - X^I (n)^{\dag})^2], \ \ {\cal N}_V = \sum_{n=1}^{n_t} \textrm{Tr} WW^{\dag}. \label{hermiticity_norm}
\end{eqnarray}
Here, $W = (V^{-1} (n))^{\dag} (I_N-V^{\dag}(n) V(n))$ and  $I_N$ is an $N \times N$ unit matrix. ${\cal N}_{X}$ and ${\cal N}_V$ vanish only if $X^I(n)$ are hermitian and $V(n)$ is unitary, respectively. After each step of solving the discretized Langevin equation (\ref{langevin_eq2X}) and (\ref{langevin_eq2a_nostat}), we perform a gauge transformation 
\begin{eqnarray}
    & & \hspace*{-5mm}  X^I (n) \to e^{\gamma_X H_X (n)} X^I (n) e^{-\gamma_X H_X (n)}, \textrm{ where } H_X (n) = - \sum_{I=1}^D [X^I(n), X^I(n)^{\dag}], \label{gauge_cooling_X} \\
    & & \hspace*{-5mm} V(n) \to e^{\gamma_V H_V(n+1)} V(n) e^{-\gamma_V H_V (n)}, \textrm{ where } H_V (n) = \sum_{a=1}^{{\cal G}} \lambda^a G^a (n), \label{gauge_cooling_V} \\
    & & \hspace*{-5mm} G^a (n) = \textrm{Tr} \lambda^a \{ (V^{-1}(n))^{\dag} V^{-1} (n-1) +V^{\dag}(n) V(n) - V(n-1) V^{\dag} (n-1)  - V^{-1} (n) (V^{-1}(n))^{\dag} \}. \label{gauge_cooling_G}
\end{eqnarray}
Here, $H_X(n)$ and $H_V (n)$ are the gradients of ${\cal N}_X$ and ${\cal N}_V$ with respect to the gauge transformation, and the real positive parameters $\gamma_X$ and $\gamma_V$ are chosen so that the norms ${\cal N}_X$ and ${\cal N}_V$ are minimized, respectively.}
Due to the invariance (\ref{a_inv}), we set the constraint $\displaystyle \textrm{Tr} A(t)=0$, and the gauge group is SU$(N)$. $\lambda^a$ is the generator of the SU$(N)$ Lie algebra, such that $\textrm{Tr} (\lambda^a \lambda^b) = \delta^{ab}$, and ${\cal G} = N^2-1$ is the dimension of SU$(N)$. $\nu^a (V(n,\sigma))$ is the drift term defined by
\begin{eqnarray}
    \nu^a (V(n,\sigma)) = \frac{d}{d\tau} S_{\textrm{lat}} [e^{i\tau \lambda^a} V(n)]|_{\tau=0} = \frac{Ni}{\Delta t} \textrm{Tr} (\lambda^a[X^I (n+1), V(n) X^I(n) V^{-1} (n)]), \label{drift_a_nostat}
\end{eqnarray}
where $\tau$ is a real number and $S_{\textrm{lat}} [e^{i\tau \lambda^a} V(n)]$ is defined by replacing $V(n)$ and $V(n)^{-1}$ in $S_{\textrm{lat}}$, as defined by Eq.~(\ref{action-BFSS-Jlat}), with $e^{i\tau \lambda^a} V(n)$ and $(e^{i\tau \lambda^a} V(n))^{-1} = V(n)^{-1} e^{-i\tau \lambda^a}$, respectively.
Also, the drift norm for the case without the static diagonal gauge is defined as
\begin{eqnarray}
    u_{A} = \sqrt{\frac{1}{N^3 n_t} \sum_{n=1}^{n_t} \sum_{a=1}^{{\cal G}} |\nu^a (V(n,\sigma))|^2}, \label{drift_normV}
\end{eqnarray}
instead of $u_{\alpha}$  in Eq.~(\ref{drift_norms}), while the drift term for $X^I$ is the same as $u_{X}$  in Eq.~(\ref{drift_norms}). 

\subsection{The fall-off of the drift norms}
\begin{figure} [t]
    \centering
    \includegraphics[width=0.42\textwidth]{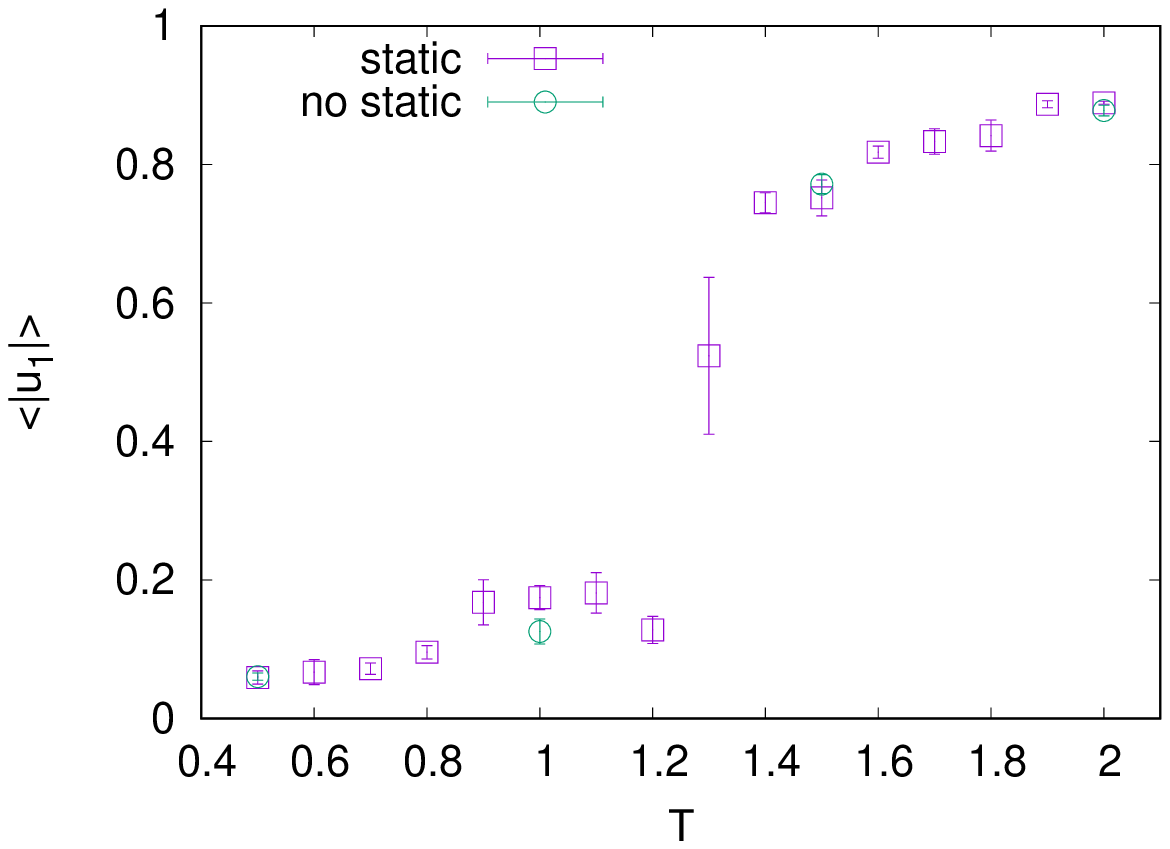}
    \includegraphics[width=0.42\textwidth]{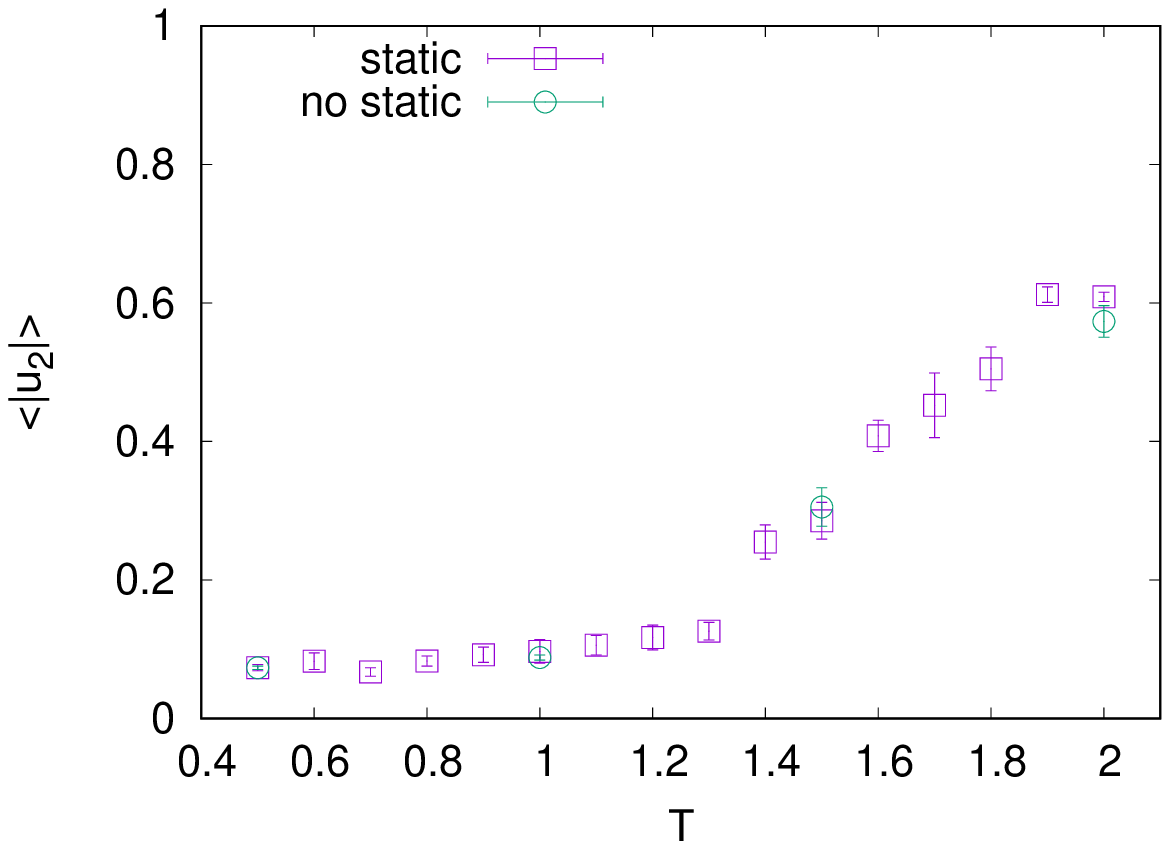}
    \includegraphics[width=0.42\textwidth]{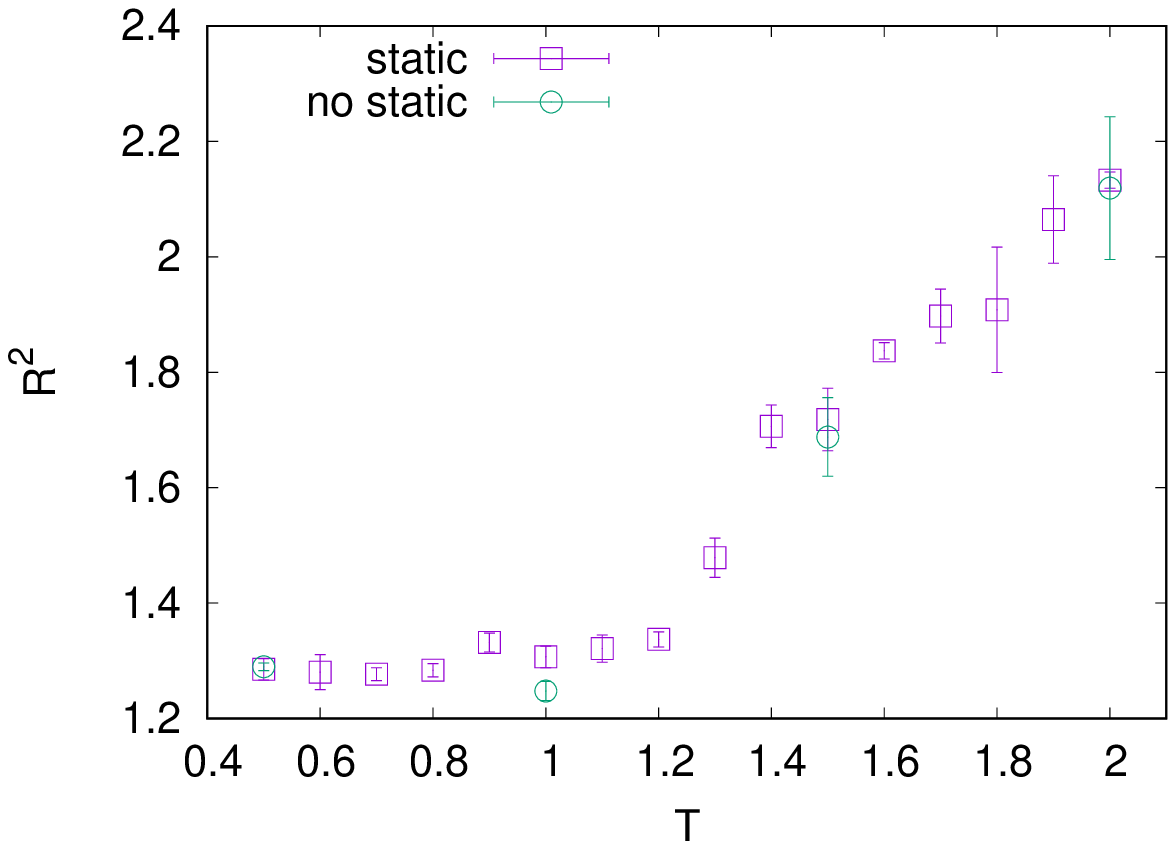}
    \includegraphics[width=0.42\textwidth]{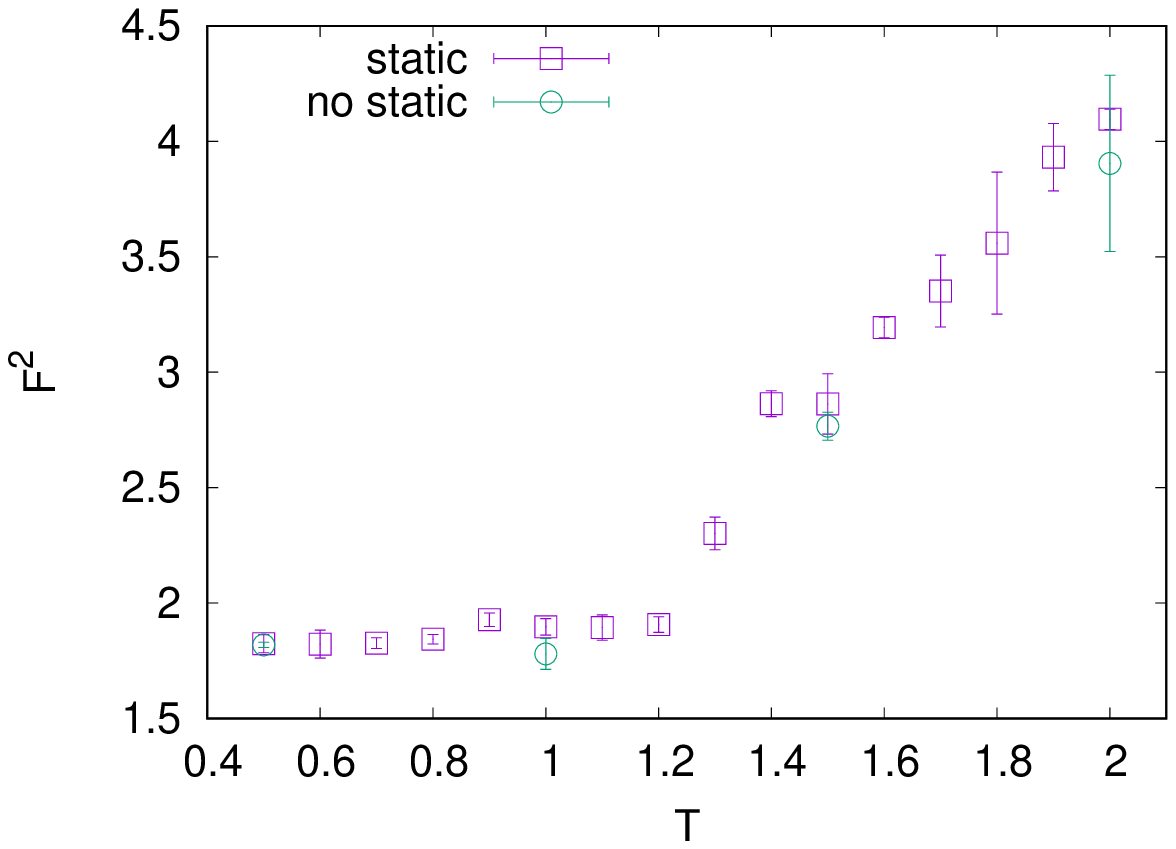}
    \caption{The observables for $D=3, N=16, \mu=0$ against $T$. ``static" and ``no static" indicate the result with and without the static diagonal gauge (\ref{gauge-diagonal}), respectively.}\label{BFSS_result_observable}
\end{figure}

We compare the result with and without the static diagonal gauge (\ref{gauge-diagonal}), for $D=3, N=16, \mu=0$. With the static diagonal gauge, we add the gauge fixing term as (\ref{S_eff}). In Fig.~\ref{BFSS_result_observable}, we compare the observables $|u_1|$, $|u_2|$ and 
\begin{eqnarray}
    R^2 &=& \frac{1}{N \beta} \int^{\beta}_{0} \textrm{Tr} \sum_{I=1}^{D} X_I^2 (t) dt, \label{R2_sq} \\
    F^2 &=& \frac{-1}{N \beta} \int^{\beta}_{0} \textrm{Tr} \sum_{I,J=1}^D [X_I (t), X_J(t)]^2 dt. \label{f2_sq}
\end{eqnarray}
$u_n$ is expressed as (\ref{polyakov_static_diag}) with the static diagonal gauge (\ref{gauge-diagonal}), while it is expressed without the static diagonal gauge as
\begin{eqnarray}
    u_n = \frac{1}{N} \textrm{Tr} V_{\textrm{P}}^n, \ \textrm{ where } V_{\textrm{P}} = V(n_t) V(n_t-1) \cdots V(2) V(1). \label{polyakov_nostat}
\end{eqnarray}
This is invariant under the gauge transformation $V(n) \to g(n+1) V(n) g^{-1} (n)$. 

\begin{figure} [t]
    \centering
    \includegraphics[width=0.42\textwidth]{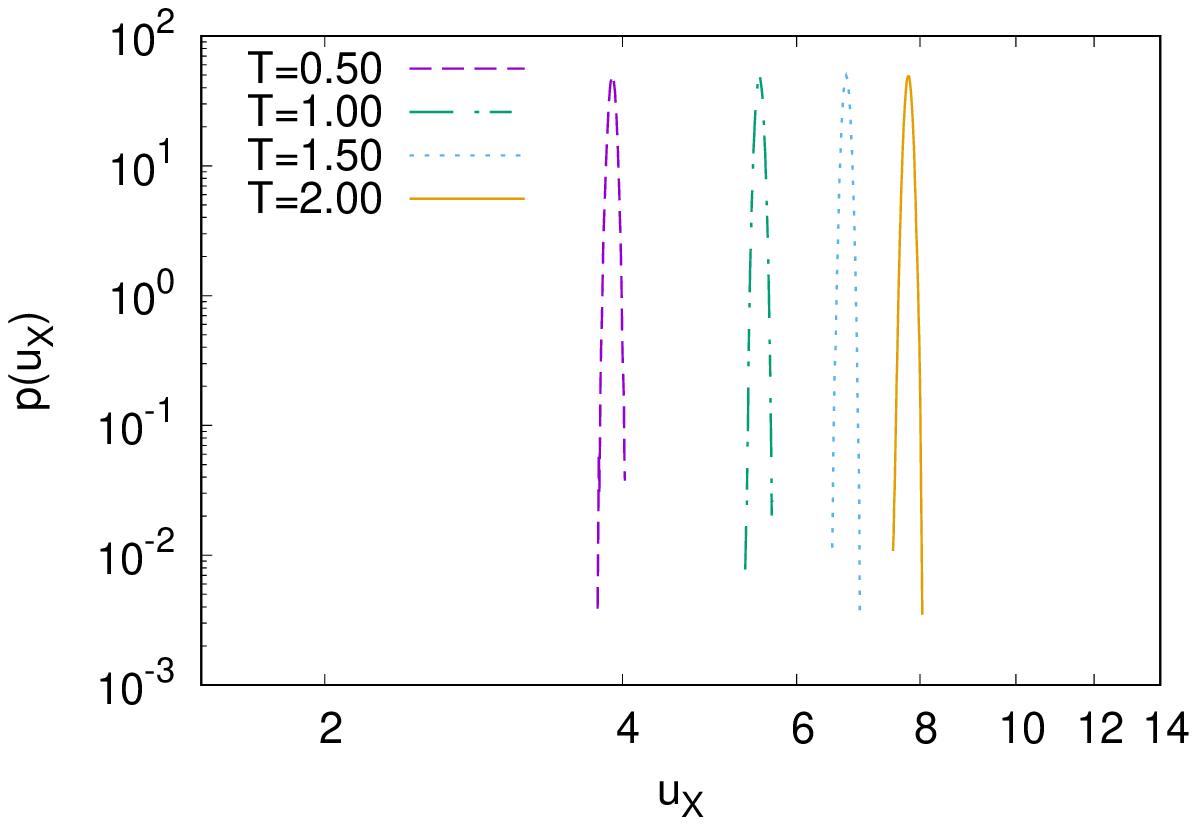}
    \includegraphics[width=0.42\textwidth]{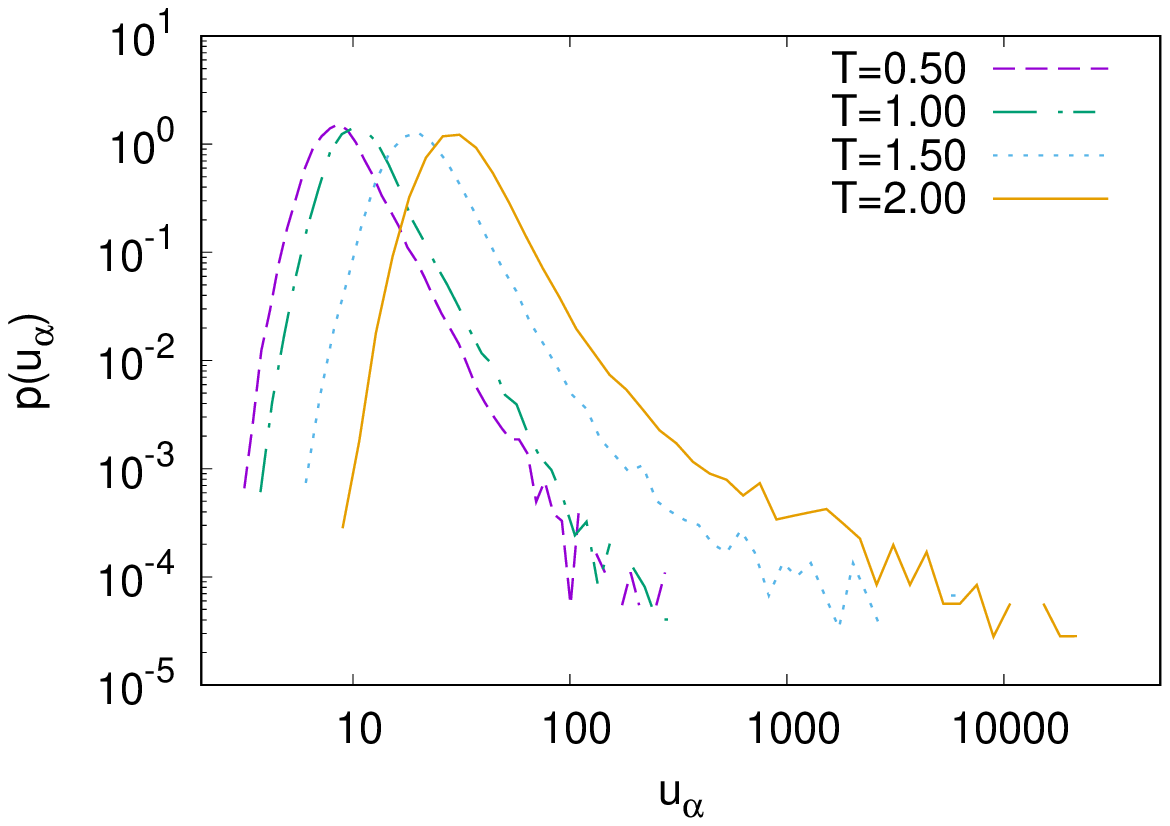}
    \caption{The log-log plot of the histogram of the drift norm $u_X$ (left) and $u_{\alpha}$ (right) with the static diagonal gauge (\ref{gauge-diagonal}), for $D=3, N=16, \mu=0, T=0.5, 1.0, 1.5, 2.0$.}\label{BFSS_result_norm_stat}
\end{figure}

\begin{figure} [t]
    \centering
    \includegraphics[width=0.42\textwidth]{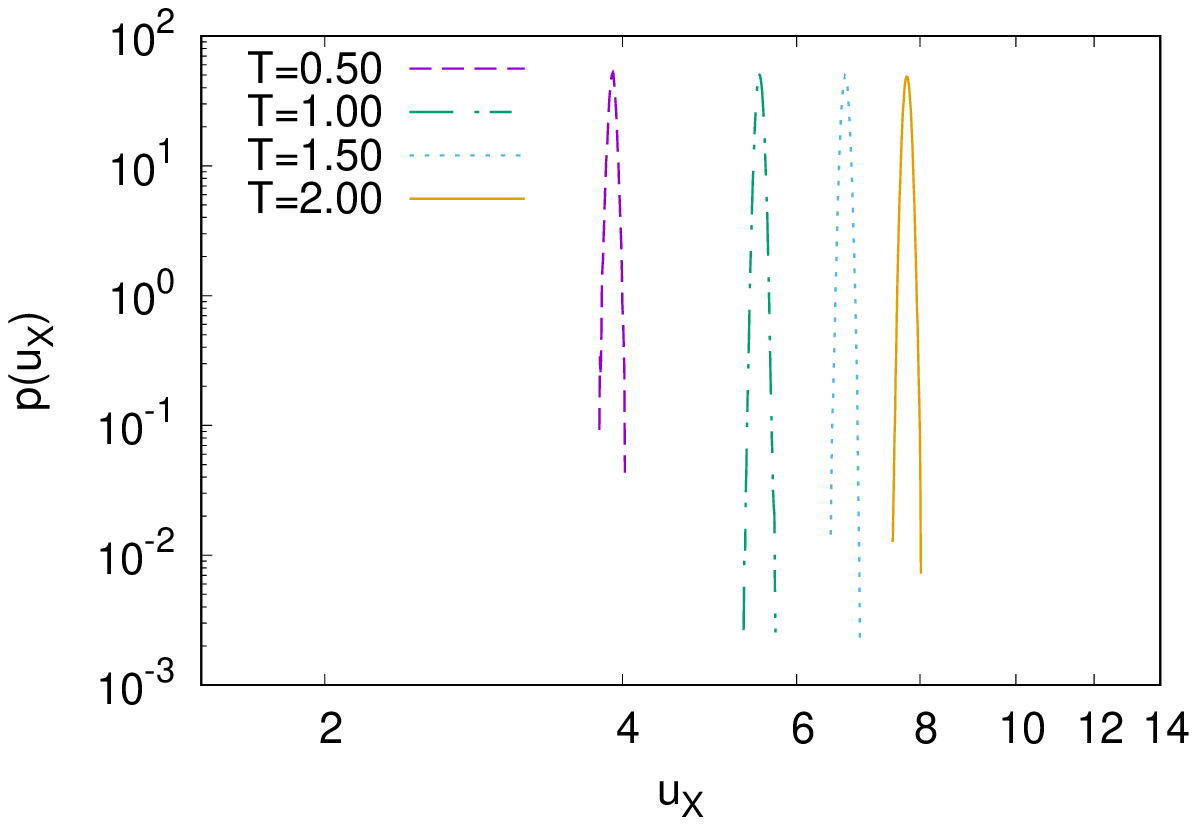}
    \includegraphics[width=0.42\textwidth]{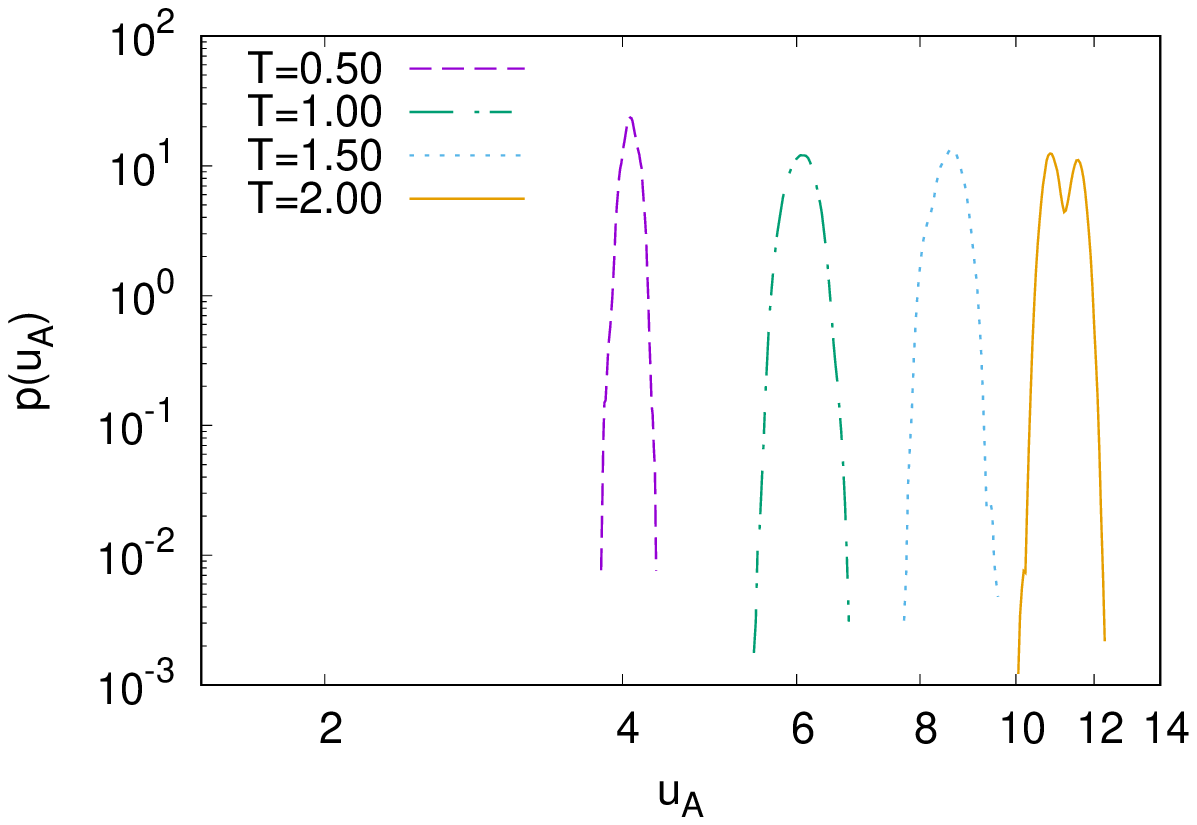}    
    \caption{The log-log plot of the histogram of the drift norm $u_X$ (left) and $u_{A}$ (right) without the static diagonal gauge (\ref{gauge-diagonal}), for $D=3, N=16, \mu=0, T=0.5, 1.0, 1.5, 2.0$.}\label{BFSS_result_norm_nostat}
\end{figure}

$|u_n|$ and $R^2$ are calculated by removing the trace part as (\ref{constraint_on_sun_discre}). The trace part does not affect $F^2$, which is expressed only in terms of the commutator of $X_I(t)$. We see that the results with and without the static diagonal gauge (\ref{gauge-diagonal}) agree. The histograms of the drift norms are plotted in Figs.~\ref{BFSS_result_norm_stat} and \ref{BFSS_result_norm_nostat} with and without the static diagonal gauge (\ref{gauge-diagonal}), respectively, for $D=3, N=16, \mu=0, T=0.5, 1.0, 1.5, 2.0$. The probability distributions $p(u_X)$ of the drift norms $u_X$ fall exponentially or faster both in Figs.~\ref{BFSS_result_norm_stat} and \ref{BFSS_result_norm_nostat} with and without the static diagonal gauge (\ref{gauge-diagonal}). However, those of the drift norms $p(u_{\alpha})$ for the gauge field $u_{\alpha}$ fall only in power law in Fig.~\ref{BFSS_result_norm_stat} with the static diagonal gauge (\ref{gauge-diagonal}), while those of $u_{A}$, which we denote as $p(u_A)$, fall  exponentially or faster without the static diagonal gauge (\ref{gauge-diagonal}). We attribute this to the singularity stemming from the derivative 
\begin{eqnarray}
    \frac{\partial}{\partial \alpha_k} S_{\textrm{g.f.}} = - \sum_{\ell \neq k} \cot \frac{\alpha_k - \alpha_{\ell}}{2}, \label{gf_diff}
\end{eqnarray}
which becomes larger when $\alpha_k$ and $\alpha_{\ell}$ approach each other, in solving the Langevin equation (\ref{langevin_eq2a}). 

{\normalsize 
\bibliographystyle{utphys}
\bibliography{bBFSS} }

\providecommand{\href}[2]{#2}\begingroup\raggedright\begin{thebibliography}{10}

\bibitem{1205_3996}
{\bfseries AuroraScience} Collaboration, M.~Cristoforetti, F.~Di~Renzo, and
  L.~Scorzato, ``{New approach to the sign problem in quantum field theories:
  High density QCD on a Lefschetz thimble},''
  \href{http://dx.doi.org/10.1103/PhysRevD.86.074506}{{\em Phys. Rev. D}
  {\bfseries 86} (2012) 074506},
  \href{http://arxiv.org/abs/1205.3996}{{\ttfamily arXiv:1205.3996 [hep-lat]}}.

\bibitem{1303_7204}
M.~Cristoforetti, F.~Di~Renzo, A.~Mukherjee, and L.~Scorzato, ``{Monte Carlo
  simulations on the Lefschetz thimble: Taming the sign problem},''
  \href{http://dx.doi.org/10.1103/PhysRevD.88.051501}{{\em Phys. Rev. D}
  {\bfseries 88} no.~5, (2013) 051501},
  \href{http://arxiv.org/abs/1303.7204}{{\ttfamily arXiv:1303.7204 [hep-lat]}}.

\bibitem{Parisi:1983mgm}
G.~Parisi, ``{On complex probabilities},''
  \href{http://dx.doi.org/10.1016/0370-2693(83)90525-7}{{\em Phys. Lett. B}
  {\bfseries 131} (1983) 393--395}.

\bibitem{Klauder:1983sp}
J.~R. Klauder, ``{Coherent State Langevin Equations for Canonical Quantum
  Systems With Applications to the Quantized Hall Effect},''
  \href{http://dx.doi.org/10.1103/PhysRevA.29.2036}{{\em Phys. Rev. A}
  {\bfseries 29} (1984) 2036--2047}.

\bibitem{1101_3270}
G.~Aarts, F.~A. James, E.~Seiler, and I.-O. Stamatescu, ``{Complex Langevin:
  Etiology and Diagnostics of its Main Problem},''
  \href{http://dx.doi.org/10.1140/epjc/s10052-011-1756-5}{{\em Eur. Phys. J. C}
  {\bfseries 71} (2011) 1756}, \href{http://arxiv.org/abs/1101.3270}{{\ttfamily
  arXiv:1101.3270 [hep-lat]}}.

\bibitem{1211_3709}
E.~Seiler, D.~Sexty, and I.-O. Stamatescu, ``{Gauge cooling in complex Langevin
  for QCD with heavy quarks},''
  \href{http://dx.doi.org/10.1016/j.physletb.2013.04.062}{{\em Phys. Lett. B}
  {\bfseries 723} (2013) 213--216},
  \href{http://arxiv.org/abs/1211.3709}{{\ttfamily arXiv:1211.3709 [hep-lat]}}.

\bibitem{1508_02377}
K.~Nagata, J.~Nishimura, and S.~Shimasaki, ``{Justification of the complex
  Langevin method with the gauge cooling procedure},''
  \href{http://dx.doi.org/10.1093/ptep/ptv173}{{\em PTEP} {\bfseries 2016}
  no.~1, (2016) 013B01}, \href{http://arxiv.org/abs/1508.02377}{{\ttfamily
  arXiv:1508.02377 [hep-lat]}}.

\bibitem{1604_07717}
K.~Nagata, J.~Nishimura, and S.~Shimasaki, ``{Gauge cooling for the
  singular-drift problem in the complex Langevin method - a test in Random
  Matrix Theory for finite density QCD},''
  \href{http://dx.doi.org/10.1007/JHEP07(2016)073}{{\em JHEP} {\bfseries 07}
  (2016) 073}, \href{http://arxiv.org/abs/1604.07717}{{\ttfamily
  arXiv:1604.07717 [hep-lat]}}.

\bibitem{1606_07627}
K.~Nagata, J.~Nishimura, and S.~Shimasaki, ``{Argument for justification of the
  complex Langevin method and the condition for correct convergence},''
  \href{http://dx.doi.org/10.1103/PhysRevD.94.114515}{{\em Phys. Rev. D}
  {\bfseries 94} no.~11, (2016) 114515},
  \href{http://arxiv.org/abs/1606.07627}{{\ttfamily arXiv:1606.07627
  [hep-lat]}}.

\bibitem{deWit:1988wri}
B.~de~Wit, J.~Hoppe, and H.~Nicolai, ``{On the Quantum Mechanics of
  Supermembranes},'' \href{http://dx.doi.org/10.1016/0550-3213(88)90116-2}{{\em
  Nucl. Phys. B} {\bfseries 305} (1988) 545}.

\bibitem{Banks:1996vh}
T.~Banks, W.~Fischler, S.~H. Shenker, and L.~Susskind, ``{M theory as a matrix
  model: A Conjecture},''
  \href{http://dx.doi.org/10.1103/PhysRevD.55.5112}{{\em Phys. Rev. D}
  {\bfseries 55} (1997) 5112--5128},
  \href{http://arxiv.org/abs/hep-th/9610043}{{\ttfamily arXiv:hep-th/9610043}}.

\bibitem{tHooft:1973alw}
G.~'t~Hooft, ``{A Planar Diagram Theory for Strong Interactions},''
  \href{http://dx.doi.org/10.1016/0550-3213(74)90154-0}{{\em Nucl. Phys. B}
  {\bfseries 72} (1974) 461}.

\bibitem{Narayanan:2003fc}
R.~Narayanan and H.~Neuberger, ``{Large N reduction in continuum},''
  \href{http://dx.doi.org/10.1103/PhysRevLett.91.081601}{{\em Phys. Rev. Lett.}
  {\bfseries 91} (2003) 081601},
  \href{http://arxiv.org/abs/hep-lat/0303023}{{\ttfamily
  arXiv:hep-lat/0303023}}.

\bibitem{Aharony:2004ig}
O.~Aharony, J.~Marsano, S.~Minwalla, and T.~Wiseman, ``{Black hole-black string
  phase transitions in thermal 1+1 dimensional supersymmetric Yang-Mills theory
  on a circle},'' \href{http://dx.doi.org/10.1088/0264-9381/21/22/010}{{\em
  Class. Quant. Grav.} {\bfseries 21} (2004) 5169--5192},
\href{http://arxiv.org/abs/hep-th/0406210}{{\ttfamily arXiv:hep-th/0406210
  [hep-th]}}.

\bibitem{Aharony:2005ew}
O.~Aharony, J.~Marsano, S.~Minwalla, K.~Papadodimas, M.~Van~Raamsdonk, and
  T.~Wiseman, ``{The Phase structure of low dimensional large N gauge theories
  on Tori},'' \href{http://dx.doi.org/10.1088/1126-6708/2006/01/140}{{\em JHEP}
  {\bfseries 01} (2006) 140},
\href{http://arxiv.org/abs/hep-th/0508077}{{\ttfamily arXiv:hep-th/0508077
  [hep-th]}}.

\bibitem{Kawahara:2007fn}
N.~Kawahara, J.~Nishimura, and S.~Takeuchi, ``{Phase structure of matrix
  quantum mechanics at finite temperature},''
  \href{http://dx.doi.org/10.1088/1126-6708/2007/10/097}{{\em JHEP} {\bfseries
  10} (2007) 097},
\href{http://arxiv.org/abs/0706.3517}{{\ttfamily arXiv:0706.3517 [hep-th]}}.

\bibitem{Azeyanagi:2009zf}
T.~Azeyanagi, M.~Hanada, T.~Hirata, and H.~Shimada, ``{On the shape of a
  D-brane bound state and its topology change},''
  \href{http://dx.doi.org/10.1088/1126-6708/2009/03/121}{{\em JHEP} {\bfseries
  03} (2009) 121},
\href{http://arxiv.org/abs/0901.4073}{{\ttfamily arXiv:0901.4073 [hep-th]}}.

\bibitem{Azuma:2012uc}
T.~Azuma, T.~Morita, and S.~Takeuchi, ``{New States of Gauge Theories on a
  Circle},'' \href{http://dx.doi.org/10.1007/JHEP10(2012)059}{{\em JHEP}
  {\bfseries 10} (2012) 059}, \href{http://arxiv.org/abs/1207.3323}{{\ttfamily
  arXiv:1207.3323 [hep-th]}}.

\bibitem{Azuma:2014cfa}
T.~Azuma, T.~Morita, and S.~Takeuchi, ``{Hagedorn Instability in Dimensionally
  Reduced Large-N Gauge Theories as Gregory-Laflamme and Rayleigh-Plateau
  Instabilities},''
  \href{http://dx.doi.org/10.1103/PhysRevLett.113.091603}{{\em Phys. Rev.
  Lett.} {\bfseries 113} (2014) 091603},
\href{http://arxiv.org/abs/1403.7764}{{\ttfamily arXiv:1403.7764 [hep-th]}}.

\bibitem{Filev:2015hia}
V.~G. Filev and D.~O'Connor, ``{The BFSS model on the lattice},''
  \href{http://dx.doi.org/10.1007/JHEP05(2016)167}{{\em JHEP} {\bfseries 05}
  (2016) 167},
\href{http://arxiv.org/abs/1506.01366}{{\ttfamily arXiv:1506.01366 [hep-th]}}.

\bibitem{Hanada:2016qbz}
M.~Hanada and P.~Romatschke, ``{Lattice Simulations of 10d Yang-Mills
  toroidally compactified to 1d, 2d and 4d},''
  \href{http://dx.doi.org/10.1103/PhysRevD.96.094502}{{\em Phys. Rev.}
  {\bfseries D96} no.~9, (2017) 094502},
\href{http://arxiv.org/abs/1612.06395}{{\ttfamily arXiv:1612.06395 [hep-th]}}.

\bibitem{Bergner:2019rca}
G.~Bergner, N.~Bodendorfer, M.~Hanada, E.~Rinaldi, A.~Sch\"afer, and P.~Vranas,
  ``{Thermal phase transition in Yang-Mills matrix model},''
  \href{http://dx.doi.org/10.1007/JHEP01(2020)053}{{\em JHEP} {\bfseries 01}
  (2020) 053}, \href{http://arxiv.org/abs/1909.04592}{{\ttfamily
  arXiv:1909.04592 [hep-th]}}.

\bibitem{Asano:2020yry}
Y.~Asano, S.~Kov\'a\v{c}ik, and D.~O'Connor, ``{The Confining Transition in the
  Bosonic BMN Matrix Model},''
  \href{http://dx.doi.org/10.1007/JHEP06(2020)174}{{\em JHEP} {\bfseries 06}
  (2020) 174}, \href{http://arxiv.org/abs/2001.03749}{{\ttfamily
  arXiv:2001.03749 [hep-th]}}.

\bibitem{Watanabe:2020ufk}
H.~Watanabe, G.~Bergner, N.~Bodendorfer, S.~Shiba~Funai, M.~Hanada, E.~Rinaldi,
  A.~Sch\"afer, and P.~Vranas, ``{Partial deconfinement at strong coupling on
  the lattice},'' \href{http://dx.doi.org/10.1007/JHEP02(2021)004}{{\em JHEP}
  {\bfseries 02} (2021) 004}, \href{http://arxiv.org/abs/2005.04103}{{\ttfamily
  arXiv:2005.04103 [hep-th]}}.

\bibitem{Dhindsa:2022uqn}
N.~S. Dhindsa, R.~G. Jha, A.~Joseph, A.~Samlodia, and D.~Schaich,
  ``{Non-perturbative phase structure of the bosonic BMN matrix model},''
  \href{http://dx.doi.org/10.1007/JHEP05(2022)169}{{\em JHEP} {\bfseries 05}
  (2022) 169}, \href{http://arxiv.org/abs/2201.08791}{{\ttfamily
  arXiv:2201.08791 [hep-lat]}}.

\bibitem{Kabat:1999hp}
D.~N. Kabat and G.~Lifschytz, ``{Approximations for strongly coupled
  supersymmetric quantum mechanics},''
  \href{http://dx.doi.org/10.1016/S0550-3213(99)00818-4}{{\em Nucl. Phys.}
  {\bfseries B571} (2000) 419--456},
\href{http://arxiv.org/abs/hep-th/9910001}{{\ttfamily arXiv:hep-th/9910001
  [hep-th]}}.

\bibitem{Hashimoto:2019wmg}
K.~Hashimoto, Y.~Matsuo, and T.~Morita, ``{Nuclear states and spectra in
  holographic QCD},'' \href{http://dx.doi.org/10.1007/JHEP12(2019)001}{{\em
  JHEP} {\bfseries 12} (2019) 001},
\href{http://arxiv.org/abs/1902.07444}{{\ttfamily arXiv:1902.07444 [hep-th]}}.

\bibitem{Morita:2020liy}
T.~Morita and H.~Yoshida, ``{Critical Dimension and Negative Specific Heat in
  One-dimensional Large-N Reduced Models},''
  \href{http://dx.doi.org/10.1103/PhysRevD.101.106010}{{\em Phys. Rev. D}
  {\bfseries 101} no.~10, (2020) 106010},
  \href{http://arxiv.org/abs/2001.02109}{{\ttfamily arXiv:2001.02109
  [hep-th]}}.

\bibitem{Brahma:2022ifx}
S.~Brahma, R.~Brandenberger, and S.~Laliberte, ``{Spontaneous symmetry breaking
  in the BFSS model: Analytical results using the Gaussian expansion method},''
  \href{http://arxiv.org/abs/2209.01255}{{\ttfamily arXiv:2209.01255
  [hep-th]}}.

\bibitem{Hotta:1998en}
T.~Hotta, J.~Nishimura, and A.~Tsuchiya, ``{Dynamical aspects of large N
  reduced models},''
  \href{http://dx.doi.org/10.1016/S0550-3213(99)00056-5}{{\em Nucl. Phys.}
  {\bfseries B545} (1999) 543--575},
\href{http://arxiv.org/abs/hep-th/9811220}{{\ttfamily arXiv:hep-th/9811220
  [hep-th]}}.

\bibitem{Mandal:2009vz}
G.~Mandal, M.~Mahato, and T.~Morita, ``{Phases of one dimensional large N gauge
  theory in a 1/D expansion},''
  \href{http://dx.doi.org/10.1007/JHEP02(2010)034}{{\em JHEP} {\bfseries 02}
  (2010) 034},
\href{http://arxiv.org/abs/0910.4526}{{\ttfamily arXiv:0910.4526 [hep-th]}}.

\bibitem{Morita:2010vi}
T.~Morita, ``{Thermodynamics of Large N Gauge Theories with Chemical Potentials
  in a 1/D Expansion},'' \href{http://dx.doi.org/10.1007/JHEP08(2010)015}{{\em
  JHEP} {\bfseries 08} (2010) 015},
  \href{http://arxiv.org/abs/1005.2181}{{\ttfamily arXiv:1005.2181 [hep-th]}}.

\bibitem{Mandal:2011hb}
G.~Mandal and T.~Morita, ``{Phases of a two dimensional large N gauge theory on
  a torus},'' \href{http://dx.doi.org/10.1103/PhysRevD.84.085007}{{\em Phys.
  Rev.} {\bfseries D84} (2011) 085007},
\href{http://arxiv.org/abs/1103.1558}{{\ttfamily arXiv:1103.1558 [hep-th]}}.

\bibitem{Eguchi:1982nm}
T.~Eguchi and H.~Kawai, ``{Reduction of Dynamical Degrees of Freedom in the
  Large N Gauge Theory},''
\href{http://dx.doi.org/10.1103/PhysRevLett.48.1063}{{\em Phys. Rev. Lett.}
  {\bfseries 48} (1982) 1063}.

\bibitem{LUSCHER1983233}
M.~L\"{u}scher, ``Some analytic results concerning the mass spectrum of
  yang-mills gauge theories on a torus,''
  \href{http://dx.doi.org/https://doi.org/10.1016/0550-3213(83)90436-4}{{\em
  Nuclear Physics B} {\bfseries 219} no.~1, (1983) 233--261}.
  \url{https://www.sciencedirect.com/science/article/pii/0550321383904364}.

\bibitem{Chernodub:2020qah}
M.~N. Chernodub, ``{Inhomogeneous confining-deconfining phases in rotating
  plasmas},'' \href{http://dx.doi.org/10.1103/PhysRevD.103.054027}{{\em Phys.
  Rev. D} {\bfseries 103} no.~5, (2021) 054027},
  \href{http://arxiv.org/abs/2012.04924}{{\ttfamily arXiv:2012.04924
  [hep-ph]}}.

\bibitem{Yamamoto:2013zwa}
A.~Yamamoto and Y.~Hirono, ``{Lattice QCD in rotating frames},''
  \href{http://dx.doi.org/10.1103/PhysRevLett.111.081601}{{\em Phys. Rev.
  Lett.} {\bfseries 111} (2013) 081601},
  \href{http://arxiv.org/abs/1303.6292}{{\ttfamily arXiv:1303.6292 [hep-lat]}}.

\bibitem{Braguta:2021jgn}
V.~V. Braguta, A.~Y. Kotov, D.~D. Kuznedelev, and A.~A. Roenko, ``{Influence of
  relativistic rotation on the confinement-deconfinement transition in
  gluodynamics},'' \href{http://dx.doi.org/10.1103/PhysRevD.103.094515}{{\em
  Phys. Rev. D} {\bfseries 103} no.~9, (2021) 094515},
  \href{http://arxiv.org/abs/2102.05084}{{\ttfamily arXiv:2102.05084
  [hep-lat]}}.

\bibitem{Chen:2022smf}
S.~Chen, K.~Fukushima, and Y.~Shimada, ``{Perturbative Confinement in Thermal
  Yang-Mills Theories Induced by Imaginary Angular Velocity},''
  \href{http://dx.doi.org/10.1103/PhysRevLett.129.242002}{{\em Phys. Rev.
  Lett.} {\bfseries 129} no.~24, (2022) 242002},
  \href{http://arxiv.org/abs/2207.12665}{{\ttfamily arXiv:2207.12665
  [hep-ph]}}.

\bibitem{Chernodub:2022wsw}
M.~N. Chernodub, ``{Instantons in rotating finite-temperature Yang-Mills
  gas},'' \href{http://arxiv.org/abs/2208.04808}{{\ttfamily arXiv:2208.04808
  [hep-th]}}.

\bibitem{Chernodub:2022veq}
M.~N. Chernodub, V.~A. Goy, and A.~V. Molochkov, ``{Inhomogeneity of a rotating
  gluon plasma and the Tolman-Ehrenfest law in imaginary time: Lattice results
  for fast imaginary rotation},''
  \href{http://dx.doi.org/10.1103/PhysRevD.107.114502}{{\em Phys. Rev. D}
  {\bfseries 107} no.~11, (2023) 114502},
  \href{http://arxiv.org/abs/2209.15534}{{\ttfamily arXiv:2209.15534
  [hep-lat]}}.

\bibitem{STAR:2017ckg}
{\bfseries STAR} Collaboration, L.~Adamczyk {\em et~al.}, ``{Global $\Lambda$
  hyperon polarization in nuclear collisions: evidence for the most vortical
  fluid},'' \href{http://dx.doi.org/10.1038/nature23004}{{\em Nature}
  {\bfseries 548} (2017) 62--65},
  \href{http://arxiv.org/abs/1701.06657}{{\ttfamily arXiv:1701.06657
  [nucl-ex]}}.

\bibitem{Maldacena:1997re}
J.~M. Maldacena, ``{The Large N limit of superconformal field theories and
  supergravity},'' \href{http://dx.doi.org/10.4310/ATMP.1998.v2.n2.a1}{{\em
  Adv. Theor. Math. Phys.} {\bfseries 2} (1998) 231--252},
  \href{http://arxiv.org/abs/hep-th/9711200}{{\ttfamily arXiv:hep-th/9711200}}.

\bibitem{Itzhaki:1998dd}
N.~Itzhaki, J.~M. Maldacena, J.~Sonnenschein, and S.~Yankielowicz,
  ``{Supergravity and the large N limit of theories with sixteen
  supercharges},'' \href{http://dx.doi.org/10.1103/PhysRevD.58.046004}{{\em
  Phys. Rev.} {\bfseries D58} (1998) 046004},
\href{http://arxiv.org/abs/hep-th/9802042}{{\ttfamily arXiv:hep-th/9802042
  [hep-th]}}.

\bibitem{Gregory:1994bj}
R.~Gregory and R.~Laflamme, ``{The Instability of charged black strings and
  p-branes},'' \href{http://dx.doi.org/10.1016/0550-3213(94)90206-2}{{\em Nucl.
  Phys.} {\bfseries B428} (1994) 399--434},
\href{http://arxiv.org/abs/hep-th/9404071}{{\ttfamily arXiv:hep-th/9404071
  [hep-th]}}.

\bibitem{Mandal:2011ws}
G.~Mandal and T.~Morita, ``{Gregory-Laflamme as the confinement/deconfinement
  transition in holographic QCD},''
  \href{http://dx.doi.org/10.1007/JHEP09(2011)073}{{\em JHEP} {\bfseries 09}
  (2011) 073},
\href{http://arxiv.org/abs/1107.4048}{{\ttfamily arXiv:1107.4048 [hep-th]}}.

\bibitem{Klebanov:1997kv}
I.~R. Klebanov and L.~Susskind, ``{Schwarzschild black holes in various
  dimensions from matrix theory},''
  \href{http://dx.doi.org/10.1016/S0370-2693(97)01318-X}{{\em Phys. Lett. B}
  {\bfseries 416} (1998) 62--66},
  \href{http://arxiv.org/abs/hep-th/9709108}{{\ttfamily arXiv:hep-th/9709108}}.

\bibitem{Sundborg:1999ue}
B.~Sundborg, ``{The Hagedorn transition, deconfinement and N=4 SYM theory},''
  \href{http://dx.doi.org/10.1016/S0550-3213(00)00044-4}{{\em Nucl. Phys. B}
  {\bfseries 573} (2000) 349--363},
  \href{http://arxiv.org/abs/hep-th/9908001}{{\ttfamily arXiv:hep-th/9908001}}.

\bibitem{Aharony:2003sx}
O.~Aharony, J.~Marsano, S.~Minwalla, K.~Papadodimas, and M.~Van~Raamsdonk,
  ``{The Hagedorn - deconfinement phase transition in weakly coupled large N
  gauge theories},'' \href{http://dx.doi.org/10.4310/ATMP.2004.v8.n4.a1}{{\em
  Adv. Theor. Math. Phys.} {\bfseries 8} (2004) 603--696},
  \href{http://arxiv.org/abs/hep-th/0310285}{{\ttfamily arXiv:hep-th/0310285}}.

\bibitem{Witten:1998zw}
E.~Witten, ``{Anti-de Sitter space, thermal phase transition, and confinement
  in gauge theories},''
  \href{http://dx.doi.org/10.4310/ATMP.1998.v2.n3.a3}{{\em Adv. Theor. Math.
  Phys.} {\bfseries 2} (1998) 505--532},
  \href{http://arxiv.org/abs/hep-th/9803131}{{\ttfamily arXiv:hep-th/9803131}}.

\bibitem{Aharony:1999ti}
O.~Aharony, S.~S. Gubser, J.~M. Maldacena, H.~Ooguri, and Y.~Oz, ``{Large N
  field theories, string theory and gravity},''
  \href{http://dx.doi.org/10.1016/S0370-1573(99)00083-6}{{\em Phys. Rept.}
  {\bfseries 323} (2000) 183--386},
  \href{http://arxiv.org/abs/hep-th/9905111}{{\ttfamily arXiv:hep-th/9905111}}.

\bibitem{Hawking:1982dh}
S.~W. Hawking and D.~N. Page, ``{Thermodynamics of Black Holes in anti-De
  Sitter Space},'' \href{http://dx.doi.org/10.1007/BF01208266}{{\em Commun.
  Math. Phys.} {\bfseries 87} (1983) 577}.

\bibitem{Gubser:1998jb}
S.~S. Gubser, ``{Thermodynamics of spinning D3-branes},''
  \href{http://dx.doi.org/10.1016/S0550-3213(99)00194-7}{{\em Nucl. Phys. B}
  {\bfseries 551} (1999) 667--684},
  \href{http://arxiv.org/abs/hep-th/9810225}{{\ttfamily arXiv:hep-th/9810225}}.

\bibitem{Chamblin:1999tk}
A.~Chamblin, R.~Emparan, C.~V. Johnson, and R.~C. Myers, ``{Charged AdS black
  holes and catastrophic holography},''
  \href{http://dx.doi.org/10.1103/PhysRevD.60.064018}{{\em Phys. Rev. D}
  {\bfseries 60} (1999) 064018},
  \href{http://arxiv.org/abs/hep-th/9902170}{{\ttfamily arXiv:hep-th/9902170}}.

\bibitem{Cvetic:1999ne}
M.~Cvetic and S.~S. Gubser, ``{Phases of R charged black holes, spinning branes
  and strongly coupled gauge theories},''
  \href{http://dx.doi.org/10.1088/1126-6708/1999/04/024}{{\em JHEP} {\bfseries
  04} (1999) 024}, \href{http://arxiv.org/abs/hep-th/9902195}{{\ttfamily
  arXiv:hep-th/9902195}}.

\bibitem{Hawking:1999dp}
S.~W. Hawking and H.~S. Reall, ``{Charged and rotating AdS black holes and
  their CFT duals},'' \href{http://dx.doi.org/10.1103/PhysRevD.61.024014}{{\em
  Phys. Rev. D} {\bfseries 61} (2000) 024014},
  \href{http://arxiv.org/abs/hep-th/9908109}{{\ttfamily arXiv:hep-th/9908109}}.

\bibitem{Gubser:2000mm}
S.~S. Gubser and I.~Mitra, ``{The Evolution of unstable black holes in anti-de
  Sitter space},'' \href{http://dx.doi.org/10.1088/1126-6708/2001/08/018}{{\em
  JHEP} {\bfseries 08} (2001) 018},
  \href{http://arxiv.org/abs/hep-th/0011127}{{\ttfamily arXiv:hep-th/0011127}}.

\bibitem{Basu:2005pj}
P.~Basu and S.~R. Wadia, ``{R-charged AdS(5) black holes and large N unitary
  matrix models},'' \href{http://dx.doi.org/10.1103/PhysRevD.73.045022}{{\em
  Phys. Rev. D} {\bfseries 73} (2006) 045022},
  \href{http://arxiv.org/abs/hep-th/0506203}{{\ttfamily arXiv:hep-th/0506203}}.

\bibitem{Yamada:2006rx}
D.~Yamada and L.~G. Yaffe, ``{Phase diagram of N=4 super-Yang-Mills theory with
  R-symmetry chemical potentials},''
  \href{http://dx.doi.org/10.1088/1126-6708/2006/09/027}{{\em JHEP} {\bfseries
  09} (2006) 027}, \href{http://arxiv.org/abs/hep-th/0602074}{{\ttfamily
  arXiv:hep-th/0602074}}.

\bibitem{Sorkin:2004qq}
E.~Sorkin, ``{A Critical dimension in the black string phase transition},''
  \href{http://dx.doi.org/10.1103/PhysRevLett.93.031601}{{\em Phys. Rev. Lett.}
  {\bfseries 93} (2004) 031601},
\href{http://arxiv.org/abs/hep-th/0402216}{{\ttfamily arXiv:hep-th/0402216
  [hep-th]}}.

\bibitem{Kudoh:2005hf}
H.~Kudoh and U.~Miyamoto, ``{On non-uniform smeared black branes},''
  \href{http://dx.doi.org/10.1088/0264-9381/22/19/004}{{\em Class. Quant.
  Grav.} {\bfseries 22} (2005) 3853--3874},
\href{http://arxiv.org/abs/hep-th/0506019}{{\ttfamily arXiv:hep-th/0506019
  [hep-th]}}.

\bibitem{Cardoso:2006ks}
V.~Cardoso and O.~J.~C. Dias, ``{Rayleigh-Plateau and Gregory-Laflamme
  instabilities of black strings},''
  \href{http://dx.doi.org/10.1103/PhysRevLett.96.181601}{{\em Phys. Rev. Lett.}
  {\bfseries 96} (2006) 181601},
\href{http://arxiv.org/abs/hep-th/0602017}{{\ttfamily arXiv:hep-th/0602017
  [hep-th]}}.

\bibitem{Miyamoto:2008rd}
U.~Miyamoto and K.-i. Maeda, ``{Liquid Bridges and Black Strings in Higher
  Dimensions},'' \href{http://dx.doi.org/10.1016/j.physletb.2008.05.010}{{\em
  Phys. Lett.} {\bfseries B664} (2008) 103--106},
\href{http://arxiv.org/abs/0803.3037}{{\ttfamily arXiv:0803.3037 [hep-th]}}.

\bibitem{hep-th/0310286}
K.~Furuuchi, E.~Schreiber, and G.~W. Semenoff, ``{Five-brane thermodynamics
  from the matrix model},''
  \href{http://arxiv.org/abs/hep-th/0310286}{{\ttfamily arXiv:hep-th/0310286}}.

\bibitem{hep-th/0601170}
N.~Kawahara, J.~Nishimura, and K.~Yoshida, ``{Dynamical aspects of the
  plane-wave matrix model at finite temperature},''
  \href{http://dx.doi.org/10.1088/1126-6708/2006/06/052}{{\em JHEP} {\bfseries
  06} (2006) 052}, \href{http://arxiv.org/abs/hep-th/0601170}{{\ttfamily
  arXiv:hep-th/0601170}}.

\bibitem{0912_3360}
G.~Aarts, E.~Seiler, and I.-O. Stamatescu, ``{The Complex Langevin method: When
  can it be trusted?},''
  \href{http://dx.doi.org/10.1103/PhysRevD.81.054508}{{\em Phys. Rev. D}
  {\bfseries 81} (2010) 054508},
  \href{http://arxiv.org/abs/0912.3360}{{\ttfamily arXiv:0912.3360 [hep-lat]}}.

\bibitem{1802_10381}
P.~Basu, K.~Jaswin, and A.~Joseph, ``{Complex Langevin Dynamics in Large $N$
  Unitary Matrix Models},''
  \href{http://dx.doi.org/10.1103/PhysRevD.98.034501}{{\em Phys. Rev. D}
  {\bfseries 98} no.~3, (2018) 034501},
  \href{http://arxiv.org/abs/1802.10381}{{\ttfamily arXiv:1802.10381
  [hep-th]}}.

\bibitem{Stevenson:1981vj}
P.~M. Stevenson, ``{Optimized Perturbation Theory},''
\href{http://dx.doi.org/10.1103/PhysRevD.23.2916}{{\em Phys. Rev.} {\bfseries
  D23} (1981) 2916}.

\bibitem{Liu:2004vy}
H.~Liu, ``{Fine structure of Hagedorn transitions},''
  \href{http://arxiv.org/abs/hep-th/0408001}{{\ttfamily arXiv:hep-th/0408001}}.

\bibitem{AlvarezGaume:2005fv}
L.~Alvarez-Gaume, C.~Gomez, H.~Liu, and S.~Wadia, ``{Finite temperature
  effective action, AdS(5) black holes, and 1/N expansion},''
  \href{http://dx.doi.org/10.1103/PhysRevD.71.124023}{{\em Phys. Rev.}
  {\bfseries D71} (2005) 124023},
\href{http://arxiv.org/abs/hep-th/0502227}{{\ttfamily arXiv:hep-th/0502227
  [hep-th]}}.

\bibitem{Rossi:1996hs}
P.~Rossi, M.~Campostrini, and E.~Vicari, ``{The Large N expansion of unitary
  matrix models},'' \href{http://dx.doi.org/10.1016/S0370-1573(98)00003-9}{{\em
  Phys. Rept.} {\bfseries 302} (1998) 143--209},
  \href{http://arxiv.org/abs/hep-lat/9609003}{{\ttfamily
  arXiv:hep-lat/9609003}}.

\bibitem{Okuyama:2017pil}
K.~Okuyama, ``{Wilson loops in unitary matrix models at finite $N$},''
  \href{http://dx.doi.org/10.1007/JHEP07(2017)030}{{\em JHEP} {\bfseries 07}
  (2017) 030}, \href{http://arxiv.org/abs/1705.06542}{{\ttfamily
  arXiv:1705.06542 [hep-th]}}.

\bibitem{Gross:1980he}
D.~J. Gross and E.~Witten, ``{Possible Third Order Phase Transition in the
  Large N Lattice Gauge Theory},''
\href{http://dx.doi.org/10.1103/PhysRevD.21.446}{{\em Phys. Rev.} {\bfseries
  D21} (1980) 446--453}.

\bibitem{Wadia:2012fr}
S.~R. Wadia, ``{A Study of U(N) Lattice Gauge Theory in 2-dimensions},''
\href{http://arxiv.org/abs/1212.2906}{{\ttfamily arXiv:1212.2906 [hep-th]}}.

\bibitem{Alvarez-Gaume:2006fwd}
L.~Alvarez-Gaume, P.~Basu, M.~Marino, and S.~R. Wadia, ``{Blackhole/String
  Transition for the Small Schwarzschild Blackhole of AdS(5)x S**5 and Critical
  Unitary Matrix Models},''
  \href{http://dx.doi.org/10.1140/epjc/s10052-006-0049-x}{{\em Eur. Phys. J. C}
  {\bfseries 48} (2006) 647--665},
  \href{http://arxiv.org/abs/hep-th/0605041}{{\ttfamily arXiv:hep-th/0605041}}.

\end{thebibliography}\endgroup

\end{document}